\pdfoutput=1
\documentclass[cernpreprint, atlasdraft=false, UKenglish, texmf, orcidlogo]{atlasdoc}
 
\usepackage{atlaspackage}
\usepackage[block=none]{atlasbiblatex}
 
\usepackage{atlasphysics}
 
\graphicspath{{logos/}{figures/}}
 
\usepackage{ANA-HIGG-2018-49-PAPER-defs}
\usepackage{nicefrac}
\usepackage{adjustbox}

\AtlasTitle{Measurements of differential cross sections of Higgs boson production through gluon fusion in the $H\to WW^{\ast}\to e\nu\mu\nu$ final state at $\sqrt{s} = 13$~TeV with the ATLAS detector}
 
\AtlasVersion{2.0}

\AtlasAbstract{
Higgs boson production via gluon--gluon fusion is measured in the $WW^{\ast} \to e\nu\mu\nu$ decay channel. The dataset utilized corresponds to an integrated luminosity of 139\,fb$^{-1}$ collected by the ATLAS detector from $\sqrt{s}=13$\,TeV proton--proton collisions delivered by the Large Hadron Collider between 2015 and 2018. Differential cross sections are measured in a fiducial phase space restricted to the production of at most one additional jet. The results are consistent with Standard Model expectations, derived using different Monte Carlo generators.
}

\author{The ATLAS Collaboration}
\journal{EPJC}
 
\AtlasJournalRef{Eur. Phys. J. C 83 (2023) 774}
\AtlasDOI{10.1140/epjc/s10052-023-11873-5}
 
\AtlasRefCode{ANA-HIGG-2018-49}
\PreprintIdNumber{CERN-EP-2022-228}

\hypersetup{pdftitle={ATLAS document},pdfauthor={The ATLAS Collaboration}}
 
\begin{document}
 
\maketitle
 
\tableofcontents

\section{Introduction}
\label{sec:intro}
 
Since the discovery of the Higgs boson in 2012, by the ATLAS~\cite{HIGG-2012-27} and CMS~\cite{CMS-HIG-12-028}
collaborations, measurements of Higgs boson properties have remained
consistent with the expectations of the Standard Model (SM). The Higgs boson mass
has been determined, measurements of the Higgs coupling to fermions and gauge
bosons have been performed (see, e.g.~Refs.~\cite{2022,2022_cms}), and
differential cross sections have been measured by the ATLAS and CMS collaborations. Since the differential cross sections of the Higgs boson can be predicted with high precision, these measurements can be used to search for deviations from the Standard Model and probe the effect of higher-order corrections in perturbation theory.
 
This paper presents measurements of differential cross sections for Higgs boson production via gluon--gluon fusion (ggF) and decay into $WW^{\ast} \to e\nu\mu\nu$. Measuring the decay of the Higgs boson in the $WW$ channel is an important component of fully characterising Higgs measurements in all of its decay states. The measurements are reported in a fiducial phase space, which minimizes extrapolations and therefore the model dependence of the results. Measurements of Higgs boson decays into $WW$ bosons are well motivated because the large branching fraction makes the measurement competitive with the cleaner $\gamma \gamma$ and $ZZ$ channels. Although the presence of neutrinos in the $WW$ final states makes it impossible to fully reconstruct the final state kinematics, the high branching fraction makes the measurement competitive with fully leptonic final states.
In addition, the Higgs boson's $WW^{\ast} \to e\nu\mu\nu$ decay mode has an advantage when compared with the decay into $b$-quark pairs, since the Higgs boson produced via ggF is not required to be highly boosted, allowing a detailed study of the full Higgs transverse momentum spectrum. Finally, the $e\nu\mu\nu$ final state is considered in order to avoid large backgrounds from Drell--Yan production that appear in the same flavour channels.
 
The analysis is performed using the full \RunTwo dataset, corresponding to an integrated luminosity of 139~\ifb, collected with the ATLAS detector during 2015--2018 at a centre-of-mass energy $\sqrt{s}= 13\,\TeV$. Final states with at most one jet are considered. Final states with two jets are not considered in order to avoid large backgrounds from top production processes. The small expected contributions from Higgs production in the vector-boson fusion (VBF) and associated production (\textit{VH}) processes are fixed to their SM expectations and considered as background.  Previous ATLAS measurements of the differential cross section in this channel were performed on 20.3~\ifb{} of data collected at a centre-of-mass energy of $\sqrt{s} = 8\,\TeV$~\cite{HIGG-2015-04}, while the inclusive cross-section measurement in this channel was performed on 139~\ifb{} of $\sqrt{s} = 13\,\TeV$ data~\cite{ATLAS:2022ooq}. The CMS Collaboration published a measurement of the differential and inclusive cross sections in this channel using 137~\ifb{} of data collected at a centre-of-mass energy of $\sqrt{s} = 13\,\TeV$~\cite{CMS-HIG-19-002}. Compared to the previous ATLAS result, the present measurement introduces a refinement in that the signal in each interval of the observable under consideration is extracted by means of a fit to a kinematic variable. Compared to the previous results obtained by subtracting expected background contributions from the data, the current method, which employs a fitting procedure gives an improved sensitivity and allows for a more comprehensive treatment of systematic uncertainties. The ATLAS and CMS collaborations have also performed differential cross-section measurements for the cases where the Higgs boson decays into $ZZ \to 4\ell$ and $\gamma\gamma$~\cite{HIGG-2022-04,CMS-HIG-17-028}.

The measurements of the differential cross sections are corrected for detector effects to a fiducial phase space defined at the particle level, through a procedure referred to as unfolding. Defining the selection criteria at the particle level removes the dependence on the theoretical modelling and facilitates direct comparison with a wide variety of theoretical predictions. The measurements are almost insensitive to the uncertainties on the ggF SM predictions to which they are compared.

The theoretical prediction of the differential cross section as a function of the Higgs boson transverse momentum ($\ptH$) is calculated at next-to-next-to-leading order (NNLO) in quantum chromodynamics (QCD). In ggF production, the transverse momentum of the Higgs boson is balanced by soft quark and gluon emission, enabling the low-momentum region of this distribution to test the non-perturbative effects, the modelling of soft-gluon emissions calculated using resummation, and loop effects sensitive to the Higgs--charm Yukawa coupling. The low $\ptH$ region is also sensitive to the interactions of the charm and bottom quarks with the Higgs boson. The high $\ptH$ region is sensitive to perturbative QCD effects, as well as possible contributions from \enquote{beyond the Standard Model} (BSM)
physics, and higher-dimensional effective operators in BSM Lagrangians~\cite{BUCHMULLER1986621, Grzadkowski_2010}.
 
\newcommand{\AtlasCoordFootnote}{
ATLAS uses a right-handed coordinate system with its origin at the nominal interaction point (IP) in the centre of the detector and the \(z\)-axis along the beam pipe. The \(x\)-axis points from the IP to the centre of the LHC ring, and the \(y\)-axis points upwards. Cylindrical coordinates \((r,\phi)\) are used in the transverse plane, \(\phi\) being the azimuthal angle around the \(z\)-axis. The pseudorapidity is defined in terms of the polar angle \(\theta\) as \(\eta = -\ln \tan(\theta/2)\). Angular distance is measured in units of \(\Delta R \equiv \sqrt{(\Delta\eta)^{2} + (\Delta\phi)^{2}}\). Rapidity \(y\) is defined in terms of four-momentum as \(y = \tfrac{1}{2}\ln((E+p_{z})/(E-p_{z}))\).}
 
The production kinematics of the Higgs boson produced in gluon fusion in a $pp$ collision can be described by jet kinematic variables that include the number of jets and the rapidity
of the leading jet ($|y_{j0}|$).\footnote{\AtlasCoordFootnote}  The $|y_{j0}|$ distribution probes the theoretical modelling of hard gluon and quark emission. Measuring the cross section as a function of the number of jets provides information about the relative contributions of different Higgs production mechanisms. Processes with at most one jet are dominated by ggF and probe the higher-order QCD contributions to this production mechanism. In addition to measuring quantities sensitive to Higgs boson
production, observables sensitive to its $WW^{\ast} \to e\nu\mu\nu$ decay are also
explored. These include leptonic kinematic observables such as the leading lepton's transverse momentum ($\pt^{\ell0}$)
and the dilepton system's transverse momentum (\ptll), invariant mass (\mll), rapidity ($y_{\ell\ell}$), azimuthal opening angle (\dphill),
and also \costhetastar, defined as $\costhetastar =
|\tanh(\tfrac{1}{2}\Delta\eta_{\ell\ell})|$ (where $\Delta\eta_{\ell\ell}$ is the lepton pseudorapidity difference).
The observable $y_{\ell\ell}$ is highly correlated with the rapidity of the reconstructed Higgs boson ($y_{H}$),
which is known to be sensitive to the parton distribution function (PDF)~\cite{Bozzi_2008}.
The observable \costhetastar is longitudinally boost-invariant and
sensitive to the spin structure of the produced diparticle pairs, as
discussed in Ref.~\cite{sleptonspin}.
 
Finally, double-differential measurements are made, in which the Higgs production cross section is explored as functions of two variables simultaneously, with one sensitive to the production kinematics and the other to the decay kinematics. Specifically, the cross section is measured as a function of 0-jet and 1-jet bins versus $\pt^{\ell0}$, \ptll, \mll, \Yll,
\dphill, and \costhetastar.

\section{ATLAS detector}
\label{sec:detector}

The ATLAS detector~\cite{PERF-2007-01} is a multipurpose detector at
the Large Hadron Collider at CERN, and almost fully covers the solid angle around the collision point. The innermost part of the detector covers the pseudorapidity
range \(|\eta| < 2.5\) and is used to gather information about the
momentum and the sign of the charge of charged particles. This inner
detector (ID) consists of a silicon Pixel detector, a silicon microstrip
tracker, and a straw-tube transition-radiation
tracker. The Pixel detector includes a new innermost layer, the insertable B-layer~\cite{ATLAS-TDR-19,PIX-2018-001}, added before \RunTwo. A thin superconducting solenoid surrounds the ID,
providing it with a \SI{2}{\tesla} axial magnetic field. The inner detector is surrounded by
electromagnetic and hadronic calorimeters. The electromagnetic (EM)
calorimeter, consisting of lead absorbers and liquid argon (LAr), provides
electromagnetic and hadronic energy measurements. It consists of a
barrel component in the centre of the detector, covering \( |\eta|< 1.475\), and
two EM endcaps covering \(1.375 <
|\eta| < 3.2 \). In addition, copper--LAr hadronic endcaps are placed
behind the EM endcaps to capture hadronic energy in the
\(1.5 < |\eta| < 3.2\) region.
 
The EM calorimeter is surrounded by the steel and
scintillator hadronic Tile calorimeter (TileCal), which measures the
energy of hadrons. The Tile calorimeter consists of a central barrel covering
\(|\eta| < 1.0\), and two extended barrels in the forward regions, covering \(
0.8 < |\eta| < 1.7\). The regions closest to the beam pipe are covered
by forward calorimeters (FCal), which cover \( 3.1< |\eta| <4.9 \)
and consist of three modules. 
The first module is made of copper--LAr and is used for
electromagnetic measurements and the beginning of the hadronic shower. The other two modules consist of
tungsten and LAr and complete the hadronic measurements.
 
The calorimeters are surrounded by a muon spectrometer (MS) with superconducting toroidal magnets
made of eight air-core barrel loops and two
endcaps. The toroid magnets produce a non-uniform magnetic field, resulting in a bending power ranging
from \num{2.0} to \SI{6.0}{\tesla\metre}. The MS
has a high-precision tracking chamber system covering \(|\eta| <2.7\) and consisting mainly of monitored
drift tubes but complemented by cathode-strip chambers in the forward directions, where the background is highest.
The MS uses resistive-plate chambers in the barrel and thin-gap chambers in the endcaps for triggering
within \(|\eta| <2.4\) and to provide track coordinates in
the non-bending plane.
 
A two-level trigger system is used to select events of interest.
The first-level trigger is implemented in hardware and uses a subset of the detector information to
accept events from the \SI{40}{\MHz} proton bunch crossings at a rate below \SI{100}{\kHz}.
This is followed by a software-based trigger, which reduces the accepted event rate
to \SI{1}{\kHz} on average depending on the data-taking conditions.
 
An exhaustive software suite~\cite{ATL-SOFT-PUB-2021-001} is used in data simulation, in the reconstruction and analysis of real and simulated data, in detector operations, and in the trigger and data acquisition systems of the experiment.

\section{Signal and background modelling}
\label{sec:SigBg}
 
The signal and background modelling is described using various
generators to model the hard-scatter process, parton shower (PS),
hadronization and underlying event (UE). The Higgs boson can be
produced through ggF processes, vector-boson fusion (VBF), in
association with a $W$ or $Z$ boson (\textit{VH})
or with a top-quark pair ($t\bar{t}H$).
There are other production modes as well, but those are not
included in this analysis as their expected contributions are negligible. These include Higgs bosons produced with a
bottom-quark pair ($b\bar{b}H$) or a single top quark ($tH$,
with either an additional $W$ boson, $tWH$,
or an additional $b$-quark and light quark, $tHqb$). In the combined 0-jet and 1-jet signal regions (see Section~\ref{EvntSel}), the non-ggF production modes make up 6.9\% of the total expected Higgs boson event yield, with 60\% of this contribution coming from VBF production.
 
The primary ggF signal process was modelled using \POWHEG~NNLOPS~\cite{Hamilton:2013fea,Hamilton:2015nsa,Alioli:2010xd,Nason:2004rx,Frixione:2007vw} interfaced with \Pythia[8]~\cite{pythia8}.  The 0-jet, 1-jet and 2-jet cross sections are predicted at NNLO, NLO and LO, respectively. To obtain NNLO accuracy for observable distributions, the Higgs boson rapidity spectrum in  \textsc{Hj}-\MINLO~\cite{Hamilton:2012np,Campbell:2012am,Hamilton:2012rf} was reweighted to that of HNNLO~\cite{Catani:2007vq}. The resulting Higgs boson transverse momentum spectrum is compatible with the fixed-order HNNLO calculation and the \HRES[2.3] calculation~\cite{Bozzi:2005wk,deFlorian:2011xf}. The \AZNLO tune~\cite{STDM-2012-23} of the parameter values was used for the PS, UE and hadronization. The ggF process is normalized to the N$^{3}$LO cross section in QCD with NLO electroweak corrections~\cite{deFlorian:2016spz,Anastasiou:2016cez,Anastasiou:2015ema,Dulat:2018rbf,Harlander:2009mq,Harlander:2009bw,Harlander:2009my,Pak:2009dg,Actis:2008ug,Actis:2008ts,Bonetti:2018ukf,Bonetti:2018ukf}. For the results, a comparison is made between the ggF process generated by \POWHEG and an alternative generation performed by \MGNLO~\cite{Alwall:2014hca} with the \NNPDF[2.3lo]~\cite{Ball:2012cx} PDF set.
 
The VBF events were generated with \POWHEG,
interfaced with \Pythia[8] and normalized to cross sections from NLO QCD
with EW corrections, with an approximate NNLO QCD correction
applied~\cite{Ciccolini:2007jr,Ciccolini:2007ec,Bolzoni:2010xr}.  The \textit{VH} production process was generated with \POWHEG \MiNLO, and interfaced with \Pythia[8]. The \textit{VH} cross section is calculated at NNLO in QCD with NLO electroweak corrections for $q\bar{q}/qg \to VH$ and at NLO and next-to-leading-logarithm accuracy in QCD for $gg \to ZH$~\cite{Ciccolini:2003jy,Brein:2003wg,Brein:2011vx,Altenkamp:2012sx,Denner:2014cla,Brein:2012ne,Harlander:2014wda}.
 
For the ggF, VBF and \textit{VH} predictions,
the \PDFforLHC[15nnlo]~\cite{Butterworth:2015oua} PDF set was used. All of the Higgs boson cross sections and branching fractions are normalized in accordance with \HDECAY~\cite{Djouadi:1997yw,Spira:1997dg,Djouadi:2006bz} and \PROPHECY~\cite{Bredenstein:2006ha,Bredenstein:2006rh,Bredenstein:2006nk} calculations, which provide NLO EW corrections and account for interference effects. The samples were generated assuming a Higgs boson mass of 125~\GeV{}. The signal predictions described are not an extensive list of all available Monte Carlo simulations, and were chosen because of their applicability to this phase space.
 
The main sources of SM background that could resemble the signal
process are the production of top quarks, dibosons, \textit{Z}+jets, \textit{W}+jets and multijets. The top-quark pair production process was simulated using \POWHEG in the \POWHEGBOX framework using the \NNPDF[3.0nlo] PDFs, and interfaced with \Pythia[8] using \NNPDF[2.3] PDFs for parton showering. The associated production of top quarks with $W$ bosons, $Wt$, was modelled using the \POWHEGBOX[v2] generator at NLO in QCD, using the five-flavour scheme and the \NNPDF[3.0nlo] set of PDFs. 
The diagram removal scheme~\cite{Frixione:2008yi} was used to remove interference between $Wt$ and \ttbar production.  The $t\bar{t}$ process is normalized using cross sections calculated at NNLO+NNLL \cite{Czakon:2013}, while the $Wt$ process is normalized to NNLO. The alternative top-quark samples used to estimate certain systematic uncertainties were produced using \MGFiveNLO~\cite{Alwall:2014hca}.
 
The $WW$ background sample was generated using \SHERPA[2.2.2] interfaced with \NNPDF[3.0nnlo] PDFs~\cite{CampbellEllis2010}. The sample was generated with up to one jet at NLO accuracy and with two or three jets at LO accuracy. The loop-induced $gg$-initiated diboson processes were simulated by \SHERPA[2.2.2] with up to one additional jet, interfaced with \SHERPA[2.2.2] for the UE and PS, and used the \NNPDF[3.0nnlo] PDF set.
 
The $WZ$ and $ZZ$ samples were generated with \SHERPA[2.2.2], using the \NNPDF[3.0nnlo] PDF set, at NLO accuracy for up to one jet and at LO accuracy for two or three jets.  Events containing three vector bosons, $VVV$,  were simulated with the \SHERPA[2.2.2] generator, using the \NNPDF[3.0nnlo] PDF set. The matrix elements are accurate to NLO for the inclusive process and to LO for up to two additional parton emissions.
 
Drell--Yan production was simulated with \SHERPA[2.2.1] using the \NNPDF[3.0nnlo] PDFs with dedicated parton shower tuning developed by the \Sherpa authors.
 
Production of $Z\gamma$ and $W\gamma$ events was modelled using \SHERPA[2.2.8] at NLO accuracy for up to one jet, using the \NNPDF[3.0nnlo] PDFs. They were interfaced with \SHERPA[2.2.8] for the UE and PS.
 
The $W$+jets process modelling is based on a purely data-driven method. However, MC samples are used to validate the estimation as well as to estimate the sample composition uncertainties. The $V$+jets processes simulated in \POWHEG \MiNLO interfaced with \Pythia[8], are used for the validation of the $W$+jets modelling. The PDF set used in \POWHEG was \CT[14nnlo] whereas the PDF set used in the parton shower was \CTEQ[6L1lo]. The alternative $V$+jets samples
used to estimate certain systematic uncertainties were produced using \MGNLO.
 
For the background processes, the AZNLO tune~\cite{STDM-2012-23} is used for the diboson processes, while the A14 tune~\cite{ATL-PHYS-PUB-2014-021} is used for other processes.

The full list of Monte Carlo
generators involved in the estimates used for this analysis is shown
in Table~\ref{tab:mcsamples}.

\begin{table}[tbp]
\centering
\caption{
Overview of simulation tools used to generate signal and background processes, as well as to model the UE and PS. The PDF sets are also summarized.
Alternative event generators or quantities varied to estimate systematic uncertainties as well as
alternative predictions used in the final-results comparison are shown in parentheses.}
\label{tab:mcsamples}
\scalebox{0.68}{
\begin{tabular}{l l l l l}
\hline\hline
Process              & Matrix element                                              & PDF set                 & UE and PS model                                           & Prediction order\\
& (alternative)                                               &                         & (alternative model)                                  &  for total cross section\\
\hline
ggF $H$              & \POWHEGBOX[v2]~\cite{Hamilton:2013fea,Hamilton:2015nsa,Alioli:2010xd,Nason:2004rx,Frixione:2007vw}
& \multirow{2}{*}{\PDFforLHC[15nnlo]~\cite{Butterworth:2015oua}} &\multirow{2}{*}{\PYTHIA[8]~\cite{Sjostrand:2014zea}} & \multirow{2}{*}{N$^{3}$LO QCD + NLO EW~\cite{deFlorian:2016spz,Anastasiou:2016cez,Anastasiou:2015ema,Dulat:2018rbf,Harlander:2009mq,Harlander:2009bw,Harlander:2009my,Pak:2009dg,Actis:2008ug,Actis:2008ts,Bonetti:2018ukf}} \\
& NNLOPS~\cite{Nason:2009ai,Hamilton:2013fea,Campbell:2012am} &                         &    & \\
& (\MGFiveNLO)~\cite{Alwall:2014hca,Frederix:2012ps}          &                         & (\Herwig[7])~\cite{Bellm:2015jjp} & \\
VBF $H$              & \POWHEGBOX[v2]~\cite{Nason:2009ai,Alioli:2010xd,Nason:2004rx,Frixione:2007vw}
& \PDFforLHC[15nlo]  & \PYTHIA[8]        & NNLO QCD + NLO EW~\cite{Ciccolini:2007jr,Ciccolini:2007ec,Bolzoni:2010xr} \\
& (\MGFiveNLO)                                                &                         & (\Herwig[7])                                        & \\
$VH$ excl.\ $gg\to ZH$ & \POWHEGBOX[v2]                                            & \PDFforLHC[15nlo]  & \PYTHIA[8]  & NNLO QCD + NLO EW~\cite{Ciccolini:2003jy,Brein:2003wg,Brein:2011vx,Denner:2014cla,Brein:2012ne} \\
\ttH                 & \POWHEGBOX[v2]                                             & \NNPDF[3.0nlo]    & \PYTHIA[8]               & NLO~\cite{deFlorian:2016spz} \\
$gg\to ZH$           & \POWHEGBOX[v2]                                             & \NNPDF[3.0nlo]  & \PYTHIA[8]               & NLO+NLL~\cite{Altenkamp:2012sx,Harlander:2014wda} \\
 
$qq \to WW$          & \SHERPA[2.2.2]~\cite{Bothmann:2019yzt}                     & \NNPDF[3.0nnlo]~\cite{Ball:2014uwa} & \SHERPA[2.2.2]~\cite{Gleisberg:2008fv,Schumann:2007mg,Hoeche:2011fd,Hoeche:2012yf,Catani:2001cc,Hoeche:2009rj} & NLO~\cite{Buccioni:2019sur,Cascioli:2011va,Denner:2016kdg} \\
& ($Q_\text{cut}$)                                            &                         & (\SHERPA[2.2.2]~\cite{Schumann:2007mg,Hoeche:2009xc}; \muQ)  \\
$qq \to WWqq$        & \MGFiveNLO~\cite{Alwall:2014hca}                             & \NNPDF[3.0nlo]    & \PYTHIA[8]                                      & LO \\
&                                                             &                         & (\Herwig[7])                                        & \\
$gg \to WW/ZZ$         & \SHERPA[2.2.2]                                             & \NNPDF[3.0nnlo]   & \SHERPA[2.2.2]                                      & LO~\cite{Caola:2015rqy}  \\
$WZ/V\gamma^{\ast}/ZZ$ & \SHERPA[2.2.2]                                             & \NNPDF[3.0nnlo]   & \SHERPA[2.2.2]                                      & NLO~\cite{Cascioli:2013gfa} \\
$V\gamma$            & \SHERPA[2.2.8]~\cite{Bothmann:2019yzt}                     & \NNPDF[3.0nnlo]   &\SHERPA[2.2.8]                                       & NLO~\cite{Cascioli:2013gfa} \\
$VVV$                  & \SHERPA[2.2.2]                                             & \NNPDF[3.0nnlo]   &\SHERPA[2.2.2]                                       & NLO  \\
$t\bar{t}$           & \POWHEGBOX[v2]                                             & \NNPDF[3.0nlo]    & \PYTHIA[8]                                          & NNLO+NNLL~\cite{Beneke:2011mq,Cacciari:2011hy,Baernreuther:2012ws,Czakon:2012zr,Czakon:2012pz,Czakon:2013goa,Czakon:2011xx} \\  
& (\MGFiveNLO)                                                &                         & (\Herwig[7])                                        & \\
$Wt$                  &\POWHEGBOX[v2]                                              & \NNPDF[3.0nlo]    & \PYTHIA[8]                         & NNLO~\cite{Kidonakis:2010ux,Kidonakis:2013zqa} \\
& (\MGFiveNLO)                                                &                         & (\Herwig[7])                                          & \\
$Z/\gamma^{\ast}$    & \SHERPA[2.2.1]                                             & \NNPDF[3.0nnlo]   & \SHERPA[2.2.1]                                      & NNLO~\cite{Anastasiou:2003ds} \\
& (\MGFiveNLO)                                                & \\
\hline\hline
\end{tabular}
}
\end{table}
All simulated samples include the effect of pile-up from multiple interactions in the same and neighbouring bunch crossings. This was achieved by overlaying each simulated hard-scatter event with minimum-bias events, simulated using \Pythia[8] with the A3 tune. The Monte Carlo (MC) events were weighted to reproduce the
distribution of the average number of interactions per bunch crossing (\(\left<\mu \right>\)) observed in the data. The \(\left<\mu \right>\) value in data was rescaled by a factor of \(1.03\pm 0.04\) to improve agreement between data and simulation in the visible inelastic proton--proton (\(pp\)) cross-section~\cite{STDM-2015-05}.

All samples were processed through the ATLAS detector simulation~\cite{SOFT-2010-01} based on \GEANT~\cite{Geant4}, and reconstructed with the standard ATLAS reconstruction software. Data are removed if relevant detector components were not fully operational, or if other data integrity issues were identified, as described in Ref.~\cite{DAPR-2018-01}.

\section{Event selection}
\label{EvntSel}
 
Events are selected using a combination of unprescaled single-lepton triggers and one $e$--$\mu$ dilepton trigger~\cite{TRIG-2018-05,TRIG-2018-01}. During the first year of data taking the single-electron (single-muon) trigger threshold was 24 (20)~\GeV, and was increased to 26~\GeV\ for the remainder of \RunTwo. The dilepton trigger had $\pt$ thresholds of 17~\GeV\ for electrons and 14~\GeV\ for muons.
At least one of the leptons reconstructed offline is required to match the online object that was triggered on. If only the dilepton trigger was used, each reconstructed lepton is required to match the triggered object. The $\pt$ of the reconstructed leptons that are matched to the trigger are required to be at least 1~\GeV above the trigger-level threshold.

The analysis utilizes both the objects identified after the full detector reconstruction and objects identified at the particle level in simulation. The particle-level and reconstructed objects are selected in two stages. An overlap removal procedure is applied to the objects resulting from the first selection stage, to ensure that no object is counted twice. After the second selection stage, the objects are used to construct fiducial, signal and control regions.

\subsection{Reconstructed objects}
\label{ObjReco}
Reconstruction-level objects are built from the signals left in the detector by traversing particles. The overlap removal procedure performed on reconstructed objects after the first
selection stage is complex, and is fully described in Ref.~\cite{ATLAS:2022ooq}. For reconstructed leptons, lepton identification is used in the signal selection and a separate lepton identification is used, with some selection criteria reversed, for the estimation of backgrounds with misidentified leptons.

Electrons are reconstructed as clusters of energy deposited in the EM calorimeter and associated tracks in the inner detector.
At the first selection stage, electrons are required to have a transverse momentum of $\pt > 10$~\GeV and $|\eta| < 2.47$. The longitudinal impact parameters must satisfy, $|z_0\sin\theta| < 0.5$~mm, while the significance of the transverse impact parameters must satisfy $|d_0|/\sigma_{d_0} < 5$ \cite{PERF-2017-01}. A likelihood-based electron selection
at the \enquote{Very Loose} operating point is used~\cite{PERF-2017-01}.
At the second selection stage, electrons are required to have  $\pt > 15$~\GeV, and to be outside the transition region between the
barrel and endcaps of the EM calorimeter, $1.37 < |\eta| < 1.52$.
Electrons with $15\, < \pt < 25\,\GeV{}$ are required to satisfy a \enquote{Tight}
operating point, while electrons with $\pt > 25$~\GeV{} must satisfy a
\enquote{Medium} operating point. Lower-energy electrons are required to satisfy a tighter operating point since they are more likely to originate from misconstructed objects.
To reduce the number of misidentified electrons, these electrons must
be well isolated from other objects, using the strictest isolation criteria as described in Ref.~\cite{EGAM-2018-01}, which employ both calorimeter and tracking information. The approximate efficiency of identifying and isolating leading and subleading electrons in the phase space of this analysis is is 81\% and 58\%, respectively
 
Muons are reconstructed using tracking information from the inner detector and the muon
spectrometer. The muon identification is defined by
making two sets of requirements on the number of detector hits, the momentum
resolution, the track goodness-of-fit, and other parameters~\cite{MUON-2018-03}. At the first selection stage, muons with $\pt > 10$~\GeV{} and $|\eta| <2.7$ are considered if they are identified with the \enquote{Loose} quality
working point~\cite{PERF-2015-10}.  These muons must satisfy $|z_0\sin\theta| < 1.5$~mm and $|d_0|/\sigma_{d_0} < 3$ requirements.
At the second selection stage, muon track segments
in the inner detector and muon spectrometer are required to match, as
defined by the \enquote{Tight} working point, which maximizes the muon purity.
Additionally, muons selected at the second stage are required to
have $\pt > 15$~\GeV{}
and pseudorapidity $|\eta| < 2.5$. These muons must also satisfy the
strictest isolation requirement~\cite{MUON-2018-03}, which employ both calorimeter and tracking information and entail that the muons be well
isolated from other objects in order to differentiate them from muons
produced in background processes containing semi-leptonic decays of heavy-flavour hadrons.  Finally, the muons selected at the second stage must satisfy $|z_0\sin\theta| < 0.5$~mm and $|d_0|/\sigma_{d_0} < 3$ requirements. The approximate efficiency of identifying and isolating leading and subleading muons in the phase space of this particular analysis is 85\% and 72\%, respectively.
 
Jets are reconstructed using the \antikt algorithm~\cite{Cacciari:2008gp}
with a radius parameter $R= 0.4$, and corrections are
applied to calibrate the jets to the particle-level energy scale, as
described in Ref.~\cite{JETM-2018-05}. Jets are then calibrated using the particle-flow algorithm as described in Ref.~\cite{PERF-2015-09}. At the first selection stage, jets are required to
have $\pt > 20$~\GeV, be located within
$|\eta| < 4.5$, and satisfy the \enquote{jet vertex tagger} (JVT) requirements
described in Ref.~\cite{PERF-2014-03}. The JVT requirements are applied to jets in the range $20 < \pt < 60$~\GeV and $|\eta| < 2.4$ in order to reduce the number of jets from pile-up. At the second selection stage, jets are required to have $\pt>30$~\GeV. Jets containing $b$-hadrons are identified using the
DL1r $b$-tagging algorithm and are required to satisfy the 85\% efficiency working point; they must also satisfy $\pt > 20$~\GeV{} and be located within $|\eta| < 2.5$~\cite{ATL-PHYS-PUB-2017-013}.
 
The missing transverse momentum, defined as the negative vector sum of the transverse
momenta of objects in the detector, measures the sum of the momenta of the neutrinos in the $WW^{\ast} \to e\nu\mu\nu$ decays. Two types
of missing transverse momentum are used in the analysis. The first,
denoted by $\met$, is described in Ref.~\cite{PERF-2016-07} and
accounts for the inclusion of electrically neutral objects by using
jets directly in the calculation of $\met$. This \met{} definition is used in the calculation of all sensitive
variables because it provides superior resolution for objects used
to discriminate signal events from background events.  The second is a
track-based variable, denoted by $\trkmet$~\cite{PERF-2016-07}, which uses tracks
matched to jets instead of the calorimeter jet energy and is used to suppress backgrounds from misidentified objects because it has
better pile-up robustness than $\met$. Both types of missing transverse momentum
use a \enquote{Tight} working point~\cite{PERF-2016-07}, which is equivalent to the condition that forward jets must satisfy the $\pt > 30$~\GeV\ requirement. The estimation of the lower energy components is based on tracks for both $\met$ definitions.
 
\subsection{Signal region selection}
\label{SigRegSel}
 
The signal region (SR) is defined using reconstructed objects at the second
selection stage, and subdivided into categories of events that contain
either no jets or one jet. The two categories share a common pre-selection and signal selection,
but have partially different background rejection selections applied to
reflect the variation in background composition between the jet
multiplicity bins.
 
The signal region contains events with exactly two opposite-sign different-flavour
leptons, with the leading (subleading)
lepton required to satisfy $\pt > 22$~\GeV{} ($\pt >
15$~\GeV{}). The invariant mass of the two leptons, $\mll$, is
required to be greater than 10~\GeV{} in order to remove low-mass
meson resonances and Drell--Yan events. To ensure
orthogonality with the \VH{} ($WH$, \ZH) phase space, events with
additional reconstructed leptons with $\pt$ greater than 15~\GeV{} are
removed. In order to suppress misreconstructed low-energy events, the
\trkmet is required to be greater than 20~\GeV{}. Events that
contain a $b$-tagged jet with $\pt > 20$~\GeV{} are removed
to suppress backgrounds from top-quark processes. Additionally,
a selection requiring a minimum value of 80~\GeV{} is applied to the transverse mass
(\mt), defined as:
\begin{equation*}
m_{\mathrm{T}} = \sqrt{ {(E_\text{T}^{\ell\ell} + E_{\mathrm{T}}^{\text{miss}})}^2 - {|\vec{p_\text{T}^{\ell\ell}} + \vec{E_{\mathrm{T}}^{\text{miss}}}|}^2 }
\end{equation*}
where $E_\text{T}^{\ell\ell} = \sqrt{|\vec{p_\text{T}^{\ell\ell}}|^2 + m_{\ell\ell}^2 }$ . This requirement suppresses non-$WW$ diboson background, as well as background from misidentified objects.
 
For the signal region category containing zero jets, a $\pt$ threshold of 30~\GeV{} is used to define the jet-counting criterion, meaning that the region has zero jets with $\pt>30$~\GeV{}. The azimuthal angle between the dilepton system and the
missing transverse momentum, $\dphillMET$, is required to be larger than
$\pi/2$. Since the leptons and $\met$ are typically back to back for
signal events, this requirement removes potential pathological events
in which the $\MET$ points in the direction of the lepton pair
because of misidentified objects. The transverse momentum of the dilepton
system, $\pTll$, is required to be at least 30~\GeV{} in order to
reject events in which a $Z$ boson decays into leptonically decaying $\tau$-leptons.
 
For the signal region category containing exactly one jet that satisfies $\pt> 30$~\GeV{}, the larger of the two lepton transverse masses, each defined as $m_{\mathrm{T}}^{\ell} = \sqrt{ 2 \pt^\ell \MET (1 - \cos ({\phi}_\ell - {\phi}_{\MET}))} $, is required to exceed 50~\GeV{}. Processes with at least one real $W$ boson typically
have a large value of $\mTl$ for at least one of the two leptons, so a
lower bound on this variable suppresses $\Zgtt$ and multi-jet
events. The 1-jet region also requires the di-$\tau$ candidate invariant mass,
$\mtautau$, as computed using the collinear approximation~\cite{Plehn_2000}, to be less than $m_Z -
25$~\GeV{} in order to remove $\Zgtt$ background events.\footnote{$m_Z$ refers to the mass of the $Z$ boson, i.e.\ 91.18~\GeV}
 
The signal region, divided into 0-jet and 1-jet categories, exploits the lepton kinematics to enhance a signal similar to that predicted by the SM.
The azimuthal opening angle between the leptons, \DPhill, is
required to be less than 1.8~radians because the charged leptons in leptonic
$\hwwlvlv$ tend to be more collimated than those from non-resonant
$WW$ background. This is due to the spin-zero initial state of the resonant
process and the $V-A$ structure of the weak interaction. Both the 0-jet and 1-jet categories also require that the dilepton invariant
mass, $\mll$, be less than 55~\GeV{}.  Table~\ref{tab:ggfselection}
summarizes all the requirements imposed on the signal region.
The overall acceptance of the signal region, as defined by the number of events in the signal region, divided by the number of events passing the pre-selection, is 28$\%$.
 
\begin{table}[t!]
\centering
\caption{
Event selection criteria used to define the signal and fiducial region in the analysis. The $\met$ definition described in Section~\ref{TruthObj} is used for the selection at particle level.
The reconstructed electrons are required to have a
pseudorapidity $|\eta| < 2.47$, excluding the transition region
between the barrel and endcaps of the EM
calorimeter, $1.37 < |\eta| < 1.52$.
}
\label{tab:ggfselection}
\begin{tabular}{c||c|c}
\hline\hline
Category          & $\NjetThirty=0$ & $\NjetThirty=1$ \\
\hline\hline
Pre-Selection     &
\multicolumn{2}{c}{
\begin{tabular}{c}
Exactly two isolated leptons ($\ell\,{=}\,e, \mu$) with opposite charge\\
$\pTlead\,{>}\,22\,\GeV$ , $\pTsublead\,{>}\,15\,\GeV$ \\
$|\eta_{e}| < 2.5$, $|\eta_{\mu}| < 2.5$, $\pT^\text{jet} >30\,\GeV$\\
$\mll\,{>}\,10\,\GeV$ \\
$\trkmet\,{>}\,20\,\GeV$       \\
\end{tabular}
}\\
\hline\hline
Background rejection    & \multicolumn{2}{c}{$\Nbjetsub=0$}     \\
& $\dphillMET > \pi/2$    &  $\max(\mT^{\ell})\,{>}\,50\,\GeV$        \\
& $\pTll\,{>}\,30\,\GeV$                  &  $\mtt\,{<}\,m_Z - 25\,\GeV$\\
& \multicolumn{2}{c}{$\mT\,{>}\,80\,\GeV$}     \\
\hline\hline
\multirow{2}{*}{\!\!\!\!\!
\begin{tabular}{l}
$\hwwlnln$ \\
topology
\end{tabular}
}
& \multicolumn{2}{c}{$\mll\,{<}\,55\,\GeV$}  \\
& \multicolumn{2}{c}{$\dphill\,{<}\,1.8$} \\
\hline\hline
\end{tabular}
\end{table}

\subsection{\enquote{Truth} objects}
\label{TruthObj}
 
The particle-level objects are simulated before particles
traverse the detector, and can be found in the signal samples after
the parton shower generation step. Collinear photons, for which the angular separation between the photon and particle-level lepton is $\Delta R < 0.1$, are added to the four-momentum of the lepton. The electrons and muons with
$|\eta|<2.7$ and transverse momentum higher than 10~\GeV{}, and jets
with $|\eta| < 4.5$ and transverse momentum higher than 20~\GeV{}, are selected at the first selection stage.
 
Simulated particle-level jets are built from all stable particles with $c\tau > 10$~mm,
including neutrinos, photons, and leptons from hadron decays or produced in the shower.
However, all decay products from the Higgs boson decay and the leptonic decays of
associated vector bosons are removed from the inputs to the jet algorithm.
The jets are reconstructed using the \antikt algorithm~\cite{Cacciari:2008gp}
with a radius parameter $R = 0.4$ implemented in the FastJet package~\cite{Fastjet}.
After this first selection stage, an overlap removal procedure is applied to objects that are near each other and mimic the SR selection.
A particle-level electron is rejected if it
is in the vicinity of a muon with $\deltaR(e,\mu) < 0.1$, or a jet
with $0.2 < \deltaR (e,j) < 0.4$. Conversely, a jet is rejected if
there is an electron within $\deltaR(e,j) < 0.2$. A muon is rejected
if there is a jet within $\deltaR(\mu,j) < 0.4$.
The MC \enquote{truth}-level $\met$ arises from neutrinos, and is calculated from preselected objects, as well as jets that are not required to satisfy the pre-selection criteria. At the second selection stage, further selection criteria are applied to the particle-level objects. A higher jet-$\pt$ requirement of 30~\GeV{} is applied, while electrons and muons are required to satisfy $\pt > 15$~\GeV{} and be
within $|\eta| <2.5$. The particle-level objects selected at the second stage are used in the definition of the fiducial regions, constructed to mimic the signal region as closely as possible.

\subsection{Fiducial region selection}
\label{subsec:FidReg}
The fiducial region (FR) selection is defined using particle-level objects,
before the detector reconstruction is applied. This selection is
designed to be as close as possible to the signal region in order to
avoid large extrapolation uncertainties. To obtain the fiducial region,
all the selections shown in Table~\ref{tab:ggfselection} are applied at
the particle level. Additionally, the absolute rapidity of the dilepton system, $|\Yll|$, is required to be less than 2.5, in order to
match the constraints due to the coverage of the ATLAS detector. As shown in Table~\ref{tab:ggfselection}, the fiducial phase space at particle level includes a veto on jets containing $b$-quarks. If a jet is matched with a weakly decaying $b$-hadron, it is labelled as a $b$-jet. 
The overall correction factor, defined as the number of reconstructed signal events in the SR divided by the number of signal events in the fiducial region, is 35\%. The number of events that pass into both the reconstructed signal region and fiducial region is 2250, the number of events that pass the reconstructed signal region and do not pass into the fiducial region is 352, the number of events that do not pass into the reconstructed signal region and pass into the fiducial region is 5117, and the number of events that fail to reach either region is 144109.

\section{Background estimation}
The contributions from SM processes that mimic the signal production are
estimated using a variety of methods.  One method involves defining a
normalization factor by comparing data with MC events in regions rich
in background, known as control regions (CRs).  The normalization factor is
a free parameter that is obtained when the fitting procedure (described in Section~\ref{sec:unfolding}) is applied
simultaneously to all signal and control regions. The background in the signal
regions is then scaled by the normalization factor, ensuring the
background modelling is normalized to the real data. The control
regions are close to the signal regions, in order to minimize the
extrapolation uncertainties, but have one or two selection criteria
reversed in order to ensure a high purity of background
events and orthogonality with respect to the signal regions. Backgrounds estimated from control regions include those containing two $W$ bosons, which closely resemble the signal and include
top-quark production (\ttbar{} and \Wt) and $WW$ production. A control
region is also used to estimate the backgrounds from the $\Zgtt$ process, which resembles the signal when the $\tau$-leptons decay leptonically.
The selections used to define the control regions are shown in
Table~\ref{tab:CRselection} for the 0-jet and 1-jet categories, and Table~\ref{tab:NFs_results01jcouplings} shows the corresponding normalization factors.
 
The presence of jets originating from $b$-quarks is used to identify processes containing top quarks.
The 0-jet top control region contains a requirement that $\dphill\,{<}\,2.8$ in order to remain alligned with the signal region, while retaining high statistics. The 1-jet top control region
removes events with low-\pt\ $b$-jets ($\Nbjetbetween$) in order
to ensure that the ratio of \ttbar{} to \Wt{} is similar to that in the signal
region and thus minimize the influence of secondary shape differences. The purity of the 0-jet (1-jet) top control region, defined as the ratio of the top-quark component to the total expectation, is 89\% (98\%).
 
\begin{table*}[t!]
\centering
\caption{
Event selection criteria used to define the control regions in the analysis.
The pre-selection requirements were applied in all six control regions.
The definitions of $\mtt$, $\mT^{\ell}$, and the jet-counting $\pT$ thresholds are the same as for the signal regions.
The selection criteria that are reversed or different in the CR
in order to make it orthogonal to the SR are marked in bold.
}
\label{tab:CRselection}
\begin{tabular}{c|| c | c }
\hline\hline
CR                         & $\NjetThirty=0$  & $\NjetThirty=1$ \\
\hline\hline
\multirow{5}{*}{$qq\rightarrow WW$}      & \multicolumn{2}{c}{$\Nbjetsub=0$} \\
& $\dphillMET > \pi/2$  & \textbf{$\mll\,{>}\,80~\GeV$} \\
& $\pTll\,{>}\,30~\GeV$         & \textbf{$|\mtt-m_Z|\,{>}\,25~\GeV$} \\
& \textbf{$55\,{<}\,\mll\,{<}\,110~\GeV$}   & $\maxmtlep\,{>}\,50~\GeV$  \\
&   \textbf{$\dphill\,{<}\,2.6$}  & \\
\hline\hline
\multirow{5}{*}{$t\bar{t}$/$Wt$} & \multirow{2}{*}{\textbf{$\Nbjetbetween>0$}}    & \textbf{$\Nbjetnom = 1$}  \\
&                                       & $\Nbjetbetween  = 0$    \\
& $\dphillMET > \pi/2$                  & $\mtt\,{<}\,m_Z - 25~\GeV$ \\
& $\pTll\, > \, 30~\GeV$                & $\maxmtlep\,{>}\,50~\GeV$   \\
& \textbf{$\dphill\,{<}\,2.8$}                   &     \\
\hline\hline
\multirow{5}{*}{$Z/\gamma^{\ast}\rightarrow\tau\tau$}    &                        \multicolumn{2}{c}{\phantom{ccccc}$\Nbjetsub=0$}   \\
& \multicolumn{2}{c}{\phantom{ccccc}\textbf{$\mll < 80~\GeV$}} \\
& \multicolumn{2}{c}{\textbf{no $\trkmet$ requirement}} \\
& \textbf{$\dphill > 2.8$}        & \textbf{$\mtt > m_Z - 25~\GeV$}\\
& &   $\maxmtlep\,{>}\,50~\GeV$ \\
\hline\hline
\end{tabular}
\end{table*}
 
In the $WW$ and $\Zgtt$ control regions the requirement that no $b$-jets with $\pt > 20\,\GeV$
be present is applied in order to reduce the top-quark contamination in
those regions.
The defining feature of regions containing $WW$
background is a selection on the
dilepton invariant mass ($\mll$), as shown in Table~\ref{tab:CRselection}.
The 0-jet $WW$ control region has a requirement that $\dphill\,{<}\,2.8$ in order to keep the region closely alligned to the signal region, while maximizing statistics. The $WW$ control region
containing one jet also reduces the $\Zgtt$ contamination by using the
same $\mtt$ and $\maxmtl$ selections that are applied in the signal
region. The $\Zgtt$ control regions are defined by requiring leptons
from the $\tau$ decay to be well separated ($\dphill\,{>}\,2.8$), and the
invariant mass of the leptons to be low ($\mll < 80\,\GeV{}$) to
eliminate contamination from $\Zeemm$ events. In addition, the $\Zgtt$
regions have no $\trkmet$ requirement. The purity of the 0-jet (1-jet) $WW$ CR is 67\% (34\%). The lower purity in the 1-jet region is acceptable because the ratio of $WW$ events to top-quark events in the control region is close to that in the SR. The purity of the $\Zgtt$ region is 94$\%$ and 76$\%$ in the 0-jet and 1-jet regions respectively. Table~\ref{tab:NFs_results01jcouplings} shows the normalization factors extracted from each control region after the fitting procedure is applied to the $\ptH$ distribution.

\begin{figure}
\includegraphics[width=.33\textwidth]{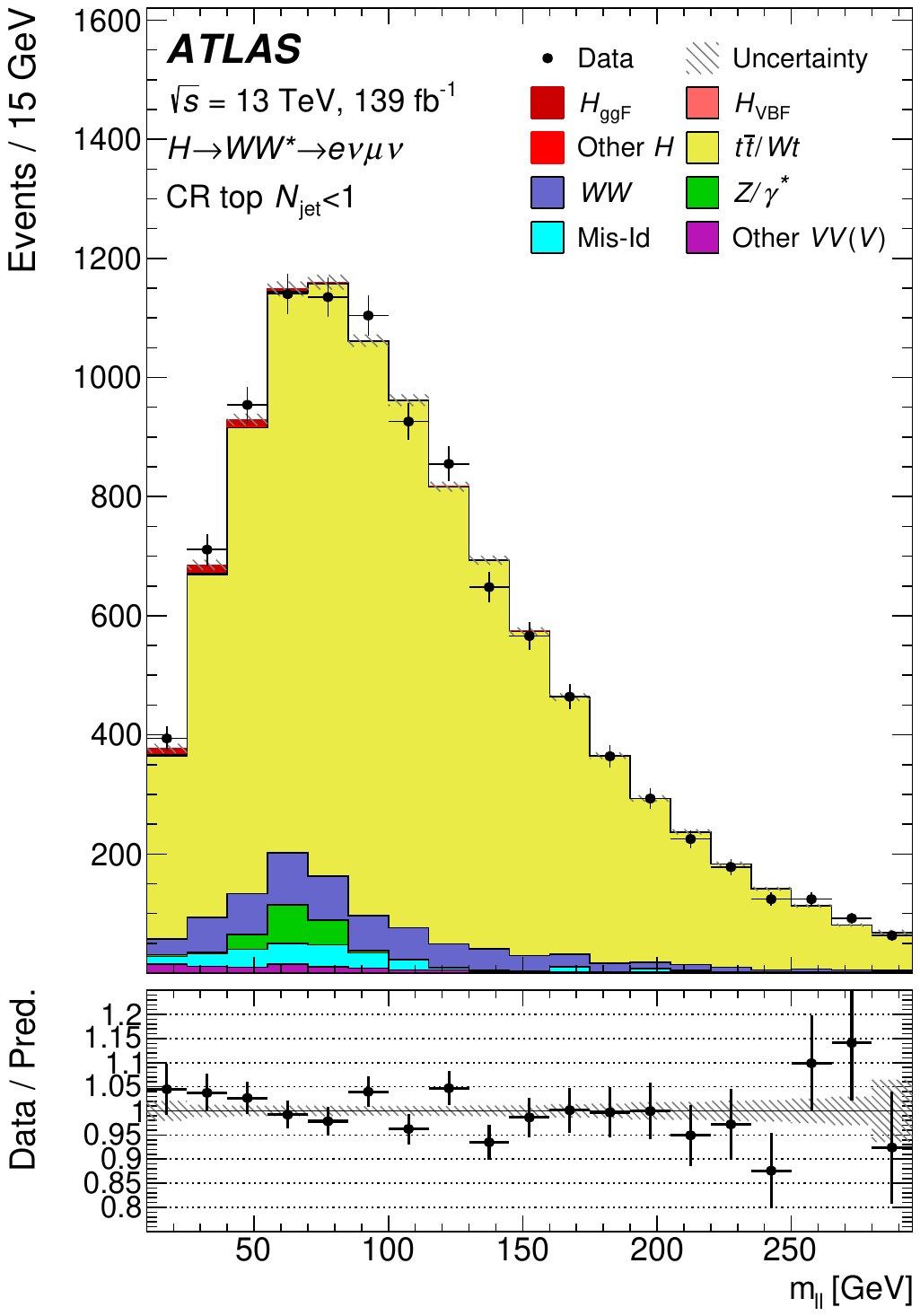}
\includegraphics[width=.33\textwidth]{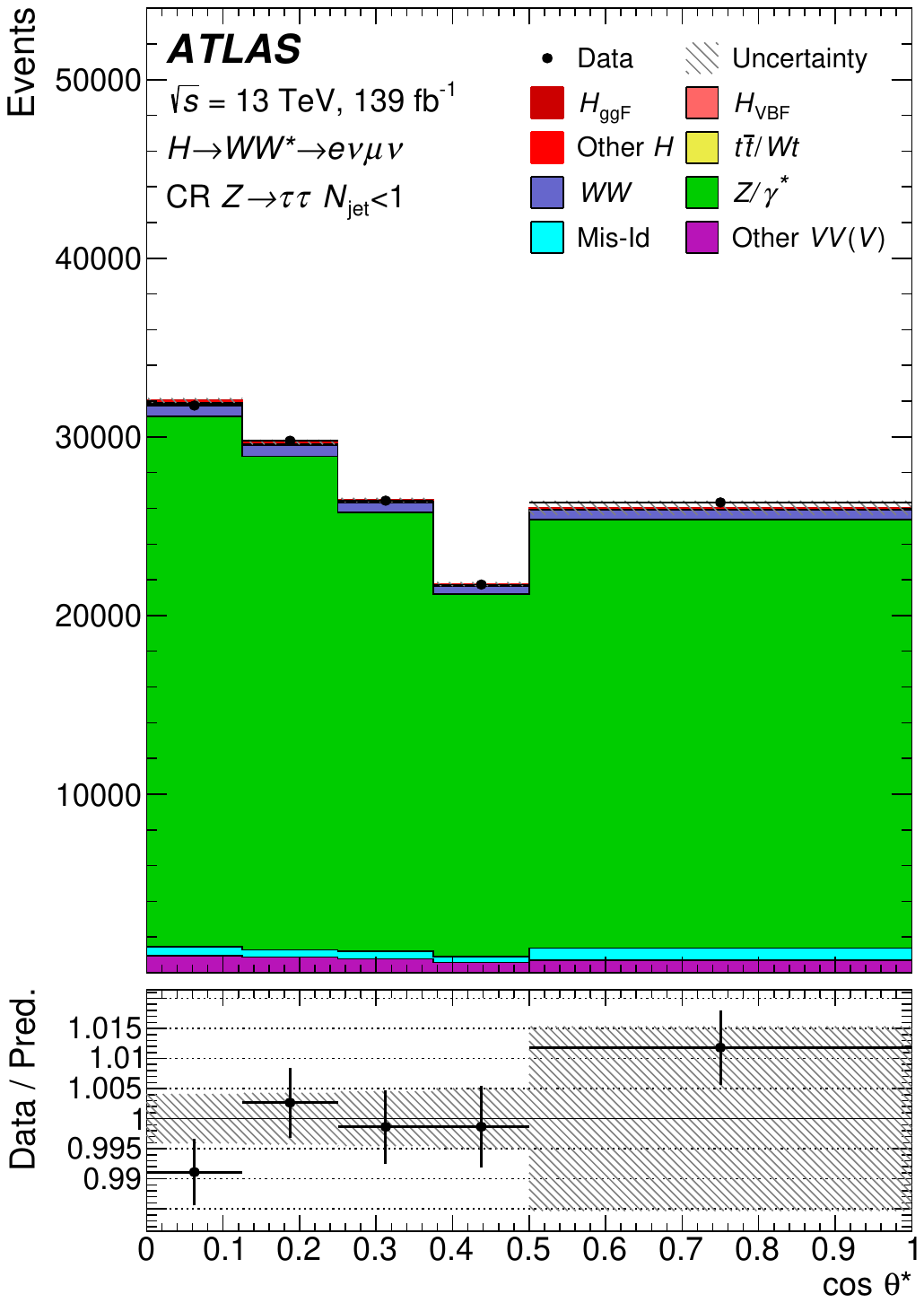}
\includegraphics[width=.33\textwidth]{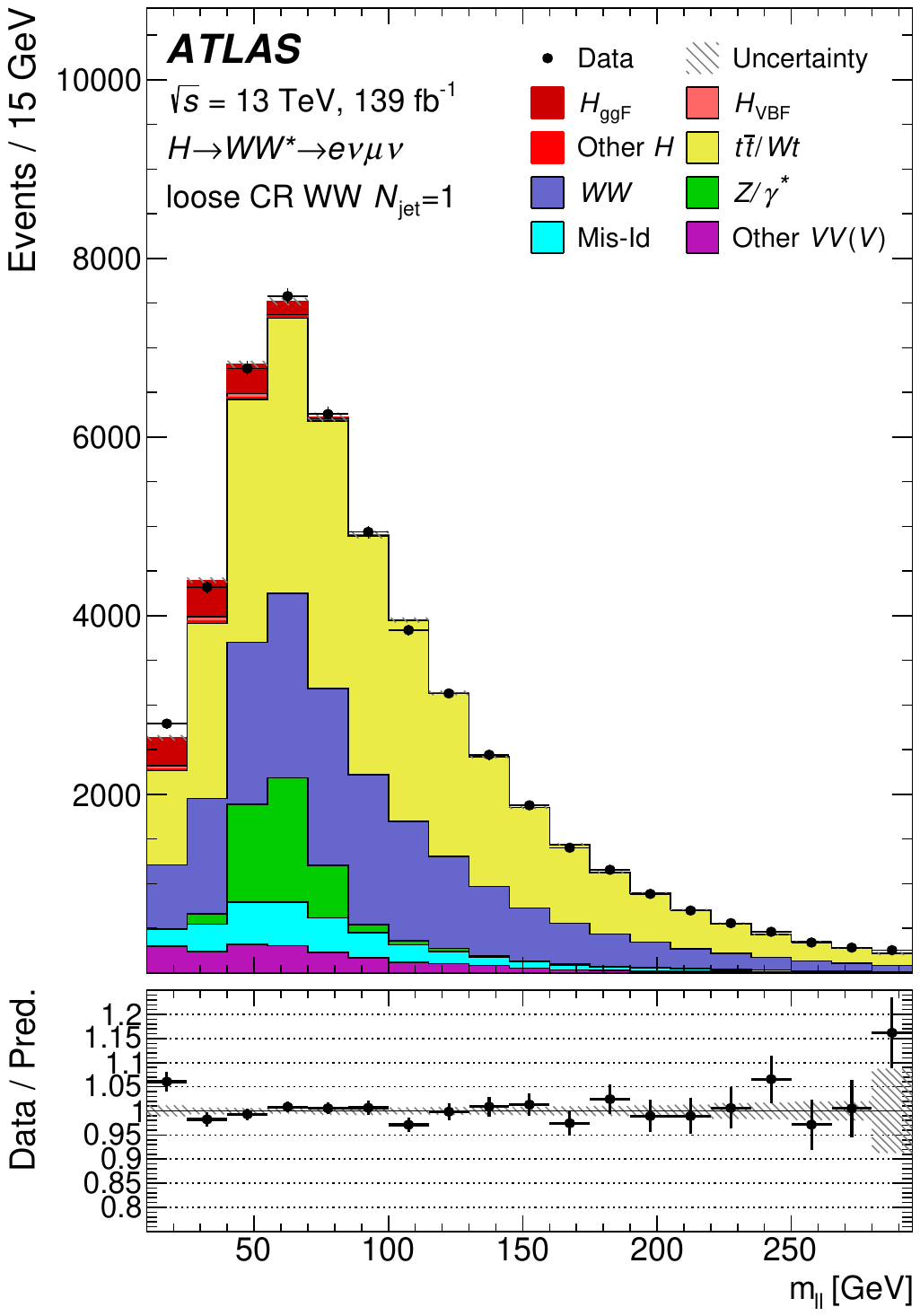}
\caption{
Distributions of $\mll$ in the top control region of the 0-jet channel (left), $\costhetastar$ in the $\Zgtt$
control regions of the 0-jet channel (centre), and $\mll$ in a loose
region including both the $WW$ control region and the signal region in
the 1-jet channel (right). The MC backgrounds are normalized by normalization factors obtained from signal plus background fits to the observable that is being plotted.  Model uncertainties shown are obtained from the fit result using a reweighting technique.
}
\label{fig:crplots}
\end{figure}
 
Distributions of $\mll$ and $\costhetastar$ in the top and $\Zgtt$
control regions of the 0-jet channel are shown in Figure~\ref{fig:crplots}, alongside the distribution of $\mll$ in a loose
region including both the $WW$ control region and the signal region in
the 1-jet channel. The normalization factors obtained from the fitting procedure are applied to distributions in the figure.

\begin{table*}[bt!]
\centering
\caption{
Normalization factors for all background contributions constrained by control regions,
as determined in the fit (described in Section~\ref{sec:unfolding}) to the $\ptH$ observable.
}
\label{tab:NFs_results01jcouplings}
\begin{tabular}{lr}
\hline\hline
Background     & Normalization factor \tabularnewline
\hline
$qqWW$ \ZeroJet  & $0.97 \pm 0.07$ \\
$qqWW$ \OneJet   & $0.91 \pm 0.13$ \\
$Z+$jets \ZeroJet& $0.91 \pm 0.07$ \\
$Z+$jets \OneJet & $1.02 \pm 0.12$ \\
Top \ZeroJet     & $1.07 \pm 0.24$ \\
Top \OneJet      & $1.03 \pm 0.18$\\
\hline\hline
\end{tabular}
\end{table*}


Backgrounds from processes such as $\Wjets$ and multi-jet production have large cross
sections, and can be mistaken for signal events in the rare cases
where a jet is misidentified as an isolated lepton.  These
backgrounds are estimated using a data-driven method known as the
extrapolation-factor method, and are labelled as \enquote{Mis-Id} in the figures. This method is described in detail in
Ref.~\cite{ATLAS:2022ooq}.
 
The remaining backgrounds include $WZ$, $W\gamma$, $ZZ$ and $VVV$, have small
cross sections, and are estimated from simulation. Finally, Higgs boson
production via processes such as vector-boson fusion, and in association with vector bosons or top-quark pairs, are fixed to their SM predictions and treated as background. This is because these additional production mechanisms have a negligible impact on the final results when treated as part of the signal model.
 
\section{Measured yields, uncertainties and distributions}

To obtain differential fiducial cross sections, the reconstruction-level
signal and background expectations are fitted to the data, and the resulting
information is corrected for detector effects to the particle level. A signal region is defined for each bin of each observable, from which the number of signal events $N_\text{S}$ is extracted from a fit to the data. A combined fit is performed to the $\mt$ distribution
in the range 80 to 160~\GeV\ in bins of 10~\GeV. This combined fit includes single-bin $WW$, top, and
$\Zgtt$ control regions. Normalization factors used to normalize the backgrounds in the signal regions are extracted from the combined fit. The fit is performed for each slice of the measured observables. Normalization factors are compatible across all observables.
 
The fit is performed using the profile likelihood method (see, e.g.,
Ref.~\cite{kendall}), where the likelihood is constructed as the
product of likelihoods from individual signal and control regions, as
well as the product of all bins in those regions. All background
estimates, both the data-driven estimate and those obtained from MC simulation,
are added to the Poisson expectations. The systematic uncertainties are included as nuisance parameters, constrained by a Gaussian term.
 
The analysis is affected by theoretical uncertainties in
the MC modelling of the signal and background, as well as experimental
uncertainties related to the detector measurements. The signal and
background expectations depend on nuisance parameters, which
correspond to systematic uncertainties.
The ggF signal uncertainties are a small contribution since the measurement is made in a fiducial region, making it independent of the type of signal model assumed. Uncertainties in the shape of the ggF signal distribution are minimized since the measurement is made in individual bins of kinematic variables. The uncertainties on the ggF cross-section predictions do not impact the measurements. The only remaining signal uncertainties are the  uncertainties that affect the detector level corrections. An example of such an uncertainty is the jet shower shape uncertainty, since changes in how jets are reconstructed can affect truth-level and reconstruction-level quantities differently. These residual uncertainties are taken into account. The uncertainties in non-ggF Higgs production processes are treated in the same way as the uncertainties in the other backgrounds, and cross-section uncertainties are accounted for.

Various uncertainties related to detector measurements are considered. Uncertainties associated with identifying, reconstructing and calibrating objects
such as electrons, muons, jets and missing transverse momentum, as well as tagging jets as originating from $b$-quarks are evaluated.
The systematic uncertainty associated with the energy/momentum scale and
resolution of objects is estimated by performing a scaling or smearing of the energy/momentum,
and observing its effect on the yields and shape of the distribution of selected events in the final state.
The systematic uncertainty is evaluated by comparing the $\pm 1\sigma$
variations with the nominal prediction. The missing transverse momentum
is constructed from various objects, as described in Section~\ref{ObjReco}, and has
uncertainties associated with those objects.   Uncertainties in the measured efficiencies for electron and muon triggers and pile-up are included as systematic uncertainties. The uncertainty in the combined 2015--2018 integrated luminosity is 1.7\% \cite{ATLAS-CONF-2019-021}, obtained using the LUCID-2 detector \cite{LUCID2} for the primary luminosity measurements.

The theoretical uncertainties in the MC modelling take into account
QCD scale uncertainties, parton shower (PS) and underlying event (UE)
modelling, and parton distribution function (PDF) uncertainties. To calculate the QCD scale uncertainties, the
renormalization (\muR) and factorization (\muF) scales are varied in order to
estimate the impact of missing higher-order corrections in fixed-order
perturbative predictions. To estimate the QCD scale uncertainty for
the Higgs boson signal, the Stewart--Tackmann method~\cite{stmethod} is
followed.
For the backgrounds originating from $WW$ bosons, top production, and $\Zgtt$, the scale uncertainty is evaluated using the envelope of seven-point \muR, \muF variations, with \muR and \muF each varying by factors of 0.5 and 2, subject to the constraint that  $0.5 \leq \muR/\muF \leq 2.0$, and the maximum variation of the event yields is taken as the uncertainty.
 
The PDF uncertainties are calculated following the standard PDF4LHC prescriptions ~\cite{PDF4LHC15}.
 
The uncertainties related to the UE and PS, as well as generator uncertainties, are estimated using alternative generators, and taken into account for the $WW$, top and $\Zgtt$ backgrounds. In addition to that, for $WW$ and top backgrounds, the uncertainties associated with
matching NLO matrix elements to parton showers are estimated. For the
$WW$ background, an uncertainty is estimated to account for the fact
that 0--1 additional jets are generated at NLO, while 2--4 additional
jets are generated at LO. An additional uncertainty applied to
single-top processes is estimated by comparing samples using different diagram removal schemes
to account for interference between $Wt$ and top-pair
production~\cite{Frixione:2008yi}.
 
A pruning procedure is used in order to improve the stability of the
likelihood maximization by neglecting sources of small
uncertainties. For this, any systematic uncertainty that induces a yield change
of less than 0.1\% for a given process in an observable bin is
neglected for that bin and that process. Furthermore, for any
systematic uncertainty which is very likely to be compatible with a pure normalization effect across
$\mT$ bins (with a $p$-value $p>95\%$), the shape component is neglected.
 
\section{Unfolding}
\label{sec:unfolding}
 
The reconstruction-level measurements are extrapolated to their particle-level
quantities using a procedure known as unfolding \cite{unfolding}. This
procedure uses Monte Carlo signal samples to define the correlation
between reconstruction-level and particle-level quantities.  These correlations
are described in terms of a response matrix $C$, such that
\begin{align*}
\vec{n}_{\textrm{reco}} = C \cdot \vec{n}_{\textrm{particle-level}} + \vec{n}_{\textrm{out-of-fiducial}},
\end{align*}
where $\vec{n}_{\textrm{reco}}$ is the vector of event counts in bins
of the reconstruction-level quantity, $\vec{n}_{\textrm{particle-level}}$ is the
vector of event counts in bins of the particle-level quantity, and
$\vec{n}_{\textrm{out-of-fiducial}}$ is the vector of signal events that pass the selection
at the reconstruction level but are not inside the fiducial region defined at the particle level.
In practice, the response matrix is derived from the
migration matrix, a two-dimensional distribution of simulated events
encoding the correlation between the particle- and reconstruction-level
quantities, by normalizing each element to the total number of events in
the corresponding particle-level bin. In this way, acceptance, resolution and efficiency
effects are encoded in the matrix.  Assuming that
the response derived from simulation is representative of the detector
response in real data, the particle-level distributions in data can be
obtained by applying the inverse of the response matrix to the measured data.
 
The practical application of this procedure faces a challenge related to the lack of regularization: the bin
migrations between particle and reconstruction levels induce
off-diagonal entries in the response matrix, leading to negative off-diagonal
entries in its inverse and ultimately to the amplification of
uncertainties.
 
Two different approaches to alleviate this are presented here. The first approach is an unfolding procedure that uses a Tikhonov-regularized \cite{tikhonov} in-likelihood
unfolding algorithm. The likelihood used extracts the reconstruction-level signal yields by means of signal strengths $\mu_{j}$ multiplying the expected event yields $s^{r,\text{exp}}_{j,m}$ in $m_{\mathrm{T}}$ bin $m$ and observable bin $j$, and is of the form
\begin{equation*}
\begin{aligned}
\mathcal{L}(\vec{\mu}, \vec{\theta}) &=
\displaystyle\prod_{j=1}^{N_{\text{bin}}}\prod_{m=1}^{N_{m_{\mathrm{T}}}} P(N_{j}| \mu_{j}s_{j,m}^{r,\text{exp}}(\vec{\theta}) + \displaystyle\sum_{n=1}^{N_{\text{bg}}}\mu^{b}_{n}b_{j,m,n}(\vec{\theta}))\\
&\qquad \times \{ \displaystyle\prod_{c=1}^{N_{\text{CR}}} P(N_{c}|s_{c}^{r,\text{exp}}(\vec{\theta}) + \displaystyle\sum_{n=1}^{N_{\text{bg}}}\mu^{b}_{n}b_{c,n}(\vec{\theta}))\}
\times
\displaystyle\prod_{k=1}^{N_{\theta}}N(\tilde{\theta}_{k}|\theta_{k}).
\end{aligned}
\end{equation*}
In the Poisson terms $P$, the quantities $N_j$ and $N_{c}$ are the numbers of observed events in bin $j$ of the signal
region, and in control region $c$, respectively;
$s_{c}^{r,\text{exp}}$ is the expected event count in CR $c$;
and $b_{j,m,n}$ and $b_{c,n}$ are the expected event counts in observable bin $j$ and $m_{\mathrm{T}}$ bin $m$ of the SR and in CR $c$, respectively, of the $n$-th background process.
Finally, $\mu^{b}_{n}$ is the normalization factor for the $n$-th background process (and is fixed to 1 for backgrounds not estimated using CRs).
The $N_{\theta}$ nuisance parameters in the Gaussian terms $N$ are represented by $\vec{\theta}$.
The SR is further subdivided into $e\mu$ and $\mu e$, depending on which of the two leptons has the higher $\pt$, but this is not shown explicitly in the above expression. Also, when not considering the $N_{\text{jet}}$ observable, the \ZeroJet{} and \OneJet{} SRs are treated separately in the fit; this further subdivision is not displayed explicitly either.
 
The signal $s^r_j = \mu_{j}s^{r,\text{exp}}_{j}$ of the reconstruction-level distribution in the $j$-th observable bin can be rewritten as
\begin{align*}
s^r_j &= \sum_{i} s^t_i \, C_{ij} + f_j
\end{align*}
where $s_i^t$ is the signal yield in the particle-level bin $i$, $f_j$ is the number of reconstructed events that are not in the fiducial region
\begin{align*}
f_j = s^{r,\text{exp}}_j - \sum_i M_{ij}
\end{align*}
and $C_{ij}$ is the response matrix
\begin{align*}
C_{ij} = M_{ij} / s^{t,\textrm{exp}}_i
\end{align*}
given by the migration matrix $M_{ij}$ and the predicted particle-level yield $s^{t,\textrm{exp}}_i$.
The quantities $s^{r,\textrm{exp}}$,
$s^{t,\textrm{exp}}$, and $M$ are obtained from Monte Carlo predictions, with full dependencies on nuisance
parameters from theory and experimental sources, i.e.\
$s^{r,\textrm{exp}}=s^{r,\textrm{exp}}(\vec\theta)$,
$s^{t,\textrm{exp}}=s^{t,\textrm{exp}}(\vec\theta)$, and
$M=M(\vec\theta)$, and these dependencies are fully maintained
throughout the likelihood maximization.
 
A Tikhonov regularization term is included in the likelihood as a penalty term, taking the form
\begin{align*}
P(\vec{x}) = \exp\left(-\tau \cdot \left( \sum_{i=2}^{N_{\text{bin}}-1} \left(\left(x_{i-1} - x_i\right) - \left(x_{i} - x_{i+1}\right)\right)^2 \right)\right)
\end{align*}
with $x$ being the quantity for which the curvature should be
regularized.  For this analysis, the measured particle-level signal
strength, $ x_i = s_i^{t}/s_i^{t,\textrm{exp}}$, is chosen as
the regularized quantity (see also, e.g., Ref.~\cite{HIGG-2018-29}).  This type
of regularization is chosen for all single-differential distributions in this analysis. For the
double-differential cross-section measurements, independent
values of $\tau$ are chosen for each of the two jet multiplicity bins. No regularization
across jet bins is performed, as two bins are not enough to define
a curvature.
 
The choice of a value for the regularization parameter $\tau$ is a
trade-off between minimizing statistical fluctuations on the one hand,
and the potential bias induced by adding an artificial constraint to
the measurement on the other. This compromise is reached by first
constructing distributions with the number of signal plus background
events in each particle-level bin, fluctuated randomly by its respective
Poisson uncertainty. The \enquote{toy} distributions are then unfolded using
the original response matrix in order to obtain the unfolded particle-level
distribution. A bias is then calculated by subtracting the original
particle-level events from the unfolded particle-level events for 1000 toy simulations for each
value of the regularization parameter $\tau$.  The average of the bias
for all these toy simulations is compared with the statistical
uncertainty for each value of $\tau$ for each SR bin, and a value for
$\tau$ is chosen so as to attain the lowest bias relative to uncertainty for most
of the bins. Values of $\tau$ near 1 were investigated since previous
studies using this technique showed this range to be optimal. The final $\tau$ values range from 0.25 to 1.5. For the example of the $\ptH$ variable, a $\tau$ value of 0.25 was chosen, and corresponds to values of the maximum bias divided by the uncertainty of 0.06, 0.01, 0.01, and 0.02 for the lowest to highest bins, respectively.

A second unfolding procedure was performed as a cross-check and yielded similar results.
This procedure extracts the reconstructed differential cross sections from the same
statistical model, but then performs an iterative Bayesian unfolding
(IBU), which uses the number of iterations as the regularization parameter.
Further details of this method and the results obtained using it are provided in the Appendix.

The statistical and systematic uncertainties are
propagated to the final result consistently, including their effects
on the detector response.  The entries of the response matrix are
subject to correlated uncertainties among the matrix elements, which are incorporated into the
statistical model for the in-likelihood procedure. For the iterative
Bayesian unfolding procedure, the post-fit values of all nuisance parameters are
propagated to the detector response, and the uncertainties
are included in the bin-by-bin covariances and checked for coverage using toy distributions.
Cross-section normalization uncertainties from theory that
affect the signal are excluded from the measurement and instead are
included in the predictions for comparison, while effects on
the shape are fully propagated alongside the experimental
uncertainties.
 
Several tests were performed to assess the robustness of the unfolding procedure. The purpose of the tests is to determine whether the unfolding procedure is biased by variations in the input samples that account for different spectral shapes, different Monte Carlo generators, new-physics contributions and systematic correlations. For the first tests, fluctuated MC truth-level distributions were created by fluctuating individual observable bins within their respective Poisson uncertainties. The fluctuated distributions are folded, and then unfolded, and the resulting unfolded distributions are found to closely match the original fluctuated truth-level distribution, showing that no biases were introduced. The same migration matrix, derived from SM ggF simulation, is used to fold and unfold the distributions. Since the migration matrix is only composed of Monte Carlo signal events, biases could be introduced by correlated nuisance parameters that only affect the signal. Thus, for the second tests, toy datasets generated with random values of the nuisance parameters demonstrated that no such biases are introduced in the construction of the migration matrix.

\section{Results}
\label{sec:result}
 
The results of the measurement are illustrated in the figures in this
section. The \mT distributions are
obtained for all bins of each observable: $|y_{j0}|$, $\ptH$, $\pt^{\ell0}$, \ptll, \mll, \Yll, \dphill,
and \costhetastar . Figures~\ref{fig:MTSysPtHB1} to \ref{fig:MTSysPtHB4}
show examples of the pre- and post-fit \mT distributions at the reconstruction level for bins of the $\ptH$ observable, which are inputs to the unfolding procedure. The number of bins and bin edges were chosen in order to maximize the Signal/$\sqrt{\text{Background}}$ significance measure, while minimizing the bin migrations and statistical errors. In the distributions obtained after the fitting procedure is performed, all systematic uncertainties are included, and the background is normalised using the normalisation factors extracted from the control regions. The Higgs boson
transverse momentum, $\ptH$, is defined as the $\pt$ of the combined two-lepton system and missing transverse momentum at the reconstruction and particle levels, and is binned in $\pt$ bins in the ranges
0--30~\GeV, 30--60~\GeV, 60--120~\GeV\ and 120--1000~\GeV. The ggF Higgs production predictions are derived from Monte Carlo simulated samples normalized to the best SM predicted cross sections from Table~\ref{tab:mcsamples}.
These figures show that the data agree with the Standard Model
expectation for the Higgs boson signal and backgrounds
within the uncertainties. After the fitting procedure is performed, there is a decrease in uncertainties due to the additional information that results from the comparison to data in both the signal and control regions. The fit induces correlations between the nuisance parameters, which reduce the total uncertainties in the post-fit distributions via the correlation matrix.
\FloatBarrier
 
\begin{figure}[tbp]
\centering
\includegraphics[width=0.43\textwidth]{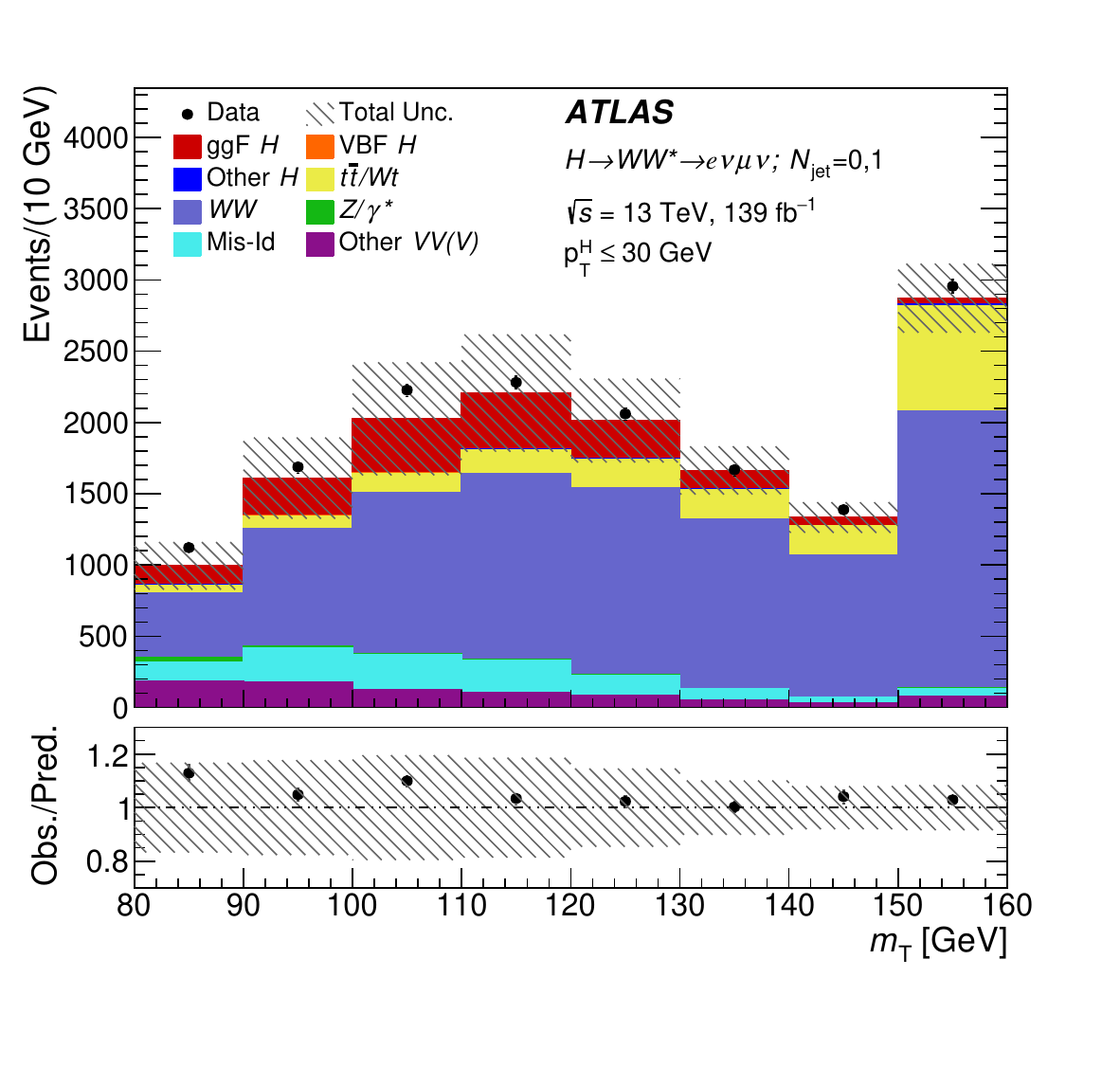}
\includegraphics[width=0.43\textwidth]{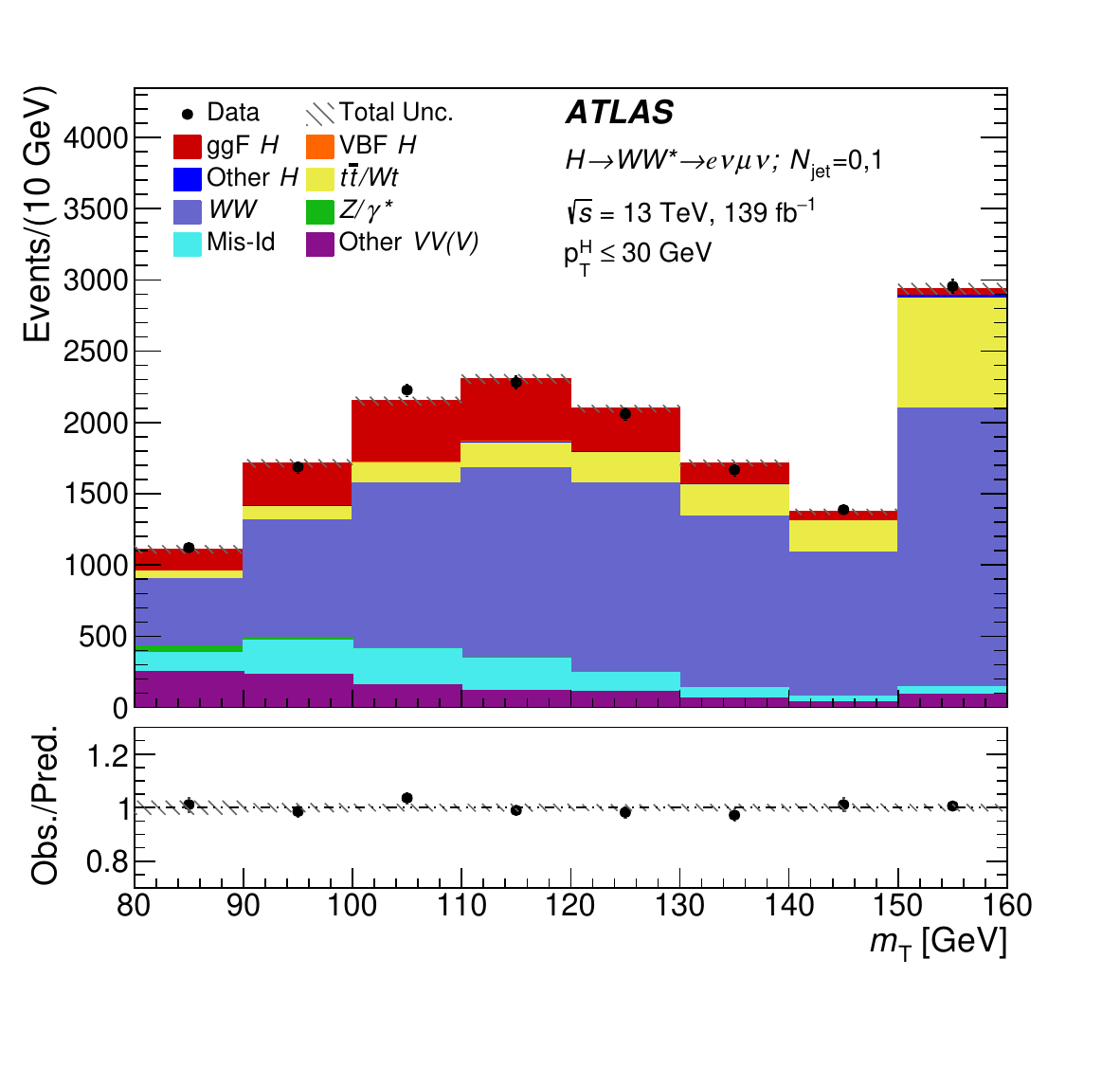}
\vspace{-8mm}
\caption{
Pre-fit (left) and post-fit (right) distribution of the transverse mass $\mT$
of the Higgs boson in the first bin of the transverse momentum $\ptH$: 0--30~\GeV for $\Njet=0,1$.
The last bin in $\mT$ contains the overflow of events with $\mT>160$~\GeV.
\label{fig:MTSysPtHB1}}
\end{figure}
 
\begin{figure}[tbp]
\centering
\includegraphics[width=0.43\textwidth]{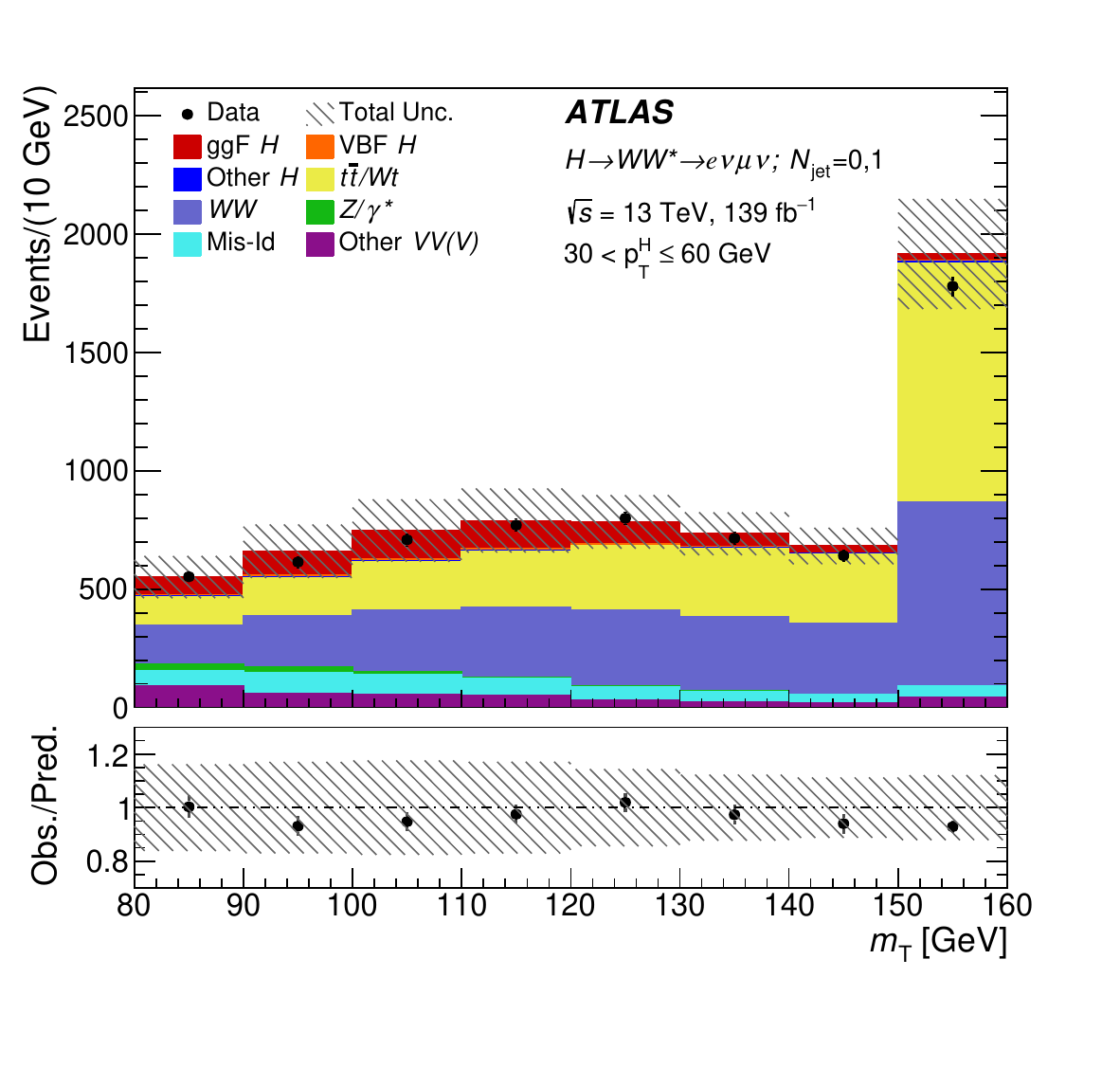}
\includegraphics[width=0.43\textwidth]{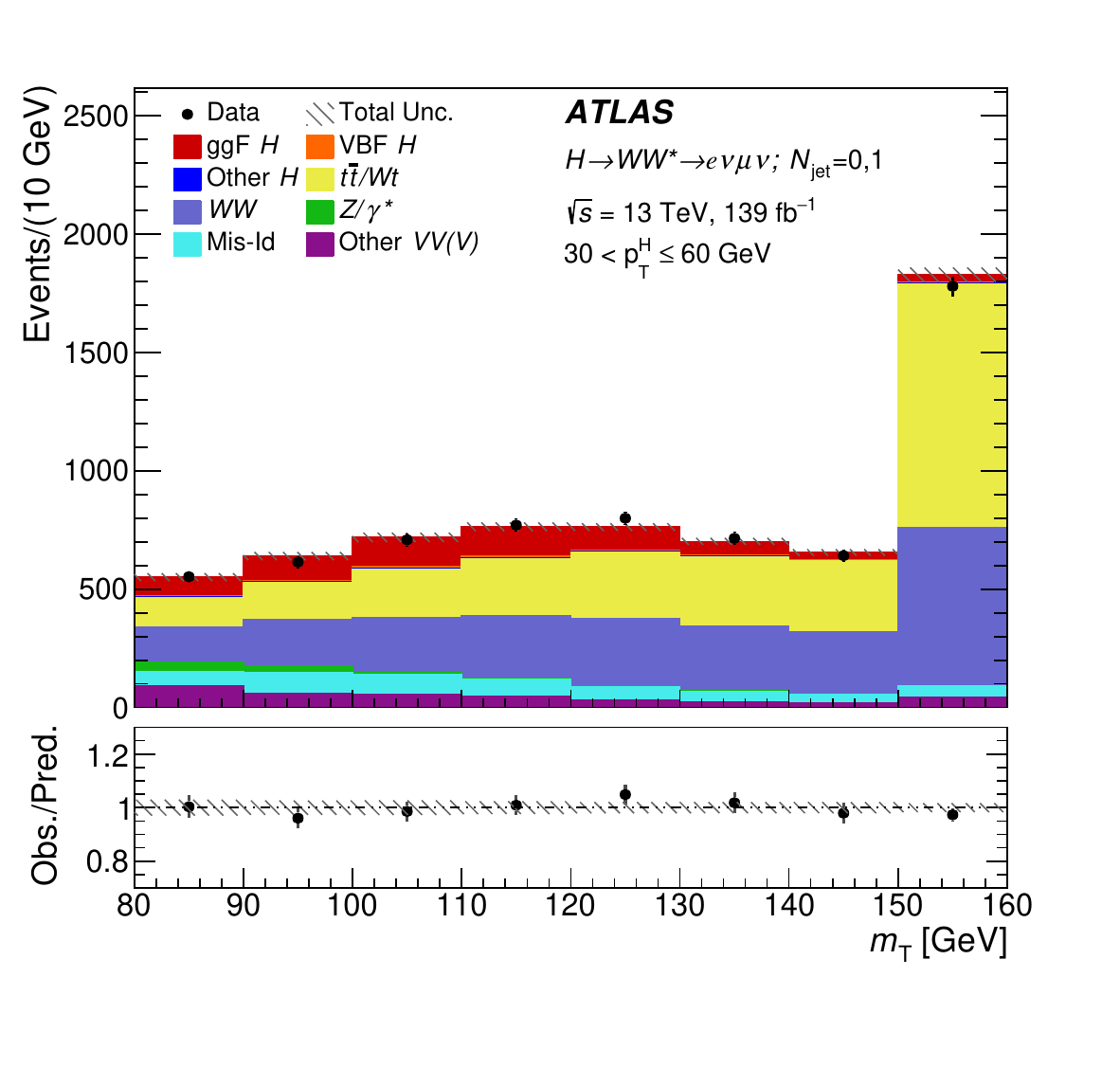}
\vspace{-8mm}
\caption{Pre-fit (left) and post-fit (right) distribution of the transverse mass $\mT$
of the Higgs boson in the second bin of the transverse momentum $\ptH$: 30--60~\GeV for $\Njet=0,1$.
The last bin in $\mT$ contains the overflow of events with $\mT>160$~\GeV.
\label{fig:MTSysPtHB2}}
\end{figure}
\begin{figure}[tbp]
\centering
\includegraphics[width=0.43\textwidth]{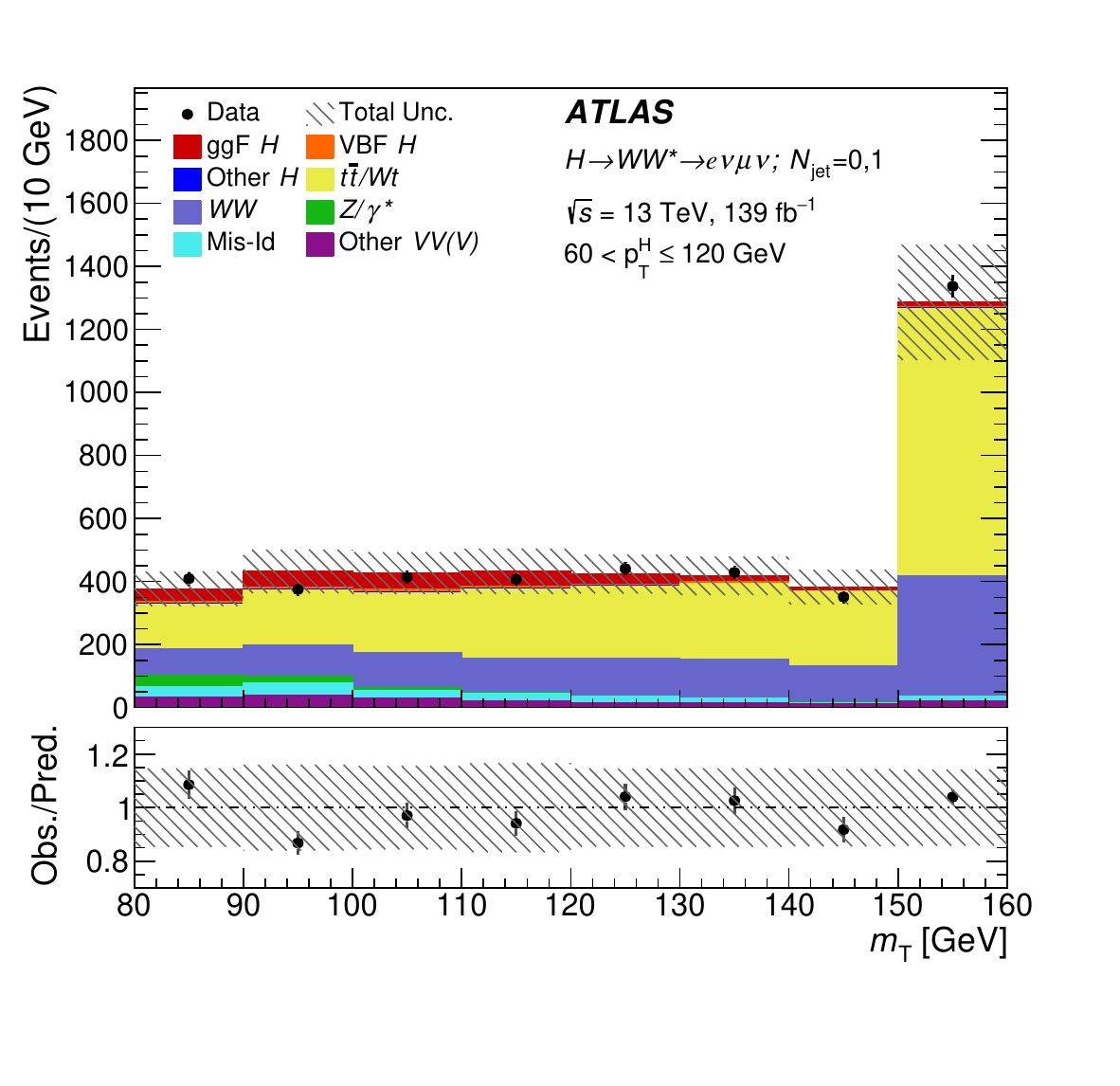}
\includegraphics[width=0.43\textwidth]{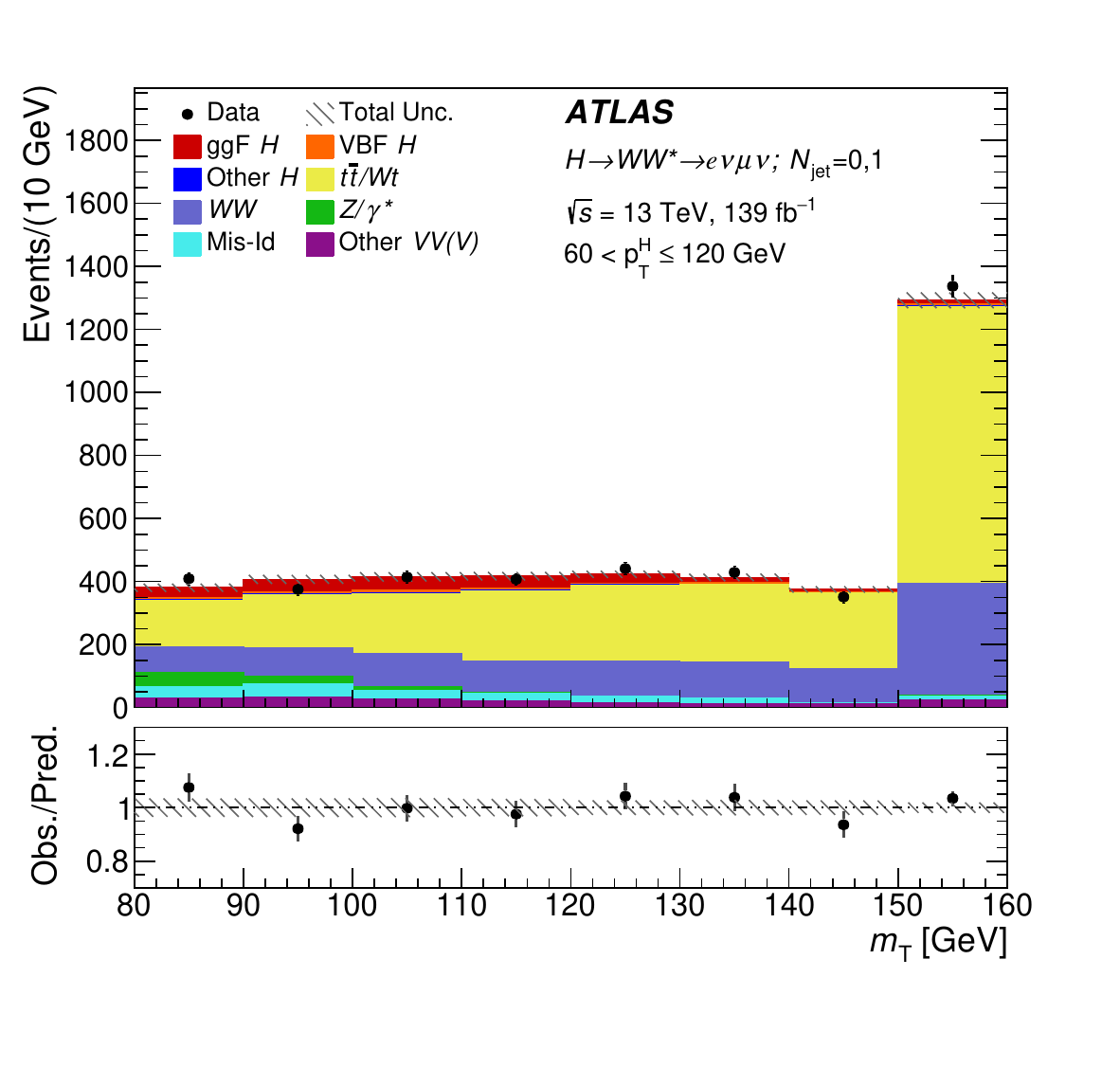}
\vspace{-8mm}
\caption{Pre-fit (left) and post-fit (right) distribution of the transverse mass $\mT$
of the Higgs boson in the third bin of the transverse momentum $\ptH$: 60--120~\GeV for $\Njet=0,1$.
The last bin in $\mT$ contains the overflow of events with $\mT>160$~\GeV.
\label{fig:MTSysPtHB3}}
\end{figure}
\begin{figure}[tbp]
\centering
\includegraphics[width=0.43\textwidth]{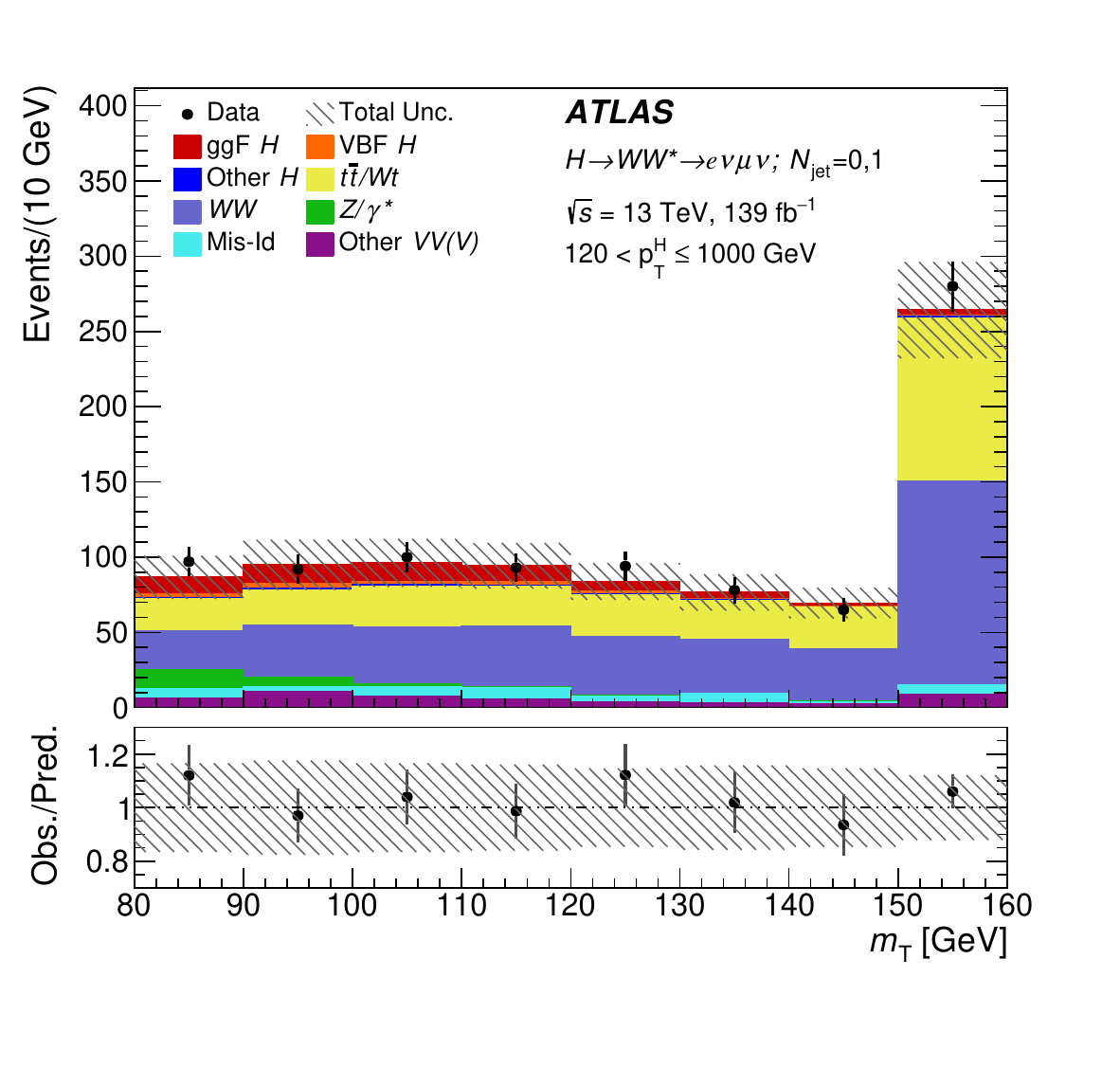}
\includegraphics[width=0.43\textwidth]{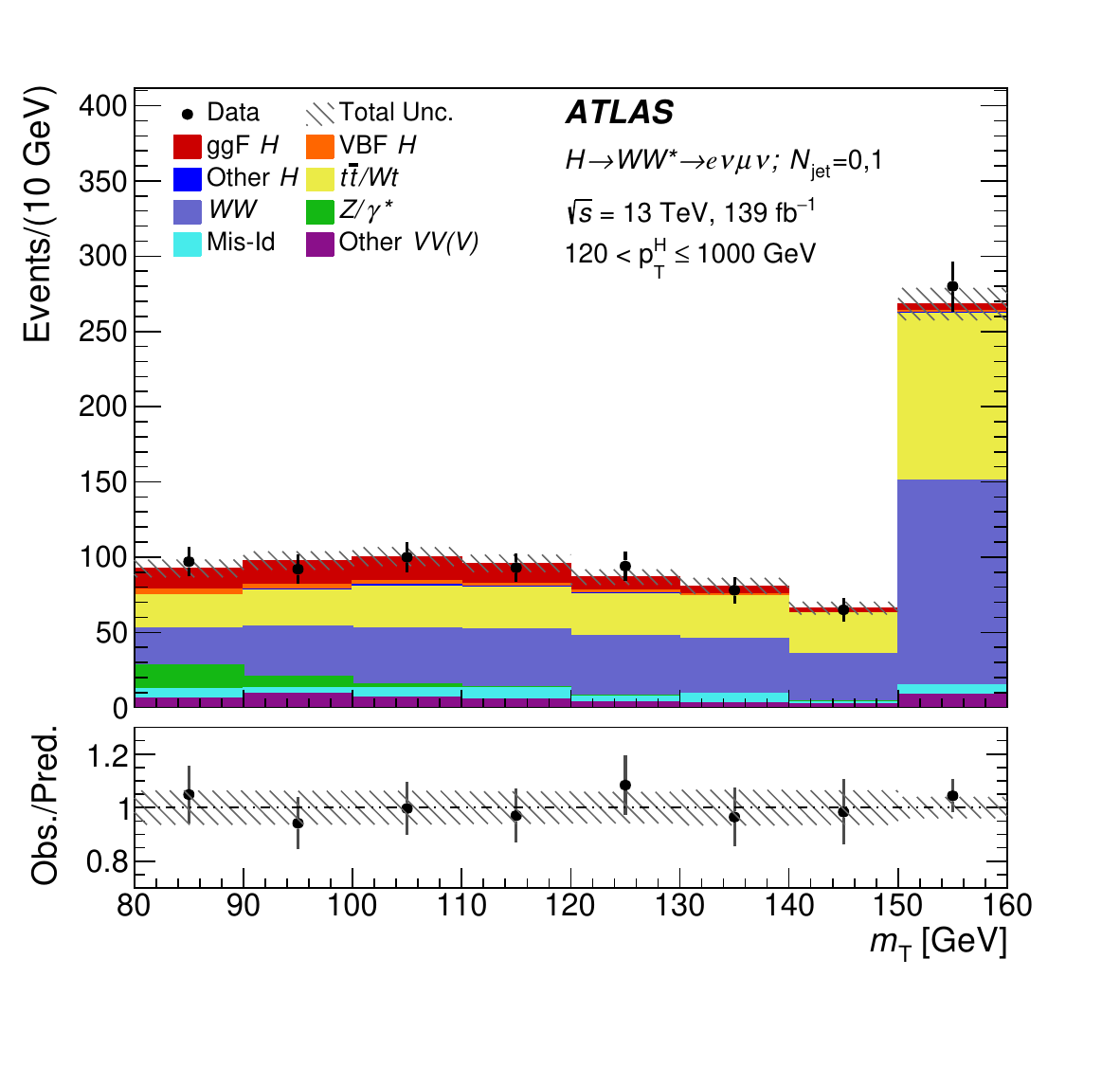}
\vspace{-8mm}
\caption{Pre-fit (left) and post-fit (right) distribution of the transverse mass $\mT$
of the Higgs boson in the last bin of the transverse momentum $\ptH$: 120--1000~\GeV for $\Njet=0,1$.
The last bin in $\mT$ contains the overflow of events with $\mT>160$~\GeV.
\label{fig:MTSysPtHB4}}
\end{figure}

The final differential fiducial cross sections for each observable,
obtained using the regularized in-likelihood unfolding method are
shown in Figures~\ref{fig:UnfoldedPtHUnblindedInLH0j1jFR} to
\ref{fig:UnfoldedYJ0UnblindedInLH0j1jFR} for the single-differential
distributions, and Figures~\ref{fig:UnfoldedDPhill_NjetUnblindedInLH0j1jFR} to
\ref{fig:Unfoldedcosthetastar_NjetUnblindedInLH0j1jFR} for the double-differential distributions.
Figure~\ref{correlations_extrasmall} shows the correlations of the unfolded cross sections in the signal regions in the different $\ptH$ intervals. The numbers in the legend on the right are shown in percentages.
 
The compatibility of the data and each set of predictions is characterized by the $p$-values, shown in the respective figure legends.
These $p$-values are computed from the fit of the respective predictions, including uncertainties in normalization and shape, to the observed data.
The total cross sections (sum over all bins) are compatible with each other for all observables.
The data and predictions agree very well, as demonstrated by the high $p$-values for all of the plots.
No significant differences are observed between the measured cross sections and their Standard Model Monte Carlo predictions.
 
The uncertainties with the largest impact on the results include
uncertainties related to jet and muon reconstruction. Theory
uncertainties associated with the top-quark and $WW$ backgrounds, and with the
difficulty of modelling $V\gamma$ processes also play a leading role. For the \POWPY[8] prediction, the
uncertainties were evaluated in several different regions of phase space and summed in quadrature as described
in Ref.~\cite{ATLAS:2022ooq}, whereas for the \POWHER[7] sample, they could only be evaluated
inclusvely for each bin, resulting in a slightly less precise evaluation.
Other leading uncertainties include ones affecting the data-driven background
estimates for misidentified objects, as well as uncertainties related to normalizing the backgrounds from control regions. The uncertainties are listed in Table~\ref{tab:UncertSummary}, which shows ranges of values that correspond the different bins of the measured observables. Table~\ref{fig:breakdowntablePtH} shows a detailed breakdown of the uncertainties for the $\ptH$ variable obtained after the final fitting procedure is performed.
 
The integrated fiducial cross section is obtained using the procedure described above but without binning in any observable (and consequently without any regularization). It is measured to be $56.0^{+10.0}_{\phantom{1}-9.5}$ fb.
 
\begin{table}[tbp]
\centering
\caption{ The statistical uncertainties of the data and Monte Carlo predictions, as well as experimental and theoretical uncertainties of the measured cross sections for each kinematic variable. The uncertainties vary depending on the \mt bin in which the cross sections are measured, as demonstrated by the range shown for each entry. The dominant sources of systematic uncertainties are discussed in the text.}
 
\label{tab:UncertSummary}
\scalebox{0.85}{
\begin{tabular}{l c c c c}
\hline\hline
 
Variable                   			& Data Statistical [\%]    	& MC Statistical [\%]    	& Experimental [\%]    	& Theory [\%]  			\\
\hline
$y_{\ell\ell}$             			& 14--22                  		& 5.3--10\phantom{.}  	& 6.9--15\phantom{.}		& 5.9--15\phantom{.}      	\\
\ptll                      			& 15--29                  		& 6.4--14\phantom{.}    	& 8.2--31\phantom{.}		& 6.8--27\phantom{.}      	\\
$\pt^{\ell0}$              			& 13--28                  		& 6.3--13\phantom{.} 	& 9.3--28\phantom{.}		& 14--34      			\\
\dphill                    			& 11--39                  		& 6.1--18\phantom{.}		& 7.8--22\phantom{.}		& 13--27      			\\
$y_{j0}$                   			& 23--51                  		& 12--26				& 21--54              		& 26--58				\\
\costhetastar    				& 11--15                  		& 5.8--7.6               		& 8.5--11\phantom{.}		& 8.9--14\phantom{.}      	\\
$\ptH$                    			& 8.5--72\phantom{.}		& 6.2--18\phantom{.}		& 10--58				& 12--27     			\\
\mll                       			& 12--25                  		& 5.6--11\phantom{.}		& 7.5--15\phantom{.}		& 7.3--20\phantom{.}      	\\
$y_{\ell\ell}$ vs $N_\text{jet}$  	& 9.0--62\phantom{.}		& 3.9--25\phantom{.}		& 8.0--20\phantom{.}		& 5.0--53\phantom{.}     	\\
\ptll vs $N_\text{jet}$           	& 9.8--36\phantom{.}		& 4.7--20\phantom{.}  	& 12--41		              	& 9.9--50\phantom{.}      	\\
$\pt^{\ell0}$ vs $N_\text{jet}$   	& 9.6--50\phantom{.}		& 5.8--20\phantom{.}		& 10--35              		& 9.4--74\phantom{.}      	\\
\dphill vs $N_\text{jet}$         	& 9.6--65\phantom{.}		& 5.6--18\phantom{.}		& 6.8--31\phantom{.}		& 14--74      			\\
\costhetastar vs $N_\text{jet}$	& 13--50                  		& 6.8--25\phantom{.}		& 7.7--39\phantom{.}		& 8.9--58\phantom{.}      	\\
\mll vs $N_\text{jet}$            	& \phantom{0}12--152	& 5.7--44\phantom{.}		& 8.9--58\phantom{.}		& 7.2--82\phantom{.}      	\\
\hline\hline
\end{tabular}
}
 
\end{table}

\begin{table}[h]
\centering
\scalebox{0.95}{
\begin{tabular}{ l | c | c | c | c }
\hline\hline
Contribution & 0 --  30 \GeV & 30 --  60 \GeV & 60 --  120 \GeV & 120 -- 1000 \GeV\\
\hline
Total relative uncertainty & 19 & 51 & 108 & 62 \tabularnewline
Total systematic uncertainty  & 17 & 41 & 81 & 45 \tabularnewline
Statistical uncertainties from data  & 8.5 & 29 & 72 & 43 \tabularnewline
Statistical uncertainties from simulation  & 6.2 & 15 & 17 & 18 \tabularnewline
Experimental systematic uncertainty  & 10 & 24 & 58 & 31 \tabularnewline
\hspace{2em}Flavour tagging  & 4.9 & 6.1 & 13 & 18 \tabularnewline
\hspace{2em}Jet Energy Scale  & 4.9 & 17 & 30 & 21 \tabularnewline
\hspace{2em}Jet Energy Resolution  & 4.8 & 8.4 & 12 & 10 \tabularnewline
\hspace{2em}Missing transverse energy  & 4.6 & 8.2 & 11 & 8.2 \tabularnewline
\hspace{2em}Muons  & 4.6 & 4.2 & 4.8 & 2.3 \tabularnewline
\hspace{2em}Electrons  & 3.2 & 2.2 & 2.5 & 1.1 \tabularnewline
\hspace{2em}Misidentified objects  & 3.9 & 10 & 3.3 & 1.3 \tabularnewline
\hspace{2em}Pile-up  & 3.8 & 1.2 & 6.9 & 4.4 \tabularnewline
\hspace{2em}Luminosity  & 2.7 & 4.5 & 3.5 & 2.8 \tabularnewline
Systematic uncertainty from theory  & 11 & 27 & 23 & 25 \tabularnewline
\hspace{2em}on gluon-fusion production  & 6.1 & 7.1 & 4.7 & 5.2 \tabularnewline
\hspace{2em}on Vector Boson Fusion production  & 2.4 & 1.8 & 2.7 & 3.2 \tabularnewline
\hspace{2em}on top quark production  & 4.3 & 20 & 18 & 21 \tabularnewline
\hspace{2em}on decays of $Z\to\tau\tau$  & 4.2 & 5.9 & 2.2 & 3.9 \tabularnewline
\hspace{2em}on the $WW$ background  & 7.3 & 10 & 12 & 6.0 \tabularnewline
\hspace{2em}on other diboson processes  & 5.9 & 14 & 7.8 & 8.8 \tabularnewline
Background normalization factors  & 2.3 & 0.7 & 0.9 & 0.6 \tabularnewline
\hline\hline
\end{tabular}
}
\caption{Relative cross section uncertainties broken down for each bin in $\pt^{H}$.
\label{fig:breakdowntablePtH}}
\end{table}

\clearpage
 
\begin{figure}[tbp]
\centering
\includegraphics[width=0.8\textwidth]{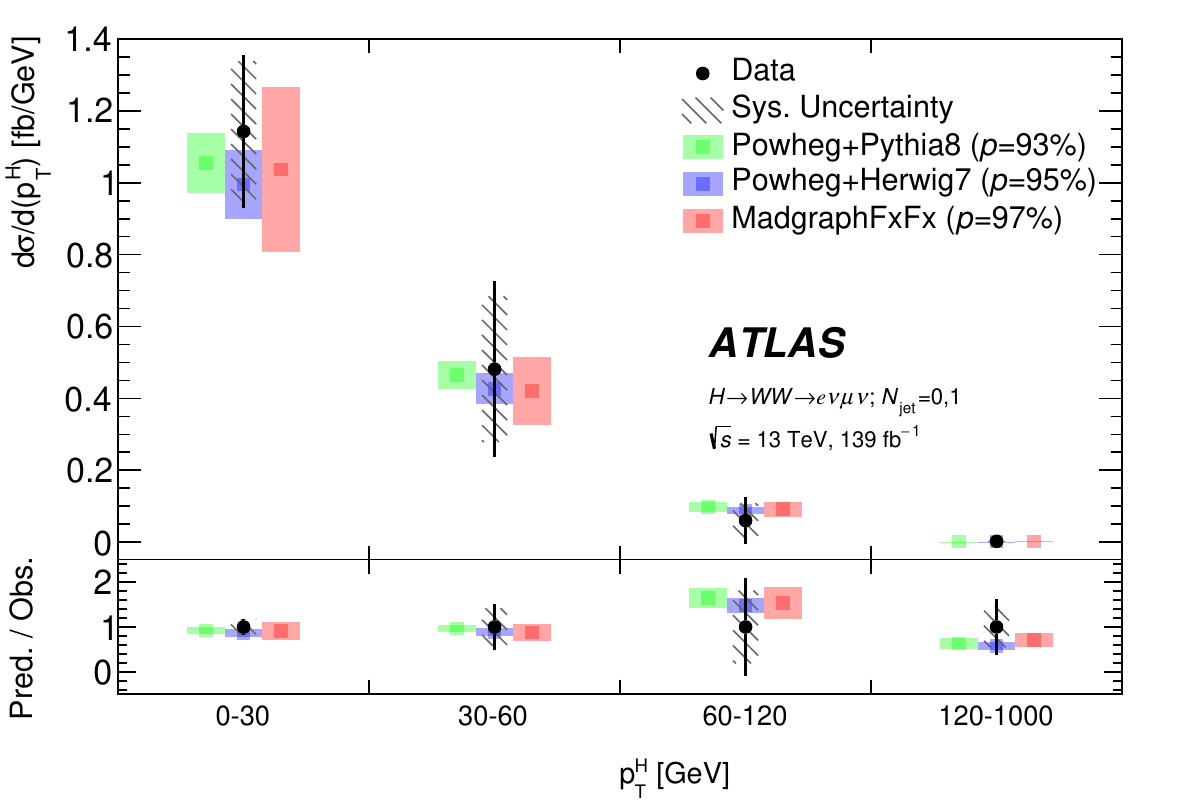}
\caption{Measured differential fiducial cross section for \ptH
in the 0+1-jet fiducial region using the regularized
in-likelihood unfolding method.  Uncertainty bars on the data
points include statistical and systematic uncertainties from
experimental and theory sources as well as background
normalization effects and shape effects from background and
signal. Uncertainty bands on the predictions shown are dominated by
normalization effects on the signal arising from showering, PDF
models, \alphas and the QCD scale. The legend includes $p$-values
quantifying the level of agreement between the data and the predictions,
including all sources of uncertainty. The systematic
uncertainties of the data are shown separately.
\label{fig:UnfoldedPtHUnblindedInLH0j1jFR}}
\end{figure}
 
\begin{figure}[tbp]
\centering
\includegraphics[width=0.49\textwidth]{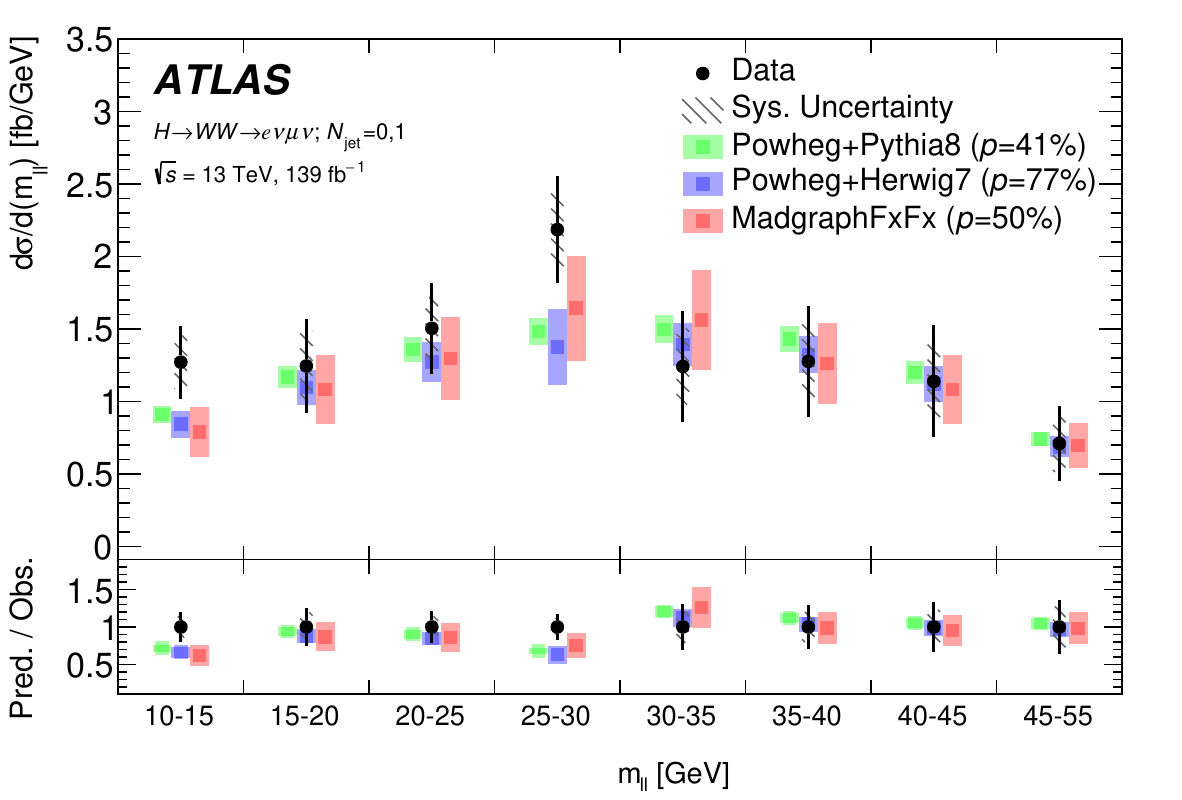}
\includegraphics[width=0.49\textwidth]{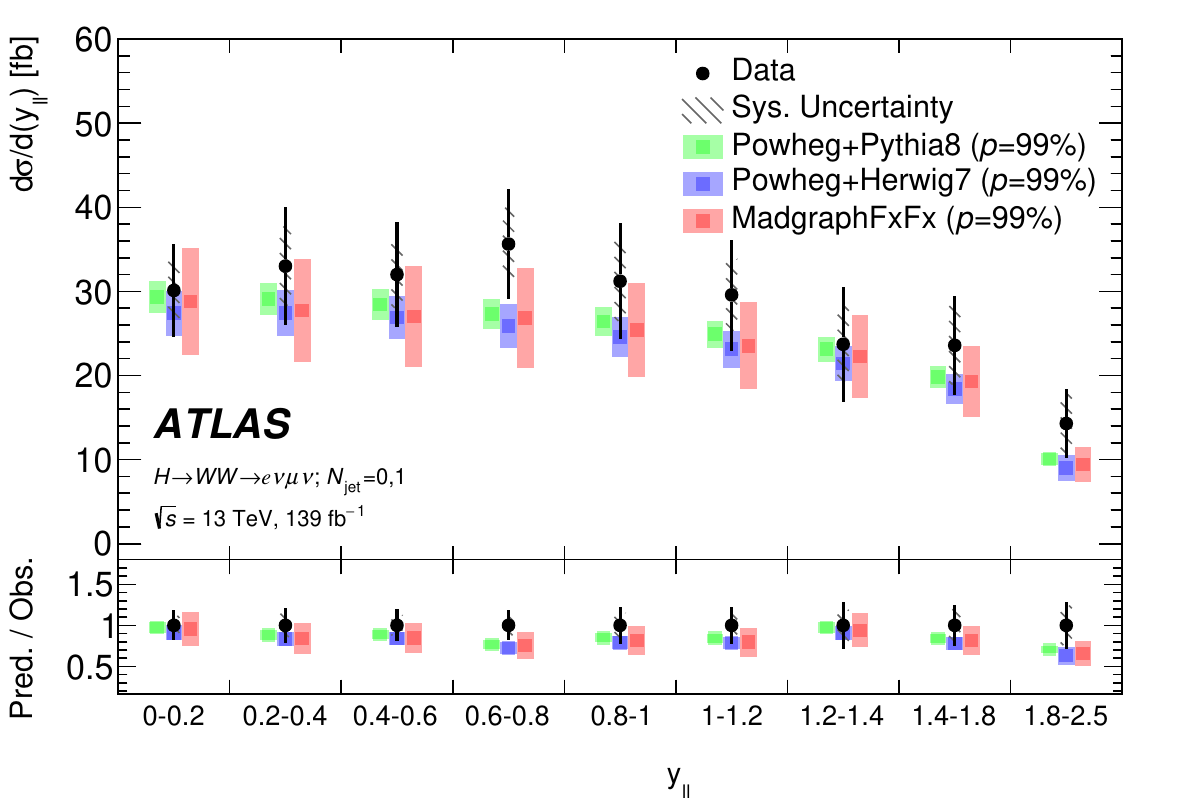}
\caption{Measured differential fiducial cross section for \mll
(left) and \Yll (right) in the 0+1-jet fiducial region
using the regularized in-likelihood unfolding method.
Uncertainty bars on the data points include statistical and
systematic uncertainties from experimental and theory sources as
well as background normalization effects and shape effects from
background and signal. Uncertainty bars on the predictions shown are
dominated by normalization effects on the signal arising from
showering, PDF models, \alphas and the QCD scale. The legend
includes $p$-values quantifying the level of agreement between the data and
the predictions, including all sources of uncertainty. The
systematic uncertainties of the data are shown separately.
\label{fig:UnfoldedMllUnblindedInLH0j1jFR}
}
\end{figure}
 
\begin{figure}[tbp]
\centering
\includegraphics[width=0.49\textwidth]{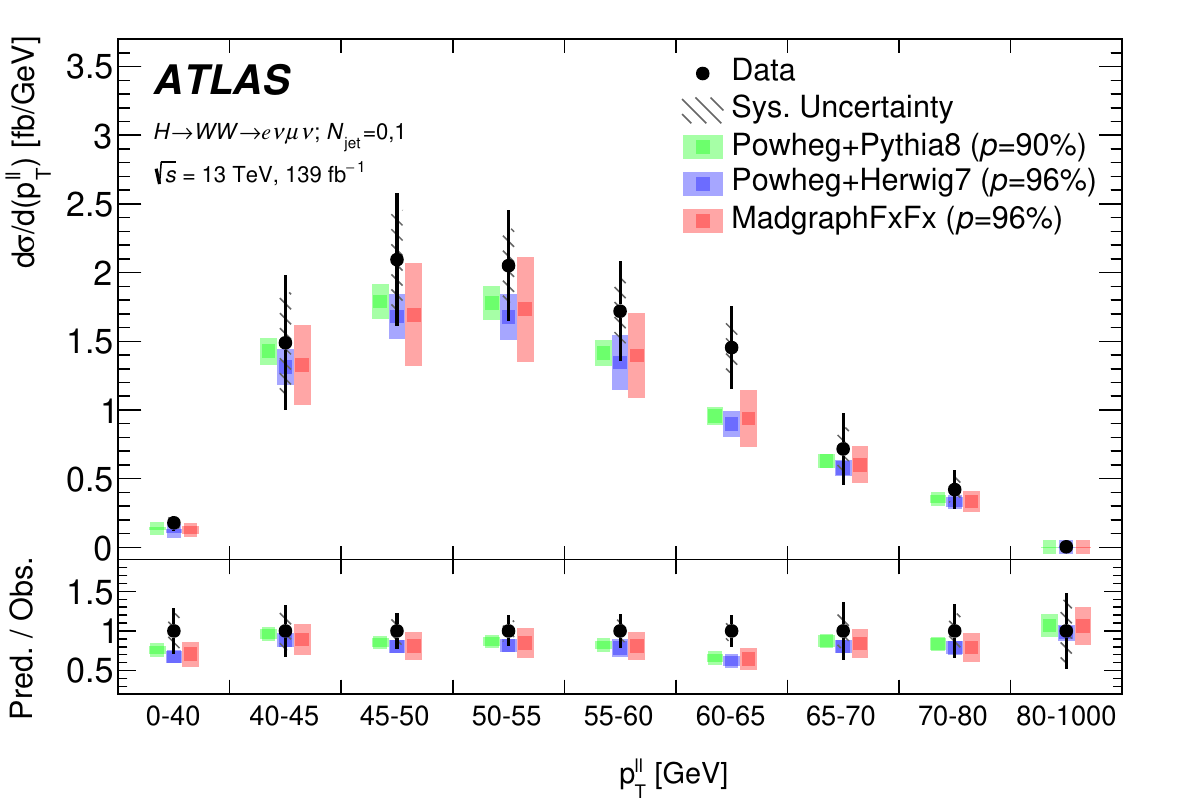}
\includegraphics[width=0.49\textwidth]{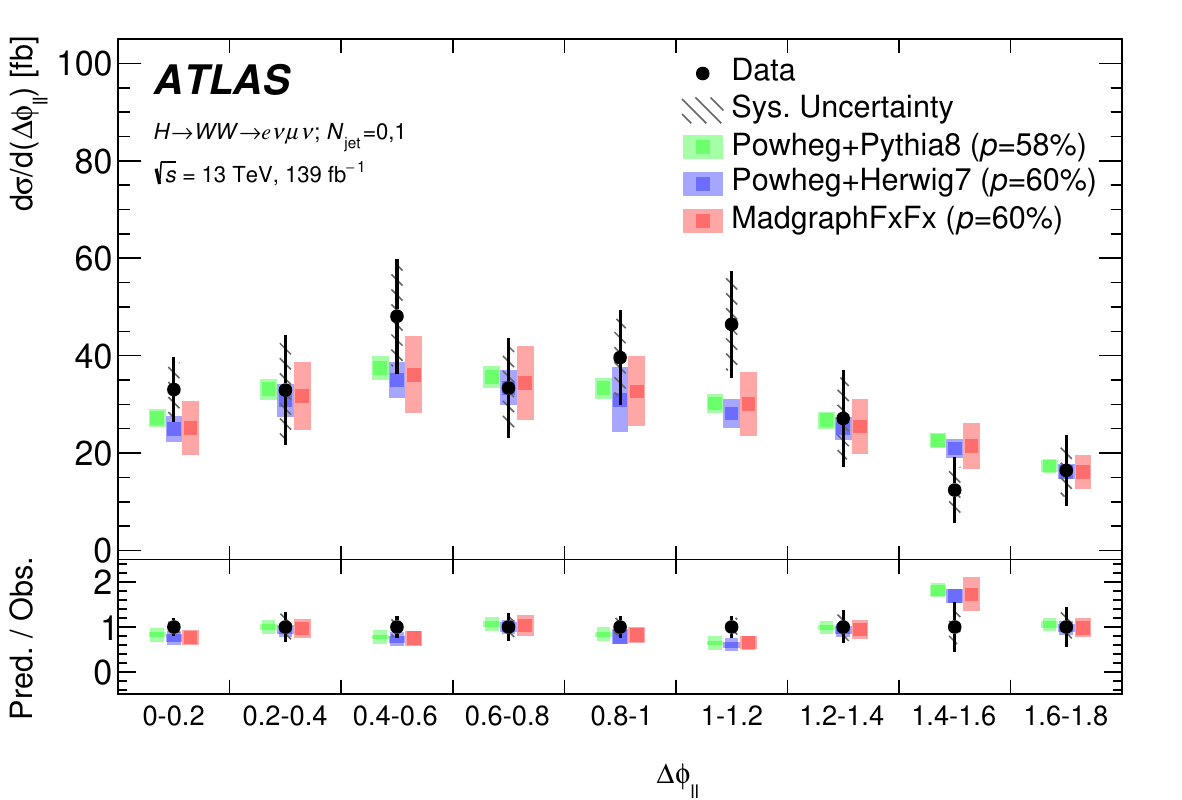}
\caption{Measured differential fiducial cross section for
\ptll (left) and \dphill (right) in the 0+1-jet
fiducial region using the regularized in-likelihood unfolding
method.  Uncertainty bars on the data points include statistical
and systematic uncertainties from experimental and theory
sources as well as background normalization effects and shape
effects from background and signal. Uncertainty bands on the
predictions shown are dominated by normalization effects on the
signal arising from showering, PDF models, \alphas and the QCD
scale. The legend includes $p$-values quantifying the level of agreement
between the data and the predictions, including all sources of
uncertainty. The systematic uncertainties of the data are shown
separately.
\label{fig:UnfoldedPtllUnblindedInLH0j1jFR}
}
\end{figure}
 
\begin{figure}[tbp]
\centering
\includegraphics[width=0.49\textwidth]{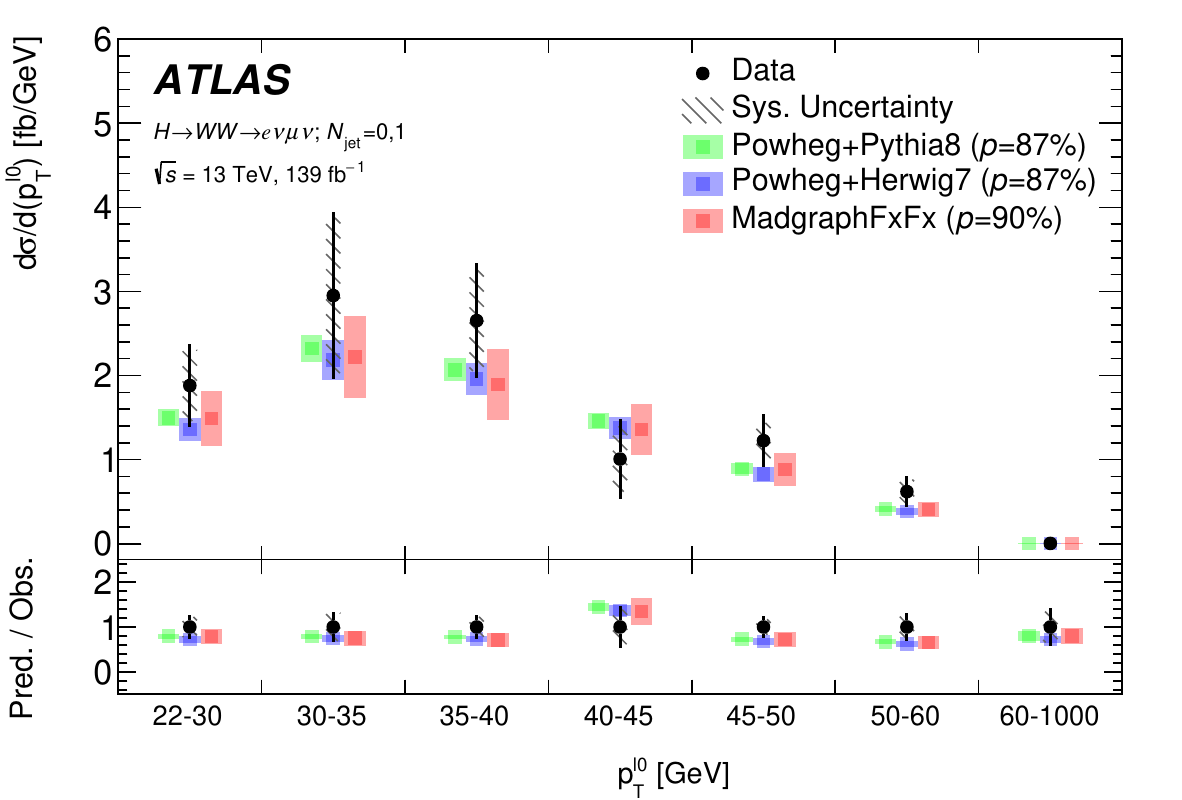}
\includegraphics[width=0.49\textwidth]{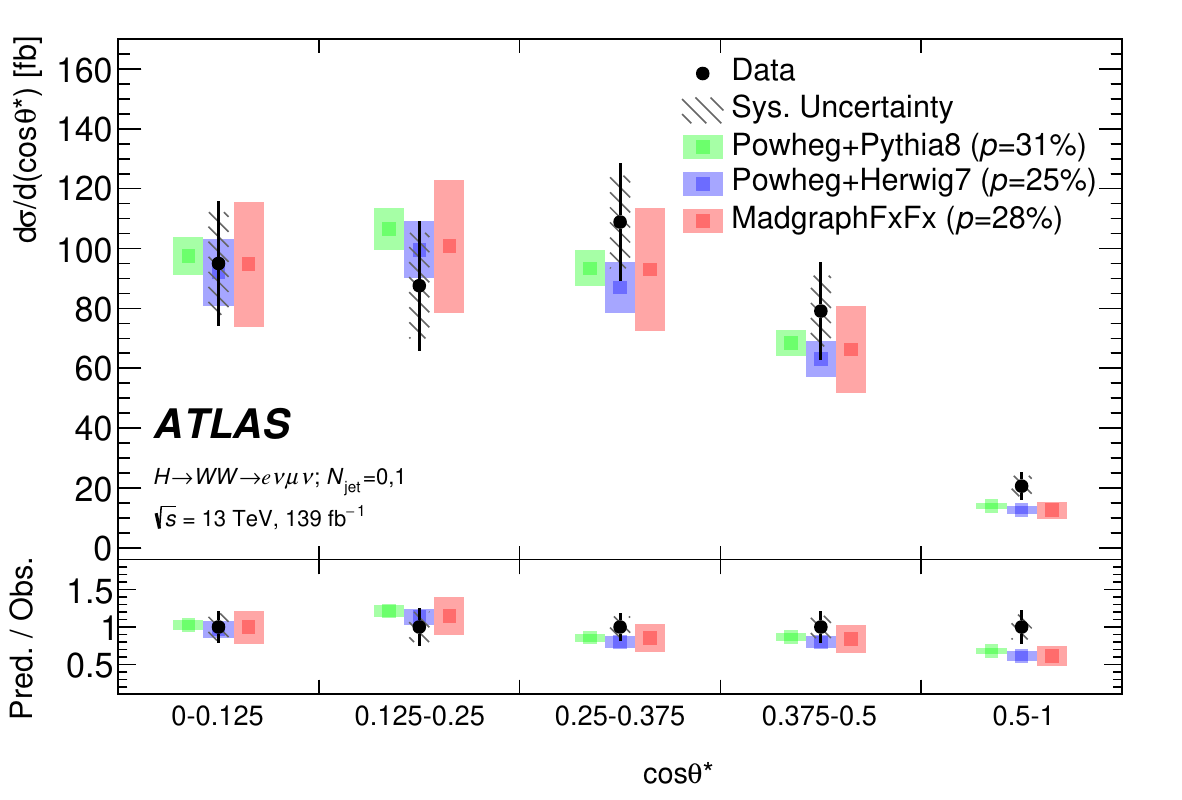}
\caption{Measured differential fiducial cross section for
$\pt^{\ell 0}$ (left) and \costhetastar (right) in the
0+1-jet fiducial region using the regularized
in-likelihood unfolding method.  Uncertainty bars on the data
points include statistical and systematic uncertainties from
experimental and theory sources as well as background
normalization effects and shape effects from background and
signal. Uncertainty bands on the predictions shown are dominated by
normalization effects on the signal arising from showering, PDF
models, \alphas and the QCD scale. The legend includes $p$-values
quantifying the level of agreement between the data and the predictions,
including all sources of uncertainty. The systematic
uncertainties of the data are shown separately.
\label{fig:UnfoldedcosthetastarUnblindedInLH0j1jFR}
}
\end{figure}
 
\begin{figure}[tbp]
\centering
\includegraphics[width=0.49\textwidth]{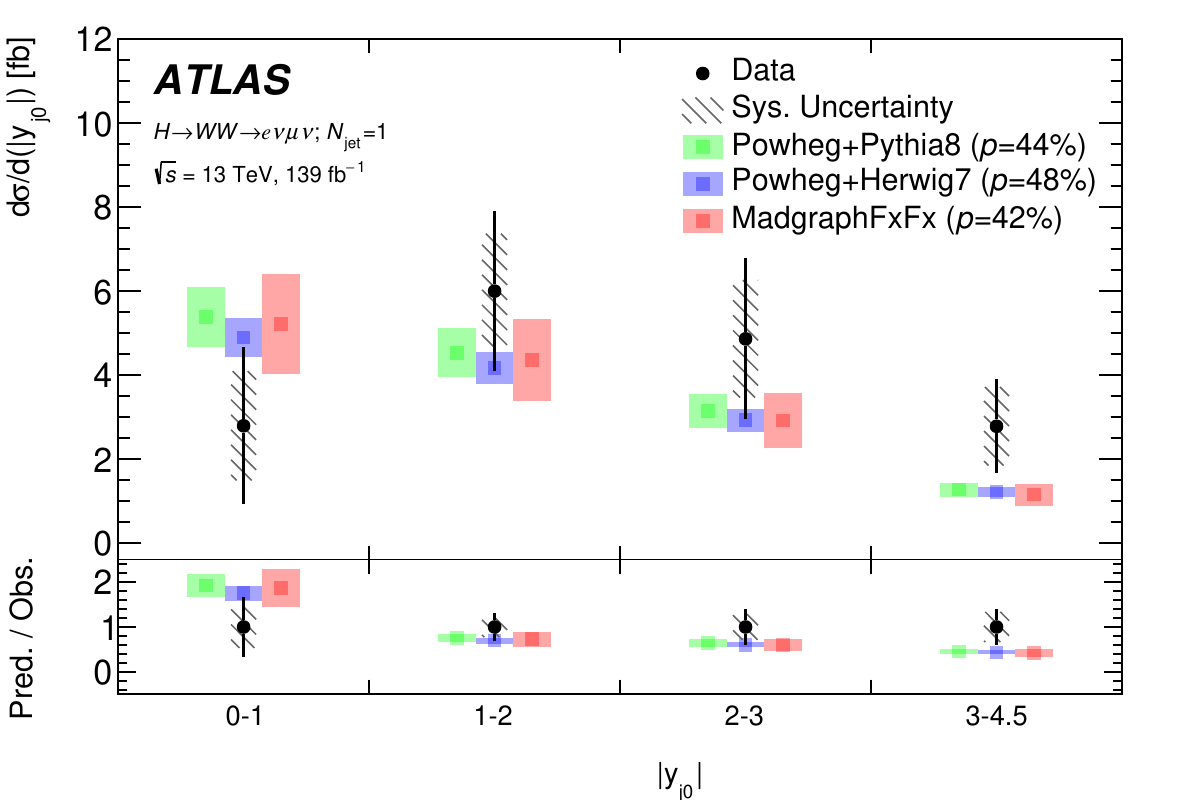}
\caption{Measured differential fiducial cross section for $|y_{j0}|$ 
in the 1-jet fiducial region
using the regularized in-likelihood unfolding method.
Uncertainty bars on the data points include statistical and systematic
uncertainties from experimental and theory sources as well as
background normalization effects and shape effects from background
and signal. Uncertainty bands on the predictions shown are dominated by
normalization effects on the signal arising from showering, PDF
models, \alphas and the QCD scale. The legend includes $p$-values
quantifying the level of agreement between the data and the predictions,
including all sources of uncertainty. The systematic uncertainties
of the data are shown separately.
\label{fig:UnfoldedYJ0UnblindedInLH0j1jFR}
}
\end{figure}
 
\begin{figure}[tbp]
\centering
\includegraphics[width=0.49\textwidth]{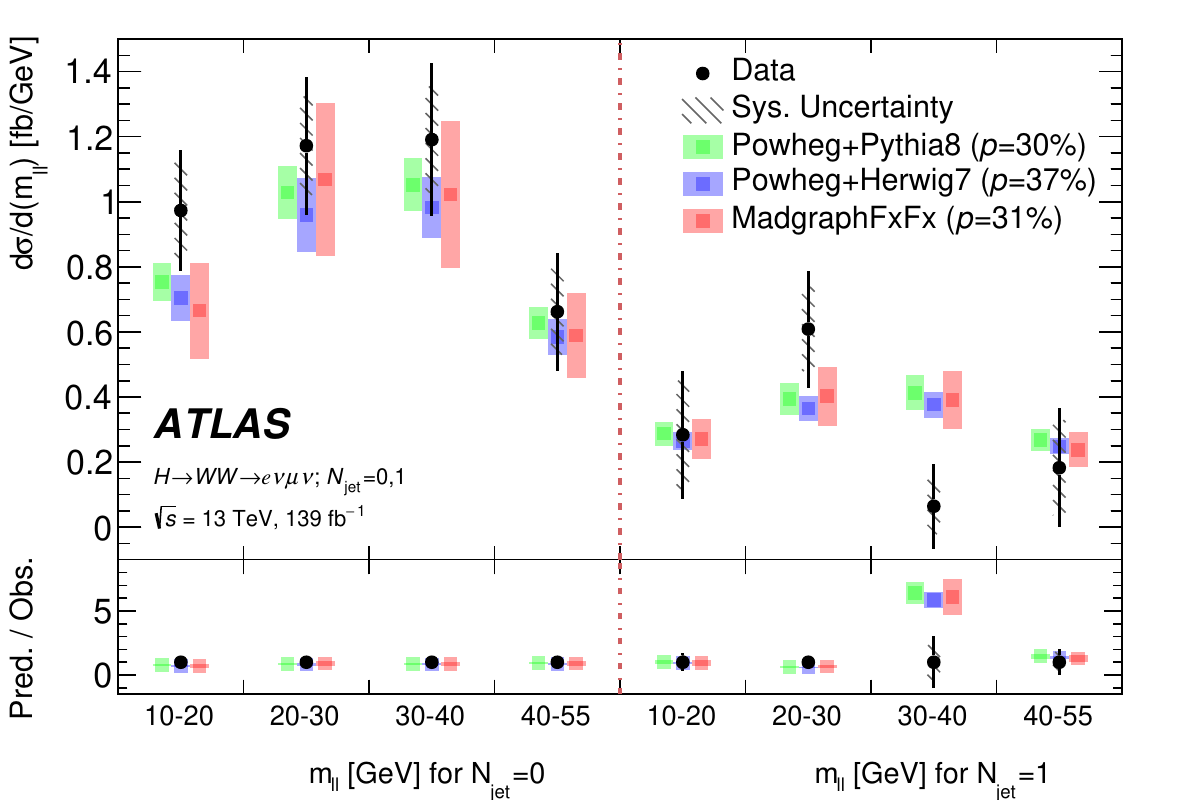}
\includegraphics[width=0.49\textwidth]{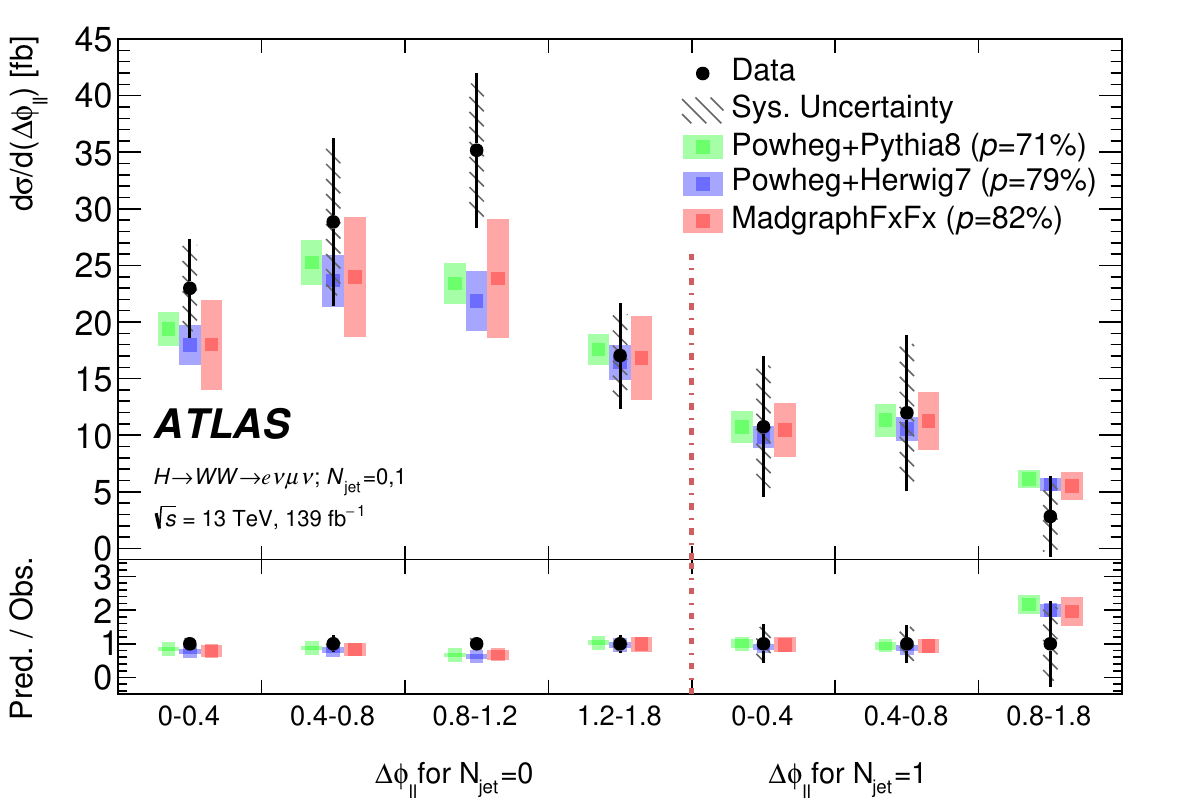}
\caption{Measured differential fiducial cross section for \mll\
(left) and \dphill (right) versus~\Njet in the 0-jet and
1-jet fiducial regions using the regularized in-likelihood unfolding
method.  Uncertainty bars on the data points include statistical and
systematic uncertainties from experimental and theory sources as
well as background normalization effects and shape effects from
background and signal. Uncertainty bands on the predictions shown are
dominated by normalization effects on the signal arising from
showering, PDF models, \alphas and the QCD scale. The legend
includes $p$-values quantifying the level of agreement between the data and
the predictions, including all sources of uncertainty. The
systematic uncertainties of the data are shown separately.
\label{fig:UnfoldedDPhill_NjetUnblindedInLH0j1jFR}
}
\end{figure}
 
\begin{figure}[tbp]
\centering
\includegraphics[width=0.49\textwidth]{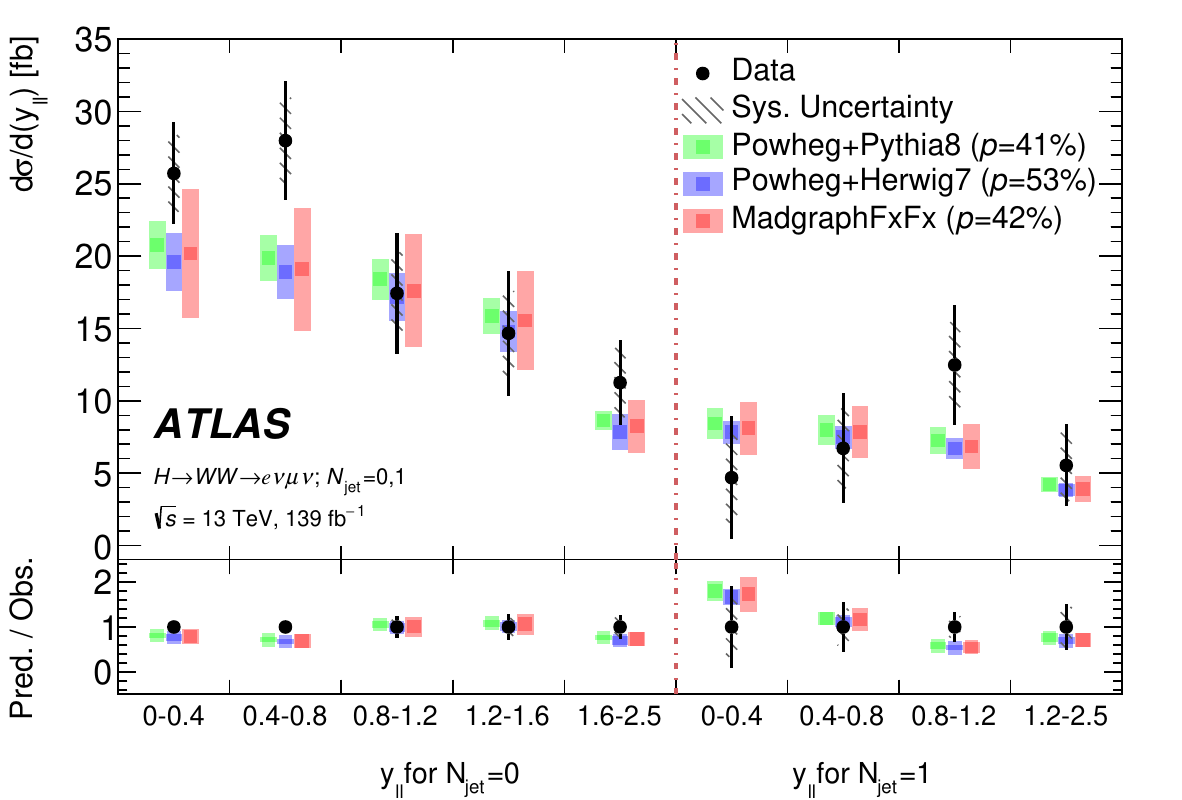}
\includegraphics[width=0.49\textwidth]{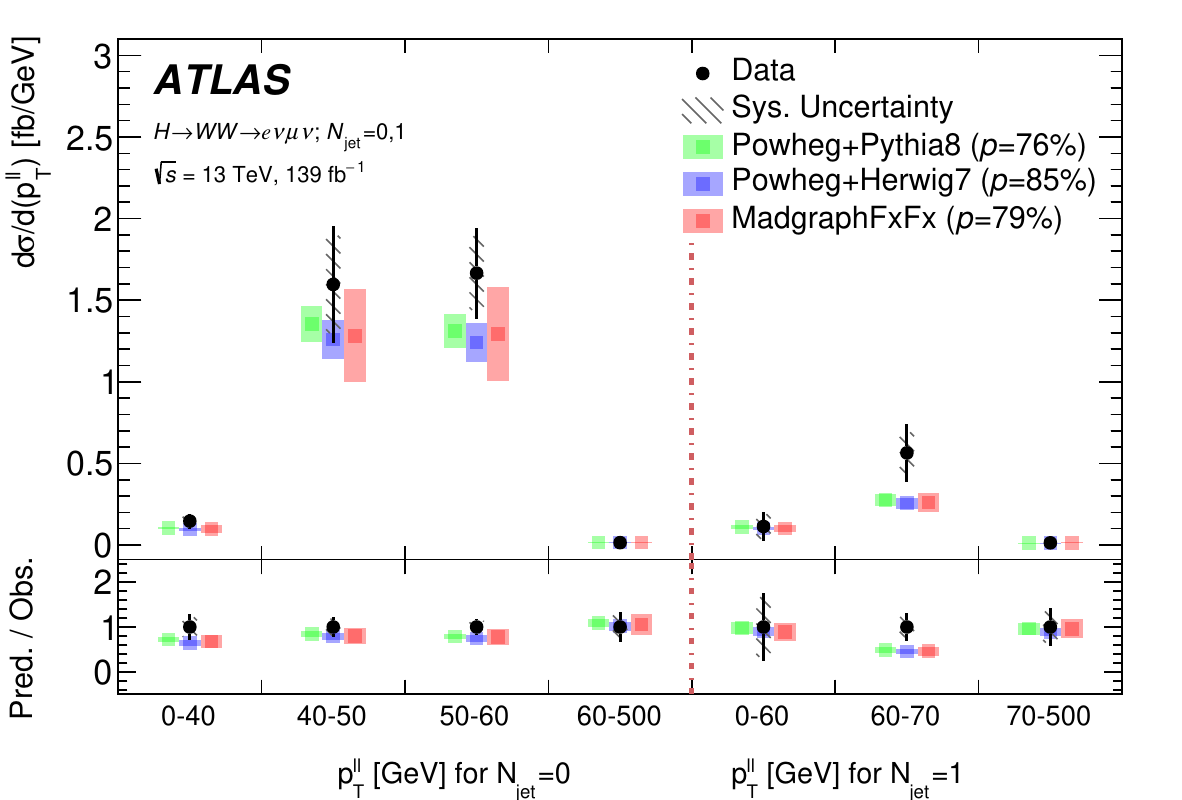}
\caption{Measured differential fiducial cross section for \Yll
(left) and \ptll (right) versus~\Njet in the 0-jet
and 1-jet fiducial regions using the regularized in-likelihood
unfolding method.  Uncertainty bars on the data points include
statistical and systematic uncertainties from experimental and
theory sources as well as background normalization effects and
shape effects from background and signal. Uncertainty bands on the
predictions shown are dominated by normalization effects on the
signal arising from showering, PDF models, \alphas and the QCD
scale. The legend includes $p$-values quantifying the level of agreement between the data and
the predictions, including all sources of uncertainty. The
systematic uncertainties of the data are shown separately.
\label{fig:UnfoldedPtll_NjetUnblindedInLH0j1jFR}
}
\end{figure}

\begin{figure}[tbp]
\centering
\includegraphics[width=0.49\textwidth]{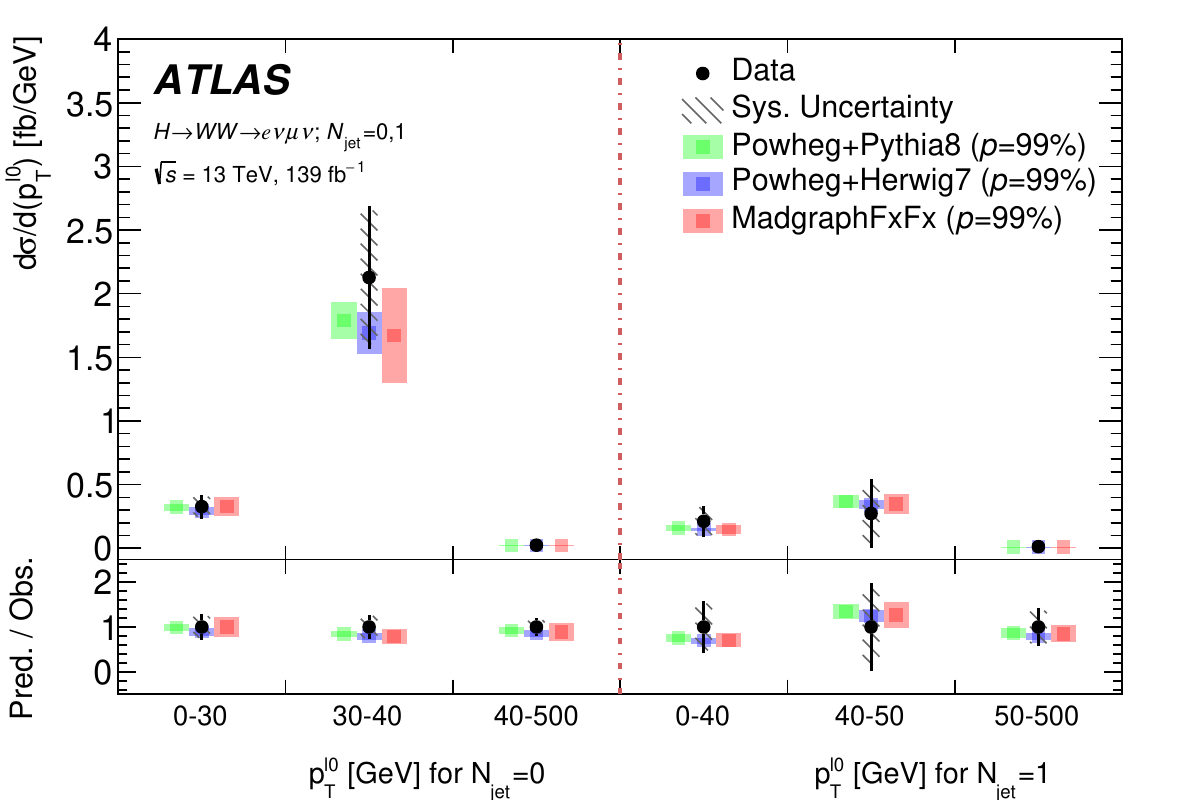}
\includegraphics[width=0.49\textwidth]{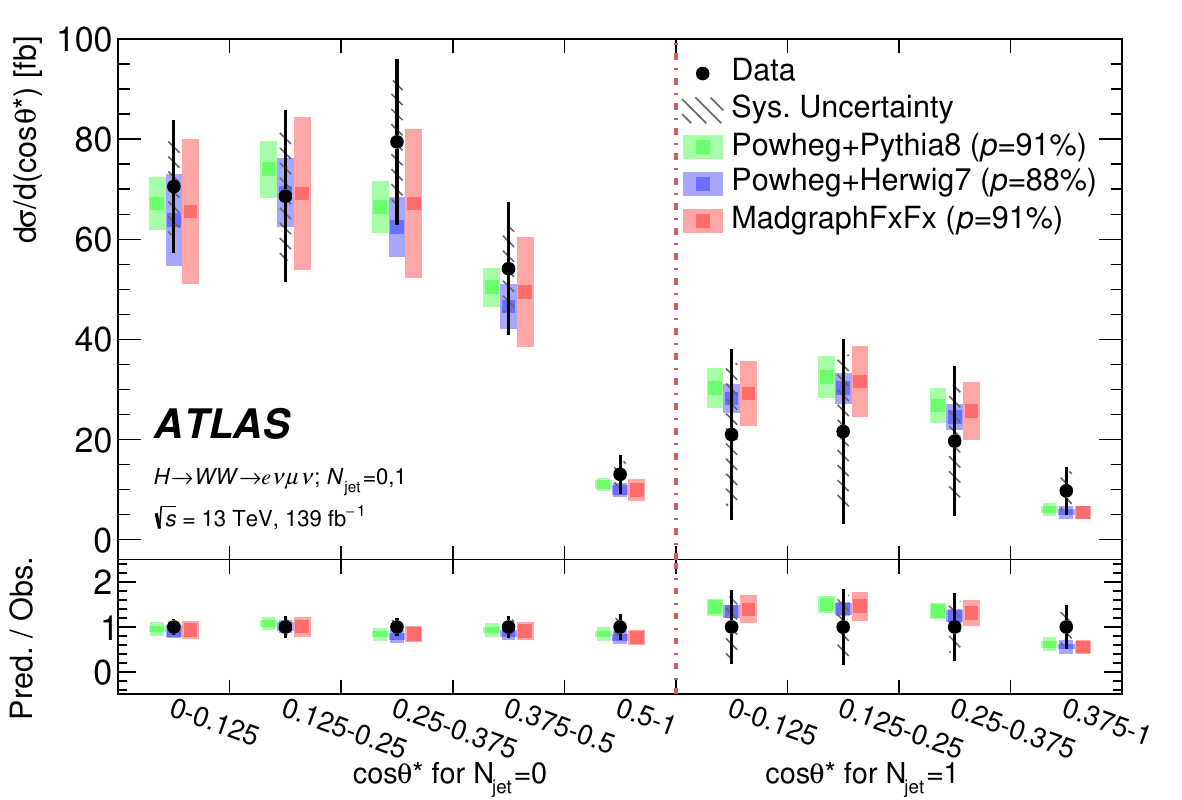}
\caption{Measured differential fiducial cross section for
$\pt^{\ell 0}$ (left) and \costhetastar (right)
versus~\Njet in the 0-jet and 1-jet fiducial regions using
the regularized in-likelihood unfolding method.  Uncertainty bars
on the data points include statistical and systematic
uncertainties from experimental and theory sources as well as
background normalization effects and shape effects from
background and signal. Uncertainty bands on the predictions shown
are dominated by normalization effects on the signal arising
from showering, PDF models, \alphas and the QCD scale. The legend
includes $p$-values quantifying the level of agreement between the data
and the predictions, including all sources of uncertainty. The
systematic uncertainties of the data are shown separately.
\label{fig:Unfoldedcosthetastar_NjetUnblindedInLH0j1jFR}
}
\end{figure}
 
\begin{figure}[tbp]
\centering
\includegraphics[width=.8\textwidth]{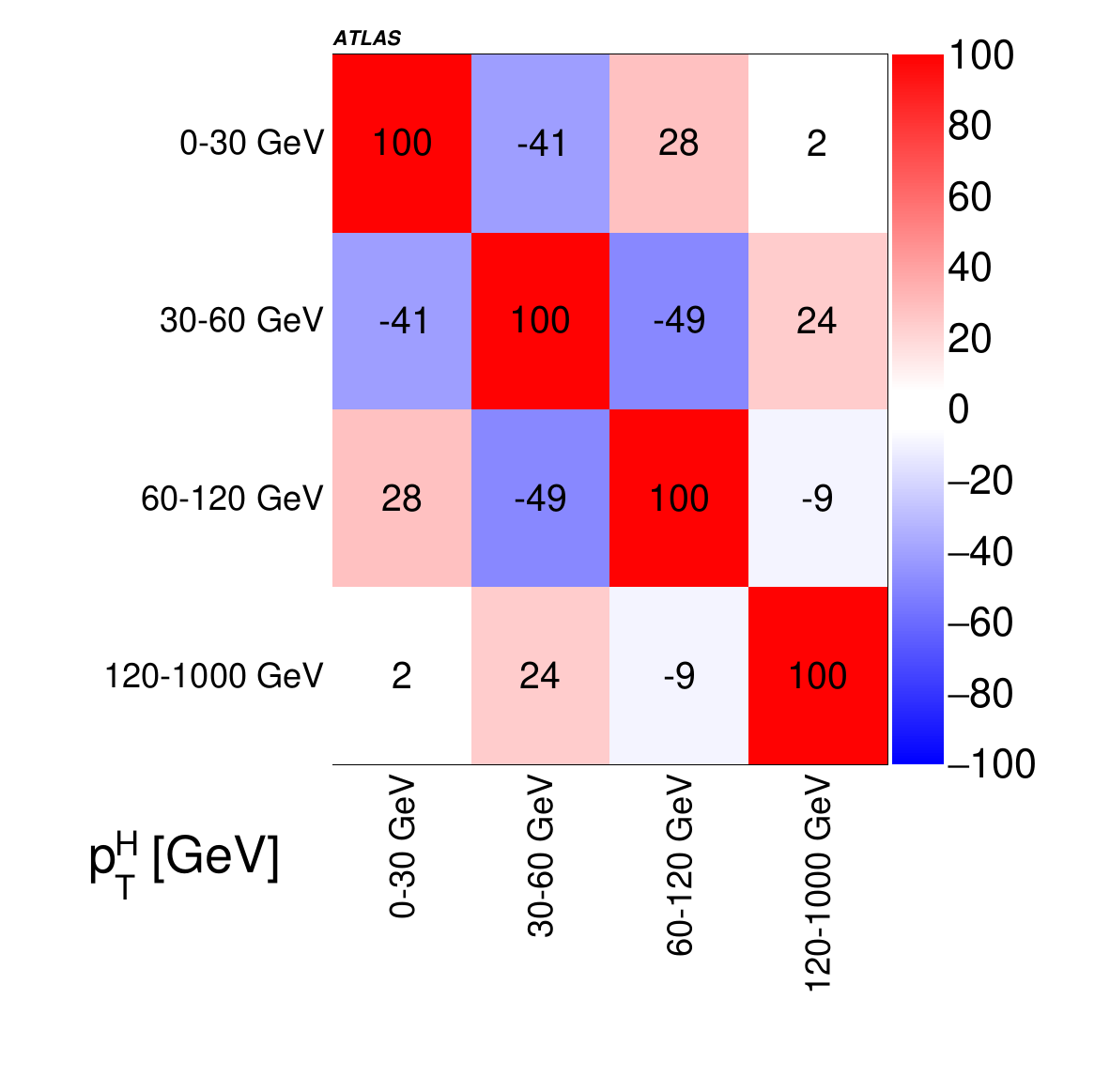}
\caption{Post-fit correlations of the observed cross sections for the bins of transverse momentum \ptH. The legend on the right shows numbers in percentages.
\label{correlations_extrasmall}
}
\end{figure}

\clearpage

\section{Conclusion}
\label{sec:conclusion}
 
Measuring the Higgs boson's differential production cross section is an important aspect
of measuring Higgs properties and further testing the Standard
Model. This analysis has measured single- and double-differential cross sections for Higgs boson production via gluon--gluon fusion and decay into $WW^{\ast} \to e\nu\mu\nu$ in bins of the final-state
transverse mass, \mT{}. The measurement was performed using the full LHC \RunTwo dataset of 13~\TeV\ proton--proton collisions collected with the ATLAS detector during 2015--2018, corresponding to an integrated luminosity of 139~\ifb. The resulting \mT{} distributions and measurements of the
differential cross sections in fiducial regions are presented for the $|y_{j0}|$, $\ptH$,
$\pt^{\ell0}$, \ptll, \mll, \Yll, \dphill, and \costhetastar
observables. Likelihood unfolding with Tikhonov regularisation is used to transform the reconstruction-level quantities to their particle-level distributions in each of these eight observables, and in each of the last six of them versus jet multiplicity.  Performing the measurement at the particle level facilitates direct comparison with theoretical predictions and minimizes the impact of the signal uncertainties on the final results. The results agree extremely well with Standard Model expectations, derived using the \POWPY[8], \POWHER[7] and \MGNLO Monte Carlo generators. The leading uncertainties are related to jet and muon reconstruction, theoretical modelling of the $WW$ and $V\gamma$ backgrounds, and data-driven background estimates for misidentified objects.  The results improve upon those previously obtained by ATLAS, mainly by using more data and analysing a larger suite of observables. In addition, these results were obtained by using a fitting procedure, unlike the previous version of the analysis, which relied on a simple subtraction of the expected background event yields from the observed event yields.
 
\section*{Acknowledgements}


We thank CERN for the very successful operation of the LHC, as well as the
support staff from our institutions without whom ATLAS could not be
operated efficiently.
 
We acknowledge the support of
ANPCyT, Argentina;
YerPhI, Armenia;
ARC, Australia;
BMWFW and FWF, Austria;
ANAS, Azerbaijan;
CNPq and FAPESP, Brazil;
NSERC, NRC and CFI, Canada;
CERN;
ANID, Chile;
CAS, MOST and NSFC, China;
Minciencias, Colombia;
MEYS CR, Czech Republic;
DNRF and DNSRC, Denmark;
IN2P3-CNRS and CEA-DRF/IRFU, France;
SRNSFG, Georgia;
BMBF, HGF and MPG, Germany;
GSRI, Greece;
RGC and Hong Kong SAR, China;
ISF and Benoziyo Center, Israel;
INFN, Italy;
MEXT and JSPS, Japan;
CNRST, Morocco;
NWO, Netherlands;
RCN, Norway;
MEiN, Poland;
FCT, Portugal;
MNE/IFA, Romania;
MESTD, Serbia;
MSSR, Slovakia;
ARRS and MIZ\v{S}, Slovenia;
DSI/NRF, South Africa;
MICINN, Spain;
SRC and Wallenberg Foundation, Sweden;
SERI, SNSF and Cantons of Bern and Geneva, Switzerland;
MOST, Taiwan;
TENMAK, T\"urkiye;
STFC, United Kingdom;
DOE and NSF, United States of America.
In addition, individual groups and members have received support from
BCKDF, CANARIE, Compute Canada and CRC, Canada;
PRIMUS 21/SCI/017 and UNCE SCI/013, Czech Republic;
COST, ERC, ERDF, Horizon 2020 and Marie Sk{\l}odowska-Curie Actions, European Union;
Investissements d'Avenir Labex, Investissements d'Avenir Idex and ANR, France;
DFG and AvH Foundation, Germany;
Herakleitos, Thales and Aristeia programmes co-financed by EU-ESF and the Greek NSRF, Greece;
BSF-NSF and MINERVA, Israel;
Norwegian Financial Mechanism 2014-2021, Norway;
NCN and NAWA, Poland;
La Caixa Banking Foundation, CERCA Programme Generalitat de Catalunya and PROMETEO and GenT Programmes Generalitat Valenciana, Spain;
G\"{o}ran Gustafssons Stiftelse, Sweden;
The Royal Society and Leverhulme Trust, United Kingdom.
 
The crucial computing support from all WLCG partners is acknowledged gratefully, in particular from CERN, the ATLAS Tier-1 facilities at TRIUMF (Canada), NDGF (Denmark, Norway, Sweden), CC-IN2P3 (France), KIT/GridKA (Germany), INFN-CNAF (Italy), NL-T1 (Netherlands), PIC (Spain), ASGC (Taiwan), RAL (UK) and BNL (USA), the Tier-2 facilities worldwide and large non-WLCG resource providers. Major contributors of computing resources are listed in Ref.~\cite{ATL-SOFT-PUB-2021-003}.


\clearpage
\appendix

\part*{Appendix: iterative Bayesian unfolding}
\addcontentsline{toc}{part}{Appendix: iterative Bayesian unfolding}

\label{sec:IBU}

A secondary unfolding method is used as a cross-check, in which the unfolded truth distribution is obtained iteratively using Bayes' theorem. Bayes' theorem can be used to estimate the probability of a specific cause, C, given an effect, E, using the equation:
 
\begin{equation*}
P(C_i|E_j,I) = \frac{P(E_j|C_i,I) \cdot P(C_i|I)}{\sum_{k=1}^MP(E_j|C_k,I) \cdot P(C_k|I)}
\end{equation*}
 
where $I$ is our prior knowledge of probabilities of causes $C_i$. If $n_j$ events are measured in bin $j$, the number of truth events in bin $i$, is then given by
 
\begin{equation*}
P(\mu_i|n_j) = \frac{P(C_i|E_j, I) \cdot n_j}{\epsilon_i}
\end{equation*}
 
where the efficiency, $\epsilon_i$, is defined as
 
\begin{equation*}
\epsilon_i = \sum_{j=1} P(E_j|C_i, I).
\end{equation*}
 
The number of truth events per bin is measured iteratively, and at each iteration the prior distribution $P(C_i, I)$, becomes the measured distribution.
The number of iterations is optimized using a procedure intended to prevent statistical fluctuations from being interpreted as features in the true distribution. As the number of iterations increases, the bias towards the truth distribution is reduced, but the statistical uncertainty is increased. To find a balance between these two factors, two-dimensional plots of \enquote{fluctuation} versus \enquote{bias} are constructed. Distributions are constructed by fluctuating the number of events in each truth bin by its respective Poisson uncertainty, and then folding it using the nominal response matrix. The true values in the fluctuated, folded distribution are intended to represent the measured distribution. These distributions are then unfolded, using the nominal response matrix. The fluctuation axis is then defined by the relation: fluctuated truth distribution minus the original truth distribution divided by the original truth distribution. The bias axis is defined by the relation: unfolded truth distribution minus the fluctuated truth distribution divided by the fluctuated truth distribution. The bias is calculated at the points where the fluctuation equals the uncertainty in the two-dimensional fluctuation-versus-bias distributions. Since this intersection occurs twice (when the fluctuation equals the negative uncertainty and when the fluctuation equals the positive uncertainty), the mean of these two values is taken as the bias. Finally, for each observable, the number of iterations is chosen such that the ratio of the bias to the uncertainty is less than a certain threshold for all bins. The number of iterations chosen varies from 2 to 18 depending on the observable.
 
This appendix contains the unfolded distributions of different observables binned to match the final binnings employed in the analysis. The differential cross sections shown allow a comparison of the results from the two unfolding procedures. The two unfolding procedures produce results that are consistent with each other, demonstrating that the choice of regularization which is different between the two methods, does not have an impact on the results.

The differential fiducial cross sections for each observable, obtained using the iterative Bayesian unfolding method, are shown in
Figures~\ref{fig:UnfoldedPtHUnblindedIBU0j1jFR} to \ref{fig:UnfoldedYJ0UnblindedIBU0j1jFR} for the single-differential distributions, and
Figures~\ref{fig:UnfoldedDPhill_NjetUnblindedIBU0j1jFR} to \ref{fig:Unfoldedcosthetastar_NjetUnblindedIBU0j1jFR} for the double-differential distributions.

\begin{figure}[tbp]
\centering
\includegraphics[width=0.8\textwidth]{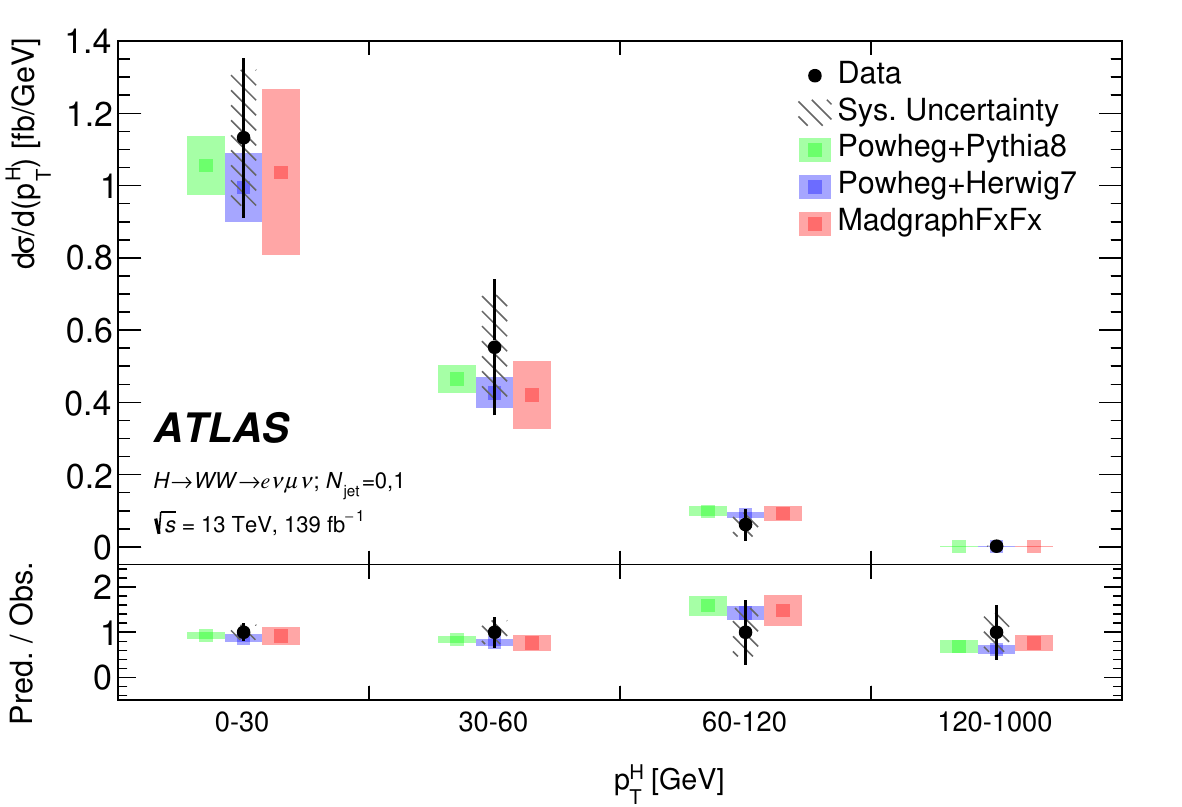}
\caption{Differential fiducial cross section for $\ptH$ in the 0+1-jet
fiducial region using the iterative Bayesian unfolding method.
Uncertainty bars on the data points include statistical and systematic
uncertainties from experimental and theory sources as well as
background normalization effects and shape effects from background
and signal. Uncertainty bands on the predictions shown are dominated by
normalization effects on the signal arising from showering, PDF
models, \alphas and the QCD scale. The systematic uncertainties of
the data are shown separately.
\label{fig:UnfoldedPtHUnblindedIBU0j1jFR}}
\end{figure}
 
\begin{figure}[tbp]
\centering
\includegraphics[width=0.49\textwidth]{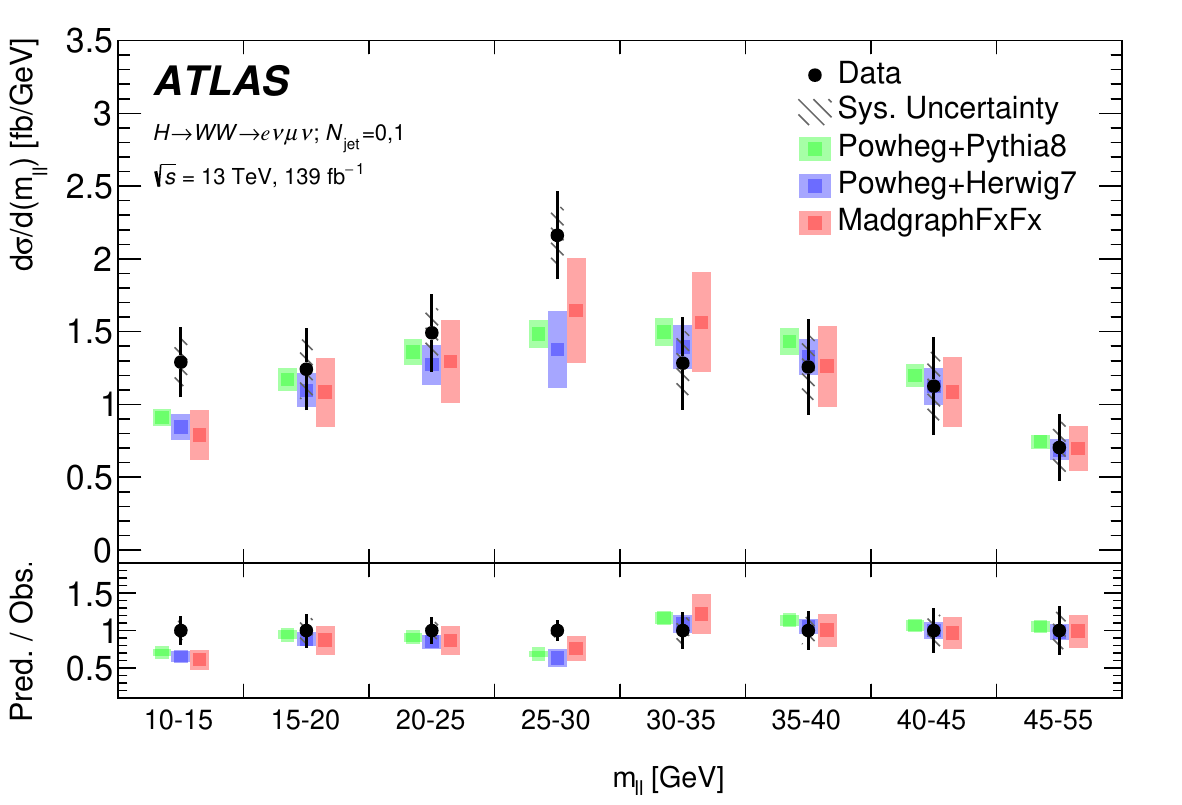}
\includegraphics[width=0.49\textwidth]{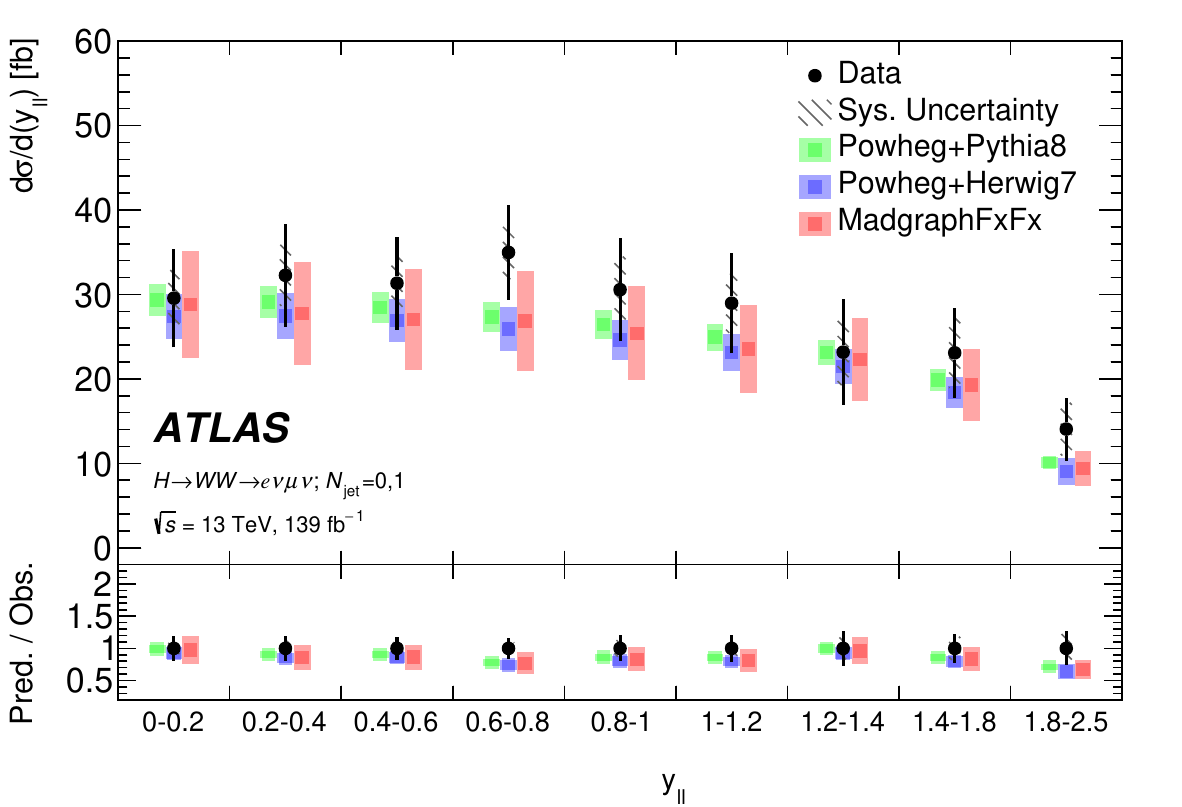}
\caption{Differential fiducial cross section for \mll (left) and
\Yll (right) in the 0+1-jet fiducial region using the iterative
Bayesian unfolding method.  Uncertainty bars on the data points
include statistical and systematic uncertainties from experimental
and theory sources as well as background normalization effects and
shape effects from background and signal. Uncertainty bands on the
predictions shown are dominated by normalization effects on the
signal arising from showering, PDF models, \alphas and the QCD
scale. The systematic uncertainties of the data are shown
separately.
\label{fig:UnfoldedMllUnblindedIBU0j1jFR}
}
\end{figure}
 
\begin{figure}[tbp]
\centering
\includegraphics[width=0.49\textwidth]{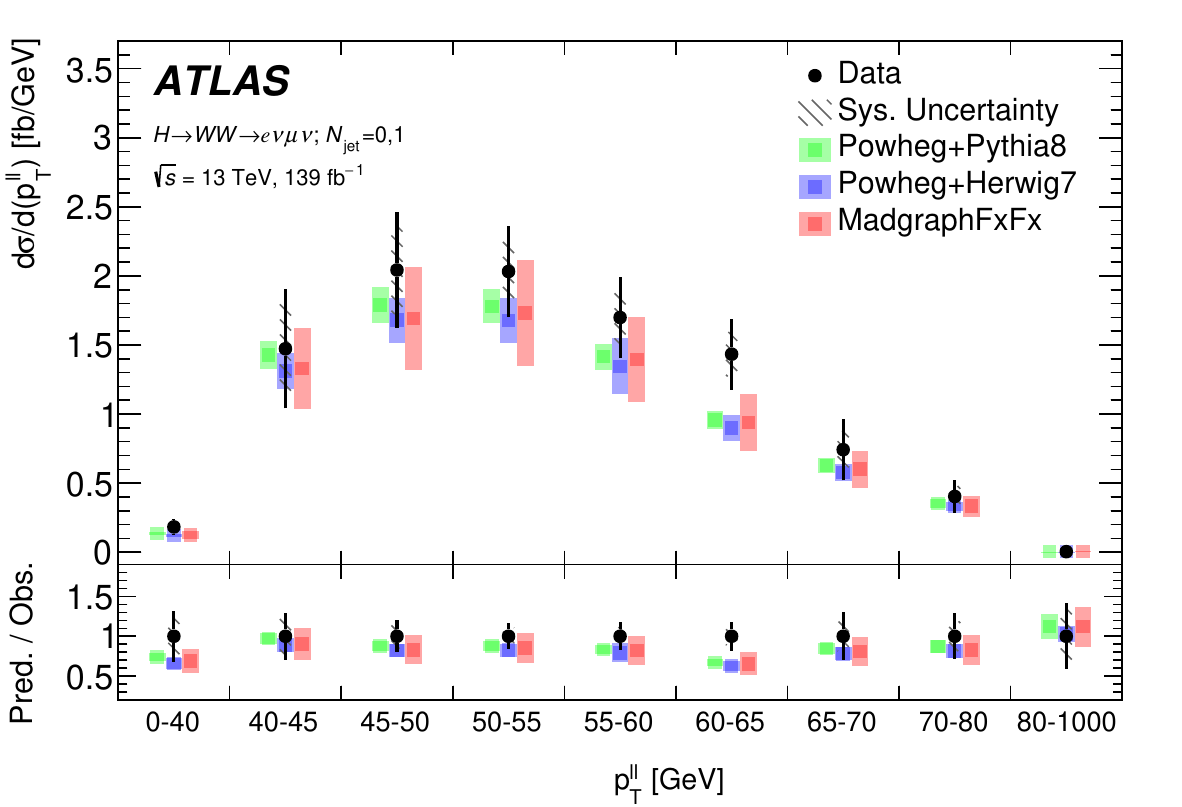}
\includegraphics[width=0.49\textwidth]{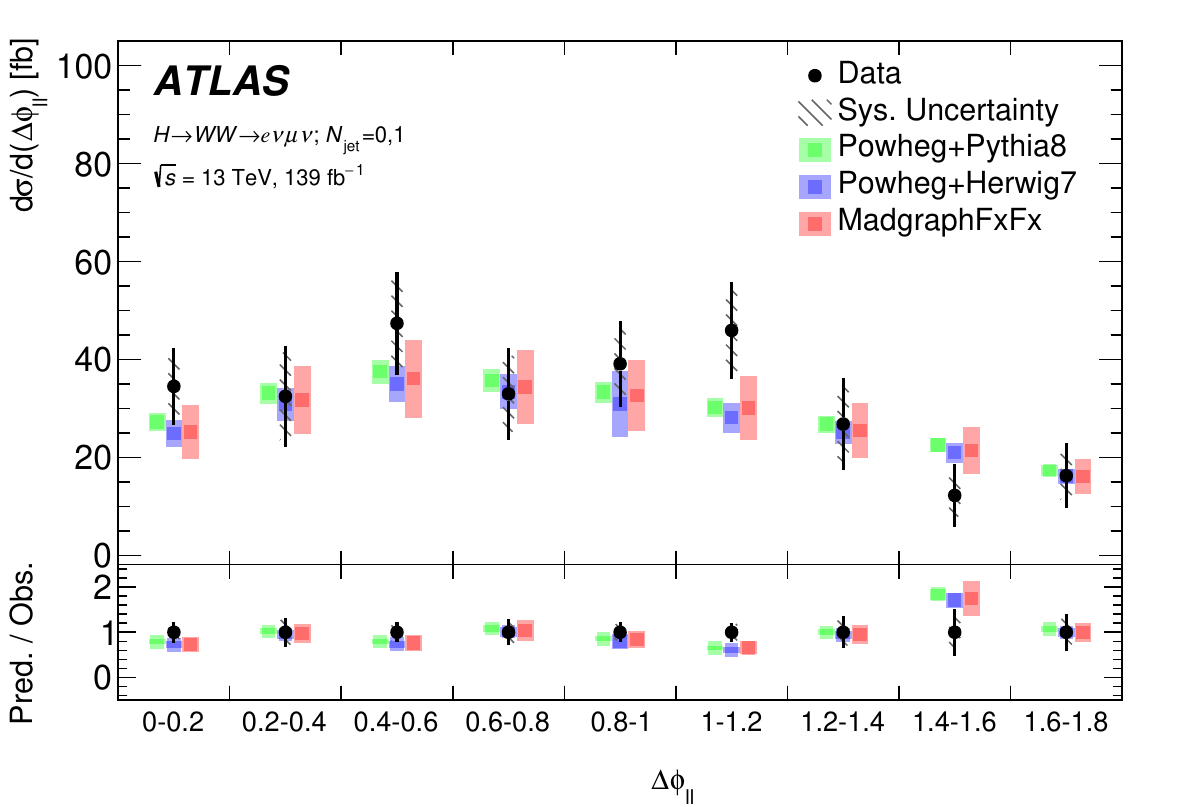}
\caption{Differential fiducial cross section for \ptll (left)
and \dphill (right) in the 0+1-jet fiducial region using the
iterative bayesian unfolding method.  Uncertainty bars on the data
points include statistical and systematic uncertainties from
experimental and theory sources as well as background
normalization effects and shape effects from background and
signal. Uncertainty bands on the predictions shown are dominated by
normalization effects on the signal arising from showering, PDF
models, \alphas and the QCD scale. The systematic uncertainties
of the data are shown separately.
\label{fig:UnfoldedPtllUnblindedIBU0j1jFR}
}
\end{figure}
 
\begin{figure}[tbp]
\centering
\includegraphics[width=0.49\textwidth]{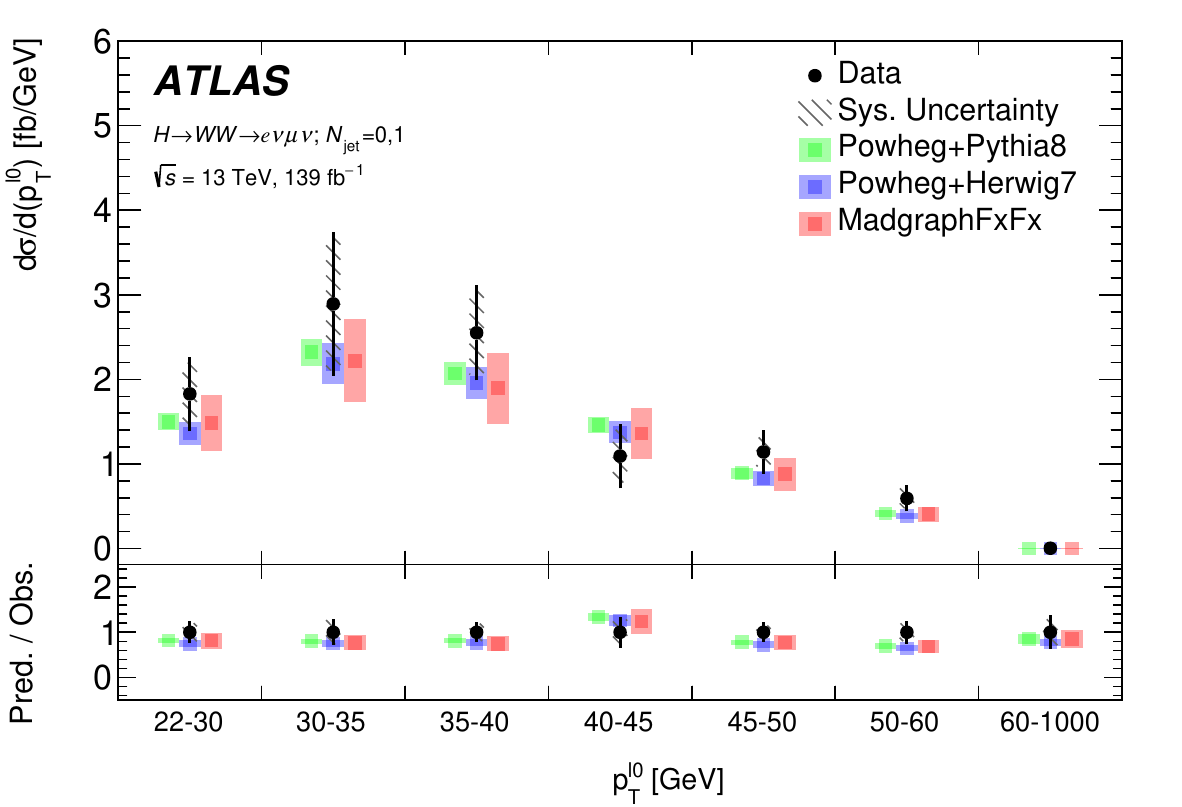}
\includegraphics[width=0.49\textwidth]{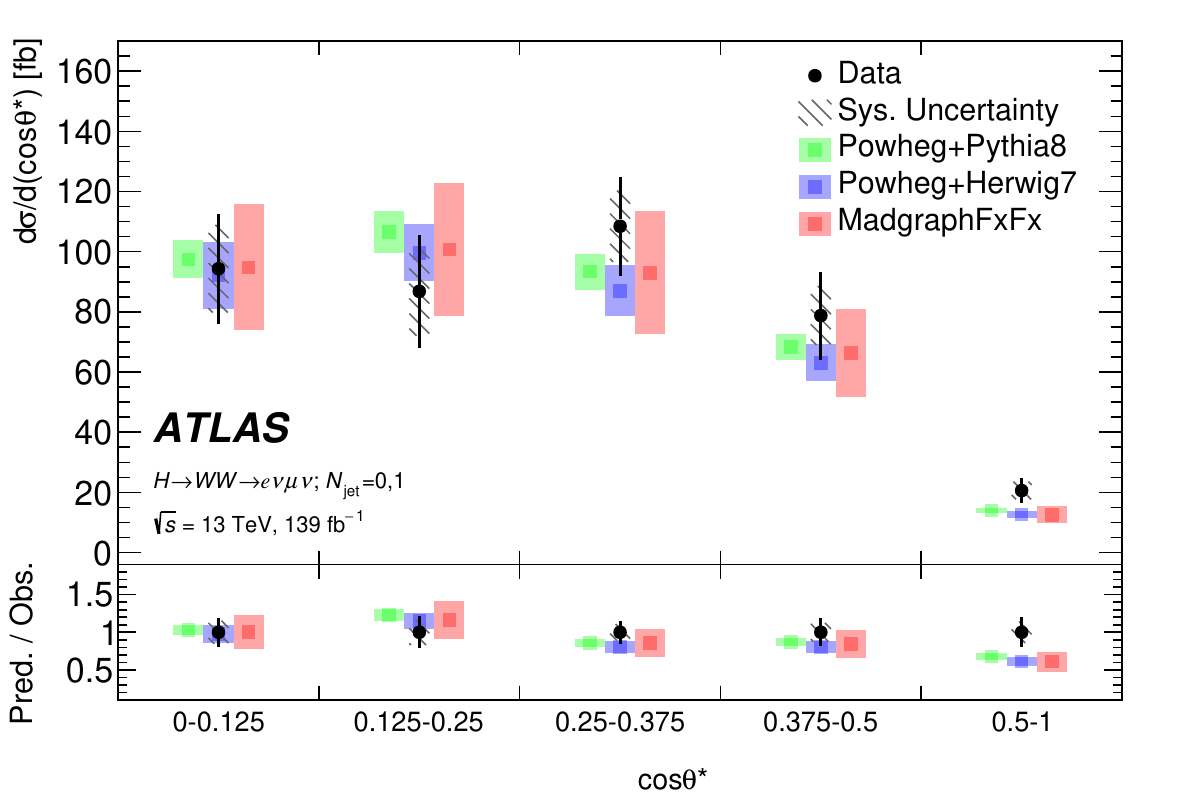}
\caption{Differential fiducial cross section for $\pt^{\ell 0}$ (left)
and $\costhetastar$ (right) in the 0+1-jet fiducial region using
the iterative Bayesian unfolding method.  Uncertainty bars on the
data points include statistical and systematic uncertainties
from experimental and theory sources as well as background
normalization effects and shape effects from background and
signal. Uncertainty bands on the predictions shown are dominated by
normalization effects on the signal arising from showering, PDF
models, \alphas and the QCD scale. The systematic uncertainties
of the data are shown separately.
\label{fig:UnfoldedcosthetastarUnblindedIBU0j1jFR}
}
\end{figure}
 
\begin{figure}[tbp]
\centering
\includegraphics[width=0.49\textwidth]{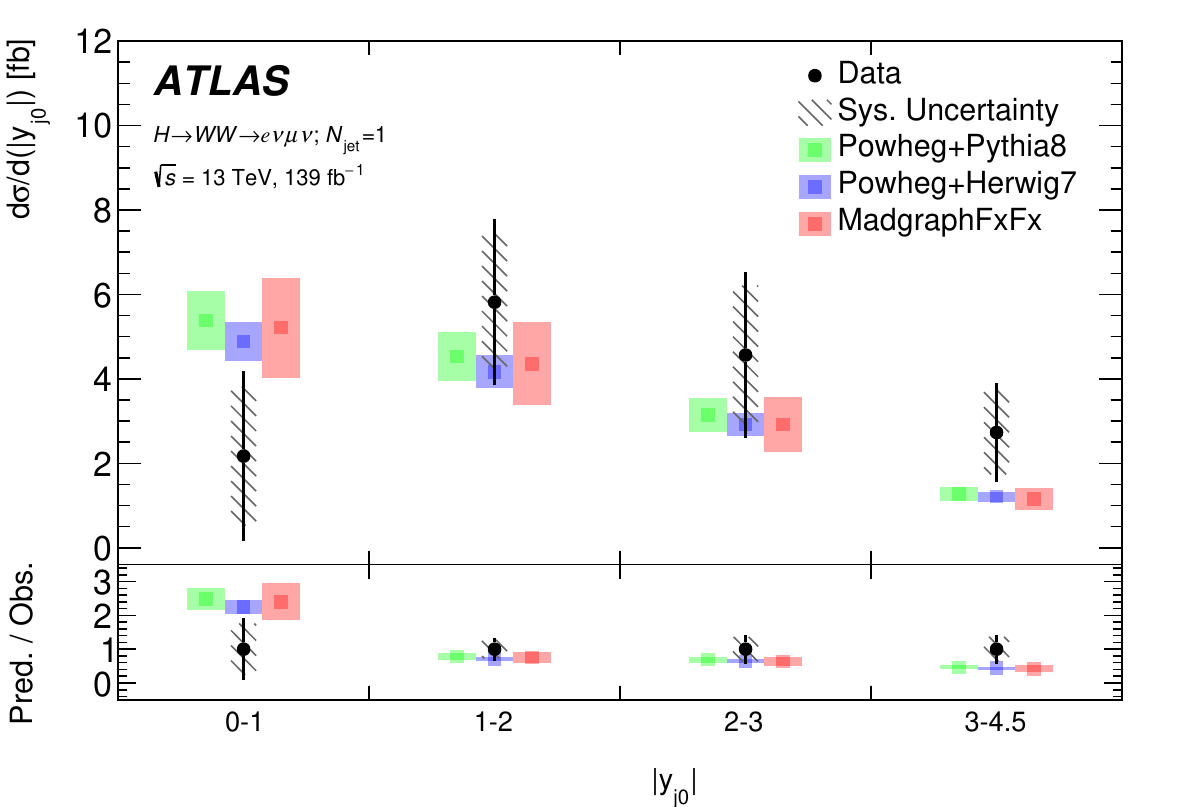}
\caption{Differential fiducial cross section for $|y_{j0}|$ in the 0+1-jet fiducial region using the
iterative Bayesian unfolding method.  Uncertainty bars on the data
points include statistical and systematic uncertainties from
experimental and theory sources as well as background normalization
effects and shape effects from background and signal. Uncertainty
bands on the predictions shown are dominated by normalization effects on
the signal arising from showering, PDF models, \alphas and the QCD
scale. The systematic uncertainties of the data are shown
separately.
\label{fig:UnfoldedYJ0UnblindedIBU0j1jFR}
}
\end{figure}
 
\begin{figure}[tbp]
\centering
\includegraphics[width=0.49\textwidth]{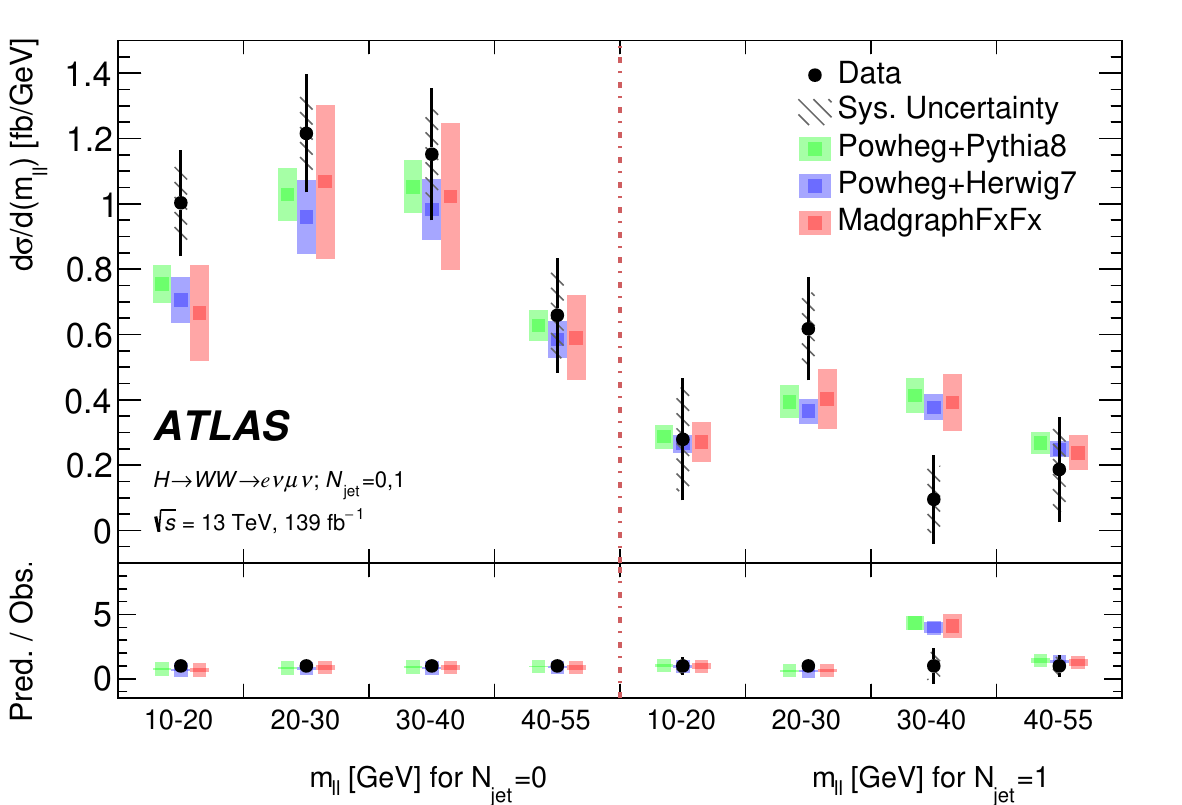}
\includegraphics[width=0.49\textwidth]{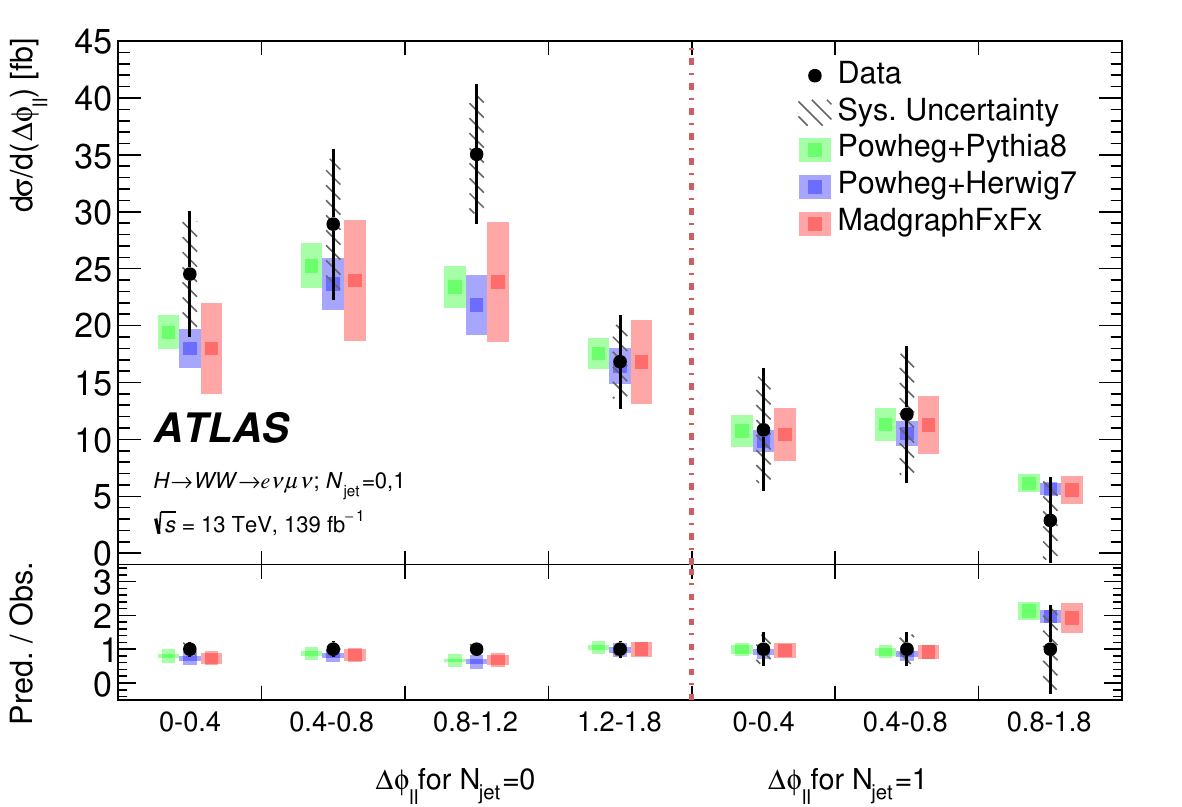}
\caption{Differential fiducial cross section for \mll (left) and
\dphill (right) versus~\Njet in the 0-jet and 1-jet fiducial
regions using the iterative Bayesian unfolding method.
Uncertainty bars on the data points include statistical and
systematic uncertainties from experimental and theory sources as
well as background normalization effects and shape effects from
background and signal. Uncertainty bands on the predictions shown are
dominated by normalization effects on the signal arising from
showering, PDF models, \alphas and the QCD scale. The systematic
uncertainties of the data are shown separately.
\label{fig:UnfoldedDPhill_NjetUnblindedIBU0j1jFR}
}
\end{figure}
 
\begin{figure}[tbp]
\centering
\includegraphics[width=0.49\textwidth]{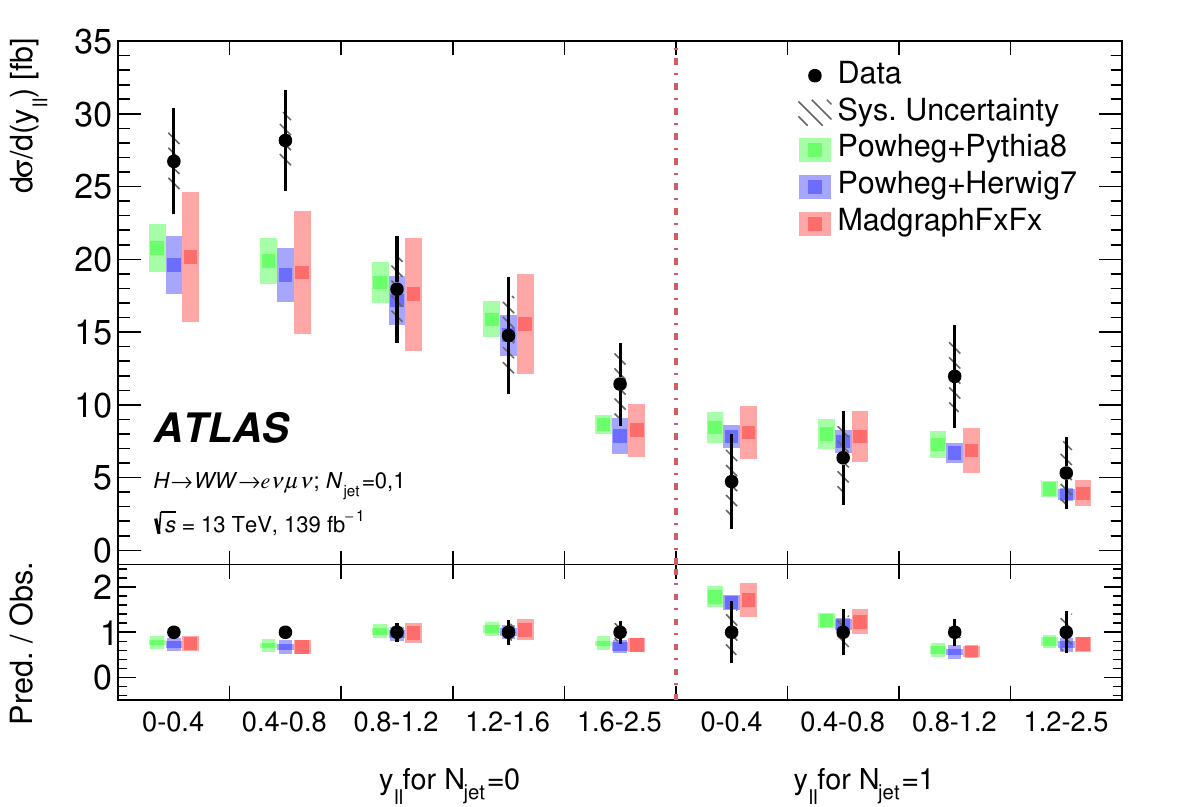}
\includegraphics[width=0.49\textwidth]{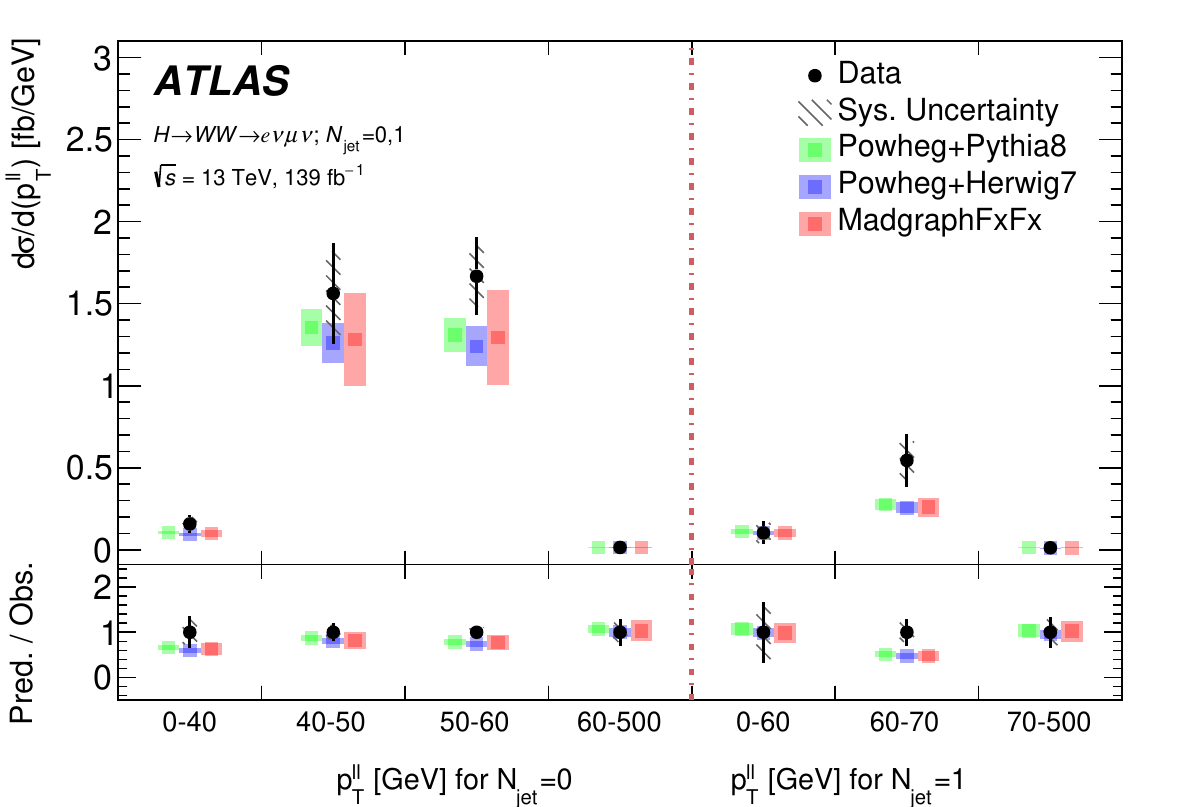}
\caption{Differential fiducial cross section for \Yll (left) and
\ptll (right) versus~\Njet in the 0-jet and 1-jet fiducial
regions using the iterative Bayesian unfolding method.
Uncertainty bars on the data points include statistical and systematic
uncertainties from experimental and theory sources as well as
background normalization effects and shape effects from background
and signal. Uncertainty bands on the predictions shown are dominated by
normalization effects on the signal arising from showering, PDF
models, \alphas and the QCD scale. The systematic uncertainties of
the data are shown separately.
\label{fig:UnfoldedPtll_NjetUnblindedIBU0j1jFR}
}
\end{figure}

\begin{figure}[tbp]
\centering
\includegraphics[width=0.49\textwidth]{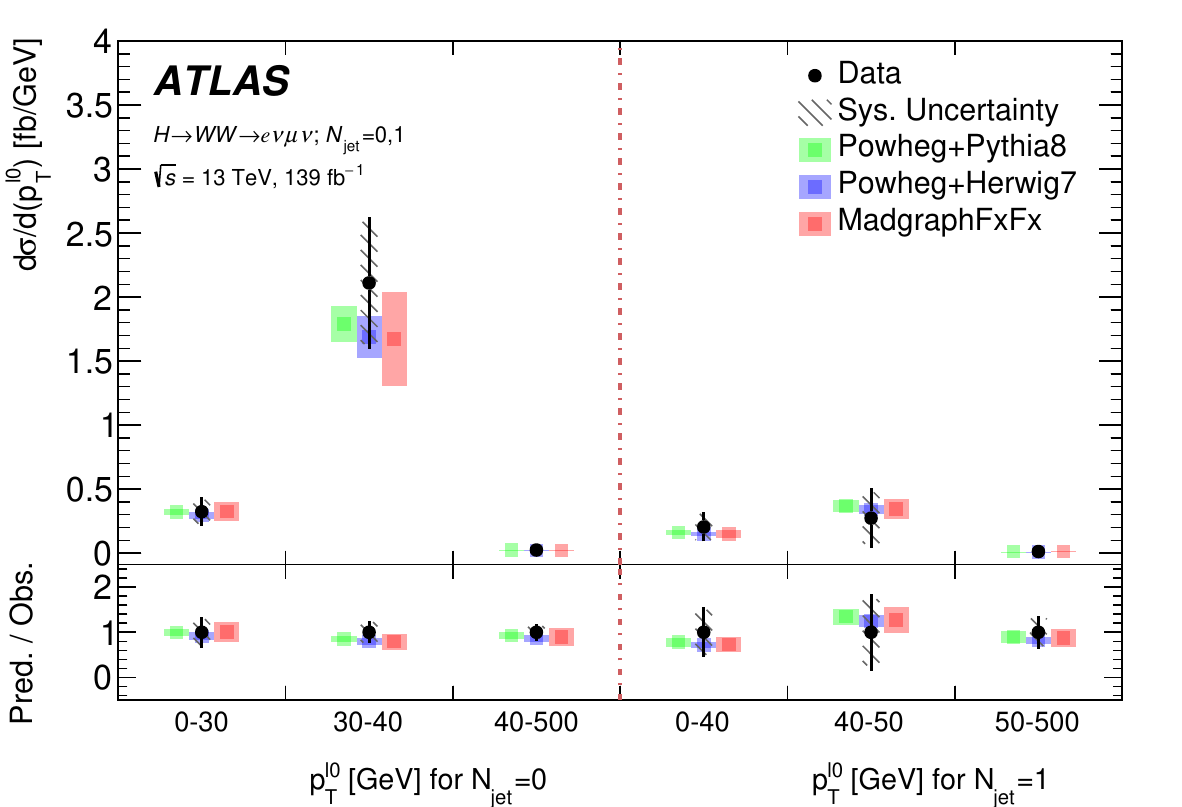}
\includegraphics[width=0.49\textwidth]{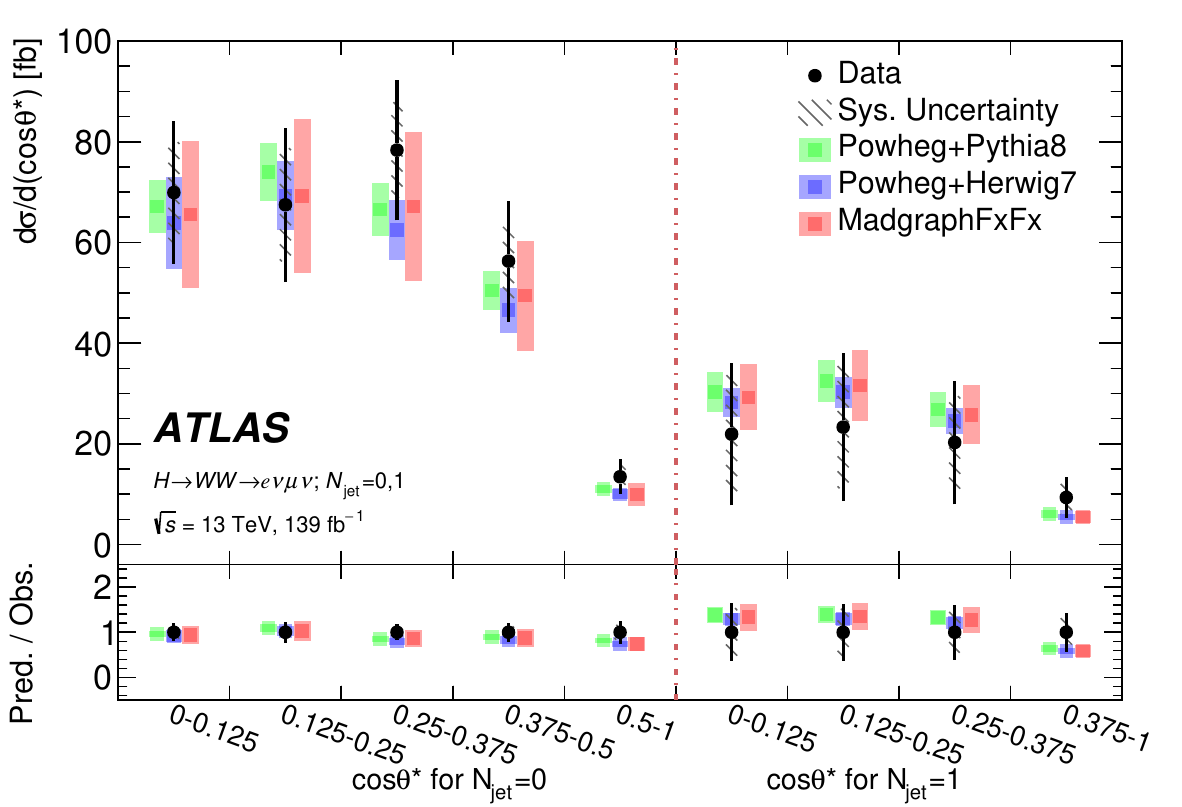}
\caption{Differential fiducial cross section for $\pt^{\ell 0}$ (left)
and $\costhetastar$ (right) versus~\Njet in the 0-jet and 1-jet
fiducial regions using the iterative Bayesian unfolding method.
Uncertainty bars on the data points include statistical and systematic
uncertainties from experimental and theory sources as well as
background normalization effects and shape effects from background
and signal. Uncertainty bands on the predictions shown are dominated by
normalization effects on the signal arising from showering, PDF
models, \alphas and the QCD scale. The systematic uncertainties of
the data are shown separately.
\label{fig:Unfoldedcosthetastar_NjetUnblindedIBU0j1jFR}}
\end{figure}


\clearpage

\printbibliography

\clearpage

\clearpage
 
 
\begin{flushleft}
\hypersetup{urlcolor=black}
{\Large The ATLAS Collaboration}

\bigskip

\AtlasOrcid[0000-0002-6665-4934]{G.~Aad}$^\textrm{\scriptsize 102}$,
\AtlasOrcid[0000-0002-5888-2734]{B.~Abbott}$^\textrm{\scriptsize 120}$,
\AtlasOrcid[0000-0002-1002-1652]{K.~Abeling}$^\textrm{\scriptsize 55}$,
\AtlasOrcid[0000-0002-8496-9294]{S.H.~Abidi}$^\textrm{\scriptsize 29}$,
\AtlasOrcid[0000-0002-9987-2292]{A.~Aboulhorma}$^\textrm{\scriptsize 35e}$,
\AtlasOrcid[0000-0001-5329-6640]{H.~Abramowicz}$^\textrm{\scriptsize 151}$,
\AtlasOrcid[0000-0002-1599-2896]{H.~Abreu}$^\textrm{\scriptsize 150}$,
\AtlasOrcid[0000-0003-0403-3697]{Y.~Abulaiti}$^\textrm{\scriptsize 117}$,
\AtlasOrcid[0000-0003-0762-7204]{A.C.~Abusleme~Hoffman}$^\textrm{\scriptsize 137a}$,
\AtlasOrcid[0000-0002-8588-9157]{B.S.~Acharya}$^\textrm{\scriptsize 69a,69b,p}$
% ...
\bigskip
\\

% ...
$^{29}$Physics Department, Brookhaven National Laboratory, Upton NY; United States of America.\\
% ...
$^{35}$$^{(a)}$Facult\'e des Sciences Ain Chock, R\'eseau Universitaire de Physique des Hautes Energies - Universit\'e Hassan II, Casablanca;$^{(b)}$Facult\'{e} des Sciences, Universit\'{e} Ibn-Tofail, K\'{e}nitra;$^{(c)}$Facult\'e des Sciences Semlalia, Universit\'e Cadi Ayyad, LPHEA-Marrakech;$^{(d)}$LPMR, Facult\'e des Sciences, Universit\'e Mohamed Premier, Oujda;$^{(e)}$Facult\'e des sciences, Universit\'e Mohammed V, Rabat;$^{(f)}$Institute of Applied Physics, Mohammed VI Polytechnic University, Ben Guerir; Morocco.\\
% ...
$^{55}$II. Physikalisches Institut, Georg-August-Universit\"{a}t G\"ottingen, G\"ottingen; Germany.\\
% ...
$^{69}$$^{(a)}$INFN Gruppo Collegato di Udine, Sezione di Trieste, Udine;$^{(b)}$ICTP, Trieste;$^{(c)}$Dipartimento Politecnico di Ingegneria e Architettura, Universit\`a di Udine, Udine; Italy.\\
% ...
$^{102}$CPPM, Aix-Marseille Universit\'e, CNRS/IN2P3, Marseille; France.\\
% ...
$^{117}$Department of Physics, New York University, New York NY; United States of America.\\
% ...
$^{120}$Homer L. Dodge Department of Physics and Astronomy, University of Oklahoma, Norman OK; United States of America.\\
% ...
$^{137}$$^{(a)}$Departamento de F\'isica, Pontificia Universidad Cat\'olica de Chile, Santiago;$^{(b)}$Millennium Institute for Subatomic physics at high energy frontier (SAPHIR), Santiago;$^{(c)}$Instituto de Investigaci\'on Multidisciplinario en Ciencia y Tecnolog\'ia, y Departamento de F\'isica, Universidad de La Serena;$^{(d)}$Universidad Andres Bello, Department of Physics, Santiago;$^{(e)}$Instituto de Alta Investigaci\'on, Universidad de Tarapac\'a, Arica;$^{(f)}$Departamento de F\'isica, Universidad T\'ecnica Federico Santa Mar\'ia, Valpara\'iso; Chile.\\
% ...
$^{150}$Department of Physics, Technion, Israel Institute of Technology, Haifa; Israel.\\
$^{151}$Raymond and Beverly Sackler School of Physics and Astronomy, Tel Aviv University, Tel Aviv; Israel.\\
% ...
$^{a}$ Also Affiliated with an institute covered by a cooperation agreement with CERN.\\
$^{b}$ Also at An-Najah National University, Nablus; Palestine.\\
$^{c}$ Also at Borough of Manhattan Community College, City University of New York, New York NY; United States of America.\\
$^{d}$ Also at Bruno Kessler Foundation, Trento; Italy.\\
$^{e}$ Also at Center for High Energy Physics, Peking University; China.\\
$^{f}$ Also at Center for Interdisciplinary Research and Innovation (CIRI-AUTH), Thessaloniki ; Greece.\\
% ...
$^{p}$ Also at Department of Physics, King's College London, London; United Kingdom.\\
% ...

\end{flushleft}


\end{document}